\documentclass[fleqn,usenatbib]{mnras}

\usepackage[T1]{fontenc}
\usepackage{ae,aecompl}

\DeclareRobustCommand{\VAN}[3]{#2}
\let\VANthebibliography\thebibliography
\def\thebibliography{\DeclareRobustCommand{\VAN}[3]{##3}\VANthebibliography}

\usepackage{graphicx}	
\usepackage{amsmath}	

\usepackage{epstopdf}
\usepackage{adjustbox}
\usepackage{textcomp, gensymb}
\usepackage{subcaption} 
\usepackage{tabu}
\usepackage{float}
\usepackage{hyperref}
\usepackage{rotfloat}
\usepackage{rotating}
\usepackage{multirow}
\usepackage{multicol}
\usepackage{pdflscape} 
\usepackage{breqn} 
\usepackage[flushleft]{threeparttable}
\usepackage{orcidlink}
\usepackage{array}
\usepackage{amsmath}
\def\farcs{\hbox{$.\!\!^{\prime\prime}$}}

\newcommand{\plus}{\scalebox{1.1}{+}}
\newcommand{\ie}{i.e.,}

\title[Short title, max. 45 characters]{Spectral Ages of Remnant Radio Galaxies}

\title{SuperMIGHTEE : Spectral Ages of Remnant Radio Galaxy Candidates in the XMM-LSS Field}

\author[Dutta et al.]
{Sushant Dutta$^{\orcidlink{0000-0002-6542-2939}}$ $^{1,2}$\thanks{E-mail: \url{sushant.dutta@uct.ac.za, sushant@idia.ac.za}},
	Veeresh Singh$^{\orcidlink{0000-0002-6040-4993}}$ $^{2}$,
	C.H. Ishwara Chandra$^{\orcidlink{0000-0001-5356-1221}}$ $^{3}$,
	Yogesh Wadadekar$^{\orcidlink{0000-0002-1345-7371}}$ $^{3}$,
	\newauthor
	Russ Taylor$^{\orcidlink{0000-0001-9885-0676}}$ $^{1,4}$,
	Mattia Vaccari$^{\orcidlink{0000-0002-6748-0577}}$ $^{1,4,5}$,
	Lucia Marchetti$^{\orcidlink{0000-0003-3948-7621}}$ $^{1,5,6}$,
	Matt Jarvis$^{\orcidlink{0000-0001-7039-9078}}$ $^{4,8}$,
	Catherine Hale$^{\orcidlink{0000-0001-6279-4772}}$ $^{7,8}$,
	\newauthor
	Solohery Randriamampandry$^{\orcidlink{0000-0001-5373-6669}}$ $^{9}$,
	and Zara Randriamanakoto$^{\orcidlink{0000-0003-2666-4158}}$ $^{9,10}$
	\\ \\
	$^{1}$The Inter-University Institute for Data Intensive Astronomy (IDIA), and University of Cape Town, Private Bag X3, Rondebosch, 7701, \\ South Africa \\
	$^{2}$Physical Research Laboratory, Navrangpura, Ahmedabad, Gujarat-380009, India \\
	$^{3}$National Centre for Radio Astrophysics, TIFR, Post Bag 3, Ganeshkhind, Pune 411007, India \\
	$^{4}$Department of Physics and Astronomy, University of the Western Cape, Robert Sobukwe Road, 7535 Bellville, Cape Town, South Africa \\
	$^{5}$INAF - Istituto di Radioastronomia, via Gobetti 101, 40129 Bologna, Italy \\
	$^{6}$Department of Astronomy, University of Cape Town, 7701 Rondebosch, Cape Town, South Africa \\
	$^{7}$School of Physics and Astronomy, Institute for Astronomy, University of Edinburgh, Royal Observatory, Blackford Hill, \\ EH9 3HJ Edinburgh, UK \\
	$^{8}$Department of Physics, Astrophysics, University of Oxford, Denys Wilkinson Building, Keble Road, Oxford OX1 3RH, UK \\
	$^{9}$A\&A, Department of Physics, Faculty of Sciences, University of Antananarivo, P.O. Box 906, Antananarivo 101, Madagascar \\
	$^{10}$South African Astronomical Observatory, P.O. Box 9, Observatory 7935, South Africa \\
}

\date{Accepted XXX. Received YYY; in original form ZZZ}

\pubyear{2023}

\begin{document}
	\label{firstpage}
	\pagerange{\pageref{firstpage}--\pageref{lastpage}}
	\maketitle
	
	\begin{abstract}
		Remnant radio galaxies, whose lobes are no longer replenished by jets from the active galactic nucleus (AGN), offer key constraints on AGN duty cycles and the timescales of radio jets. We present a spectral-ageing study of 14 candidate remnant radio galaxies in the XMM$-$LSS field, combining new broad-band data from the MeerKAT MIGHTEE ($L-$band) and uGMRT superMIGHTEE (band-3 and band-4) surveys with complementary observations from LOFAR, GMRT, and JVLA, covering 144 MHz$-$1.5 GHz. Spectral modeling confirms 12 sources as genuine remnants, while two are reclassified as active, emphasising the importance of sensitive, multi-frequency coverage for robust remnant identification. Pixel-based spectral age maps yield results ($\sim$3-43 Myr) broadly consistent with integrated estimates, revealing relatively short spectral ages ($\sim$8-42 Myr). These ages likely reflect enhanced inverse-Compton losses at higher redshifts (0.35 $\leq$ $z$ $\leq$ 2.85; median $z$ = 1.25) and possible rapid lobe expansion in low-density environments. The ratios of remnant to total source ages (t$_{\rm OFF}$/t$_{\rm s}$) span 0.04-0.83, indicating that the sample traces a broad range of evolutionary stages. Our findings reveal a previously underrepresented population of faint, rapidly fading remnants, suggesting that the remnant phase may be shorter and more dynamic than previously thought. This study highlights the crucial role of MIGHTEE and superMIGHTEE surveys in reliably classifying genuine remnants and provides a framework for constraining AGN life cycles in preparation for forthcoming SKA surveys.
	\end{abstract}
	
	
	
	
	\section{Introduction}
	\label{sec:intro}
	Radio galaxies, as a subclass of Active Galactic Nuclei (AGN), 
	are characterized by distinctive radio morphologies that consist of a central AGN (radio core), a pair of relativistic bipolar jets emanating from the core and terminating into radio lobes. The end-to-end sizes of radio galaxies span ranges of hundreds of kpc to even a few Mpc \citep[e.g.][]{Sebastian18,Oei22,Sethi22,Andernach25,Sethi25}, suggesting that the AGN activity can last for several millions of years \citep{Hardcastle20,Saikia22}. During this time, the AGN's jets interact with the intergalactic environment, enriching the surrounding medium with relativistic plasma and influencing star formation and galaxy evolution via AGN feedback mechanisms. 
	\par
	It is well known that AGN activity is a phase, also termed as `active phase' \citep[][]{Parma07,DK09,Saripalli12}, which is followed by a quiescent (or `remnant') phase  \citep{Murgia11}. 
	The remnant phase is the final stage of the evolution of a radio galaxy, during which the radio lobes fade as the plasma within them cools and disperses into the IGM. The cessation of jet activity leads to a decline in the radio emission, and the lobes may eventually become undetectable. During this phase, the spectrum of the radio emission steepens, reflecting the ageing of the plasma as high-energy electrons lose energy much faster than the low-energy electrons \citep{Parma07,Murgia11}. The fading of the radio lobes can provide important clues about the time elapsed since the cessation of jet activity, offering a probe into the duration of the AGN's active phase. The morphological and spectral characteristics of radio galaxies therefore provide a direct probe to the history of AGN activity. For instance, double-double radio galaxies \citep[DDRGs;][]{Schoenmakers00} provide clear evidence for the cessation and recurrence of AGN activity \citep{Saikia09,Nandi12,Kuzmicz17,Morganti17,Mahatma19,Jurlin24}. A DDRG is a radio galaxy that exhibits two distinct pairs of radio lobes aligned majorly along the same axis, produced by successive episodes of jet activity from the central AGN. 
	\par
	Radio galaxies with no recurrent AGN activity which only exhibit evidence for switched off AGN activity are named as `remnant radio galaxies' (RRGs) which can be detected for a period of time before lobes completely fade away due to radiative and dynamical losses \citep[see review papers by][]{Mahatma23,Morganti24}.
	It is unclear if a RRG necessarily goes through another episode of AGN activity and on what timescales. Therefore, DDRGs can be treated as a subset of evolved RRGs which have gone through another episode of AGN activity before lobes from the previous episode of activity fade away. One of the intriguing observational facts is that RRGs are even rarer than DDRGs \citep{Mahatma18,Jurlin20} which can only be understood if timescales of the remnant phase and its dependence of various parameters is known for a statistically large sample. The remnant fraction ($f_{\rm rem}$; the ratio of the number of remnant sources to the total number of radio galaxies in the sample), even in deep fields, is found only to be 5$-$8 percent and depends on the flux density at which they are observed \citep[see][]{Jurlin21,Dutta23}. Due to their paucity, the evolution of radio galaxies in their remnant phase is not well understood.
	\par
    In recent years, efforts to identify additional RRGs have intensified, particularly through radio continuum surveys targeting deep fields. These include studies in the Lockman Hole \citep{Brienza17,Jurlin21}, Herschel Astrophysical Terahertz Large Area Survey \citep[Herschel-ATLAS;][]{Eales10,Mahatma18}, Galaxy And Mass Assembly (GAMA)-23 \citep{Quici21}, XMM{\em -Newton} Large Scale Survey \citep[XMM-LSS;][]{Dutta23}, and Hobby-Eberly Telescope Dark Energy Experiment (HETDEX), where machine learning techniques have also been applied \citep{Gebhardt21,Mostert23}. Additionally, several individual RRGs have been discovered, such as B2 0924+30 \citep{Cordey87,Jamrozy04,Shulevski17}, J021659-044920 \citep{Tamhane15}, blob1 \citep{Brienza16}, Arp 187 \citep{Ichikawa16}, NGC 1534 \citep{Duchesne19}, J1615+5452 \cite{Randriamanakoto20}, WISEA J152228.01+274141.3 \citep{Lal21} and Abell 1318 \citep{Shulevski24}. The searches in deep fields are advantageous due to the availability of deep multi-frequency radio observations which in turn can allow us to detect remnants more reliably even at the fainter flux densities and higher redshifts. 
	%
	\par
	Furthermore, the plasma age can be estimated using the energy-dependent synchrotron loss rate, which manifests as spectral steepening in the radio spectrum—a technique widely referred to as the `spectral ageing' method in the literature \citep{Myers85,Carilli91,Jamrozy04,Harwood13,Shulevski17,Mahatma20,Wolnik24}. Spectral age analysis is a powerful diagnostic tool for understanding the lifecycle of radio galaxies. By quantifying the synchrotron radiative losses of relativistic electrons, it provides estimates of the time elapsed since particles were last accelerated. This enables robust determination of source ages, particularly in the absence of identifiable hotspots or cores. Additionally, the spectral age maps are especially important in identifying RRGs, where active particle injection has ceased, and in diagnosing restarted AGN activity through the presence of age discontinuities or spectral curvature. Furthermore, they allow for direct comparisons with predictions from dynamical and evolutionary models, helping constrain the physical conditions of the source environment ({\em e.g.,} magnetic fields, ambient density). Spectral ageing analysis also contributes to estimating duty cycles of AGN activity and to understanding the timescales over which radio-mode AGN feedback operates. 
	\par
	Recent studies have highlighted significant diversity in the active and remnant timescales of radio sources, with these timescales depending on a range of factors such as source size, magnetic field strength in the lobes, and the large-scale environment \citep{Turner18,Dutta22,Stewart25}. Therefore, it is important to probe the active as well as remnant timescales and large-scale environments of a variety of remnant sources. This diversity is essential for understanding the life cycle of radio sources, as the active phase typically involves ongoing particle acceleration, while the remnant phase corresponds to an evolved state where particle ageing dominates \citep{Harwood16}. As the timescales for both active and remnant phases are sensitive to both intrinsic properties and external conditions, it is crucial to investigate the spectral ages of a large sample of remnant sources across different environments.  
	\par
	In this paper we perform spectral ageing of a sample of remnants by modelling their radio spectral energy distributions (SEDs) with physically-motivated models. For the SED modeling of our sources we used Broadband Radio Astronomy ToolS \citep[{\sc BRATS\footnote{\url{https://www.askanastronomer.co.uk/brats/}}};][]{Harwood13,Harwood15}. We used 144 MHz LOw Frequency ARray \citep[LOFAR;][]{Lazio99}, 323 MHz Giant Metrewave Radio Telescope \citep[GMRT;][]{Swarup91}, band-3 (400 MHz) and band-4 (650 MHz) upgraded GMRT \citep{Nityananda09}, 1284 MHz MeerKAT \citep{Jonas09} and 1.5 GHz Jansky Very Large Array \citep[JVLA;][]{Heeschen75} radio observations to model the radio SEDs for our sample sources.
	\par
	This paper is organized as follows. Section~\ref{sec:observations} describes the multi-frequency radio observations, and Section~\ref{sec:sample} outlines the sample selection and criteria. The modeling of radio SEDs, spectral age estimates, and the key factors influencing spectral ageing analysis are discussed in Section~\ref{sec:modelling}. Details of the spectral age maps are presented in Section~\ref{sec:spec-age-maps}, followed by the main results in Section~\ref{sec:results}. Section~\ref{sec:Discussion} discusses the implications of the analysis in relation to various physical parameters, and the conclusions are summarized in Section~\ref{sec:conclusions}.
	\par
	In this paper, we use the cosmological parameters $H_{0}$ = 69.6 km s$^{-1}$ Mpc$^{-1}$, $\Omega_{\rm M}$ = 0.286 and $\Omega_{\rm \Lambda}$ = 0.714 \citep{Bennett14}. We define radio spectral index $\alpha$ as S$_{\nu}$ $\propto$ $\nu^{\alpha}$.

	\section{Multi-frequency radio continuum observations in the XMM-LSS field}
	\label{sec:observations}
	\subsection{SuperMIGHTEE (Band-3 and Band-4 uGMRT Observations)}
	The XMM-LSS field has been surveyed with the upgraded Giant Meterwave Radio Telescope (uGMRT) at band-3 (250$-$500 MHz) and band-4 (550$-$850 MHz). These observations were conducted under `superMIGHTEE project' during cycle 36 and cycle 37 in 2019 $-$ 2020 \citep{Taylor19,Lal25}. 
	\par
\begin{figure}
		\centering
		\includegraphics[scale=0.35]{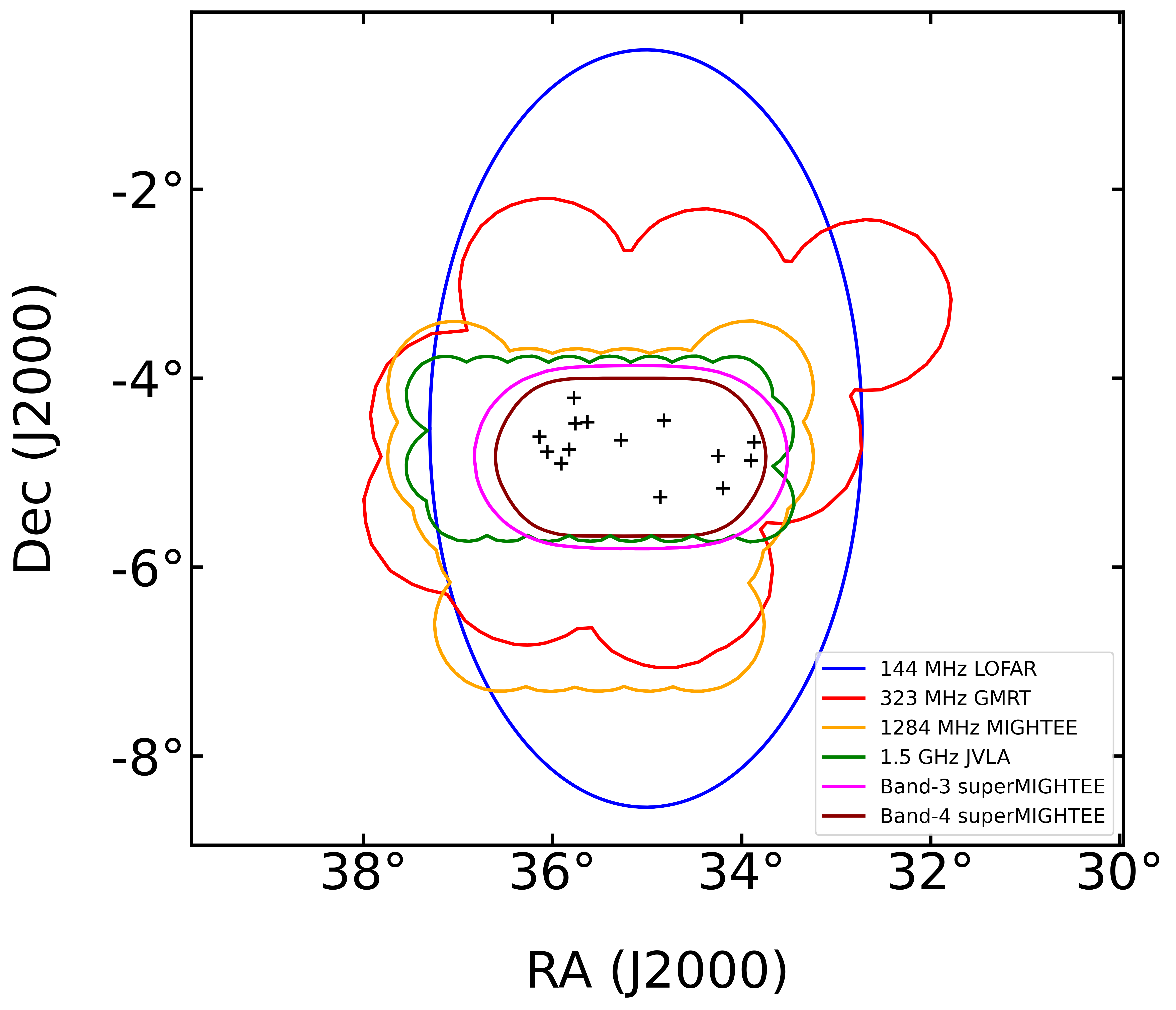}
		\caption{Footprints of different radio surveys. The positions of remnant candidates are marked with `+' symbols.}
		\label{fig:Footprints}
	\end{figure}

	The first data release \citep[DR1;][]{Lal25} provides mosaiced images of four pointings covering a total sky area of 6.87 deg$^{2}$ in band-3 and 19 pointings in band-4 covering a sky-area of only 5.02 deg$^{2}$ (see Figure~\ref{fig:Footprints}). The band-3 observations were performed in full Stokes mode with 8192 channels across the frequency range of 300 to 480 MHz. The band-4 observations were also taken in full Stokes mode with bandwidth of 560 to 816 MHz divided into 8192 channels. 
	\par
	%
	\par
	In the DR1 band-4 mosaiced image, produced using a robust weighting parameter of $-$0.4, the median noise RMS—measured in regions free from bright radio sources—is approximately 8 $\mu$Jy beam$^{-1}$. The image achieves an angular resolution of 5\farcs0 $\times$ 4\farcs7, with a position angle of 52$^{\circ}$. 
	Similarly, the band-3 mosaiced image, also generated with a robust parameter of $-$0.4, exhibits a median noise RMS of approximately 32 $\mu$Jy beam$^{-1}$ and reaches an angular resolution of 6\farcs9 $\times$ 4\farcs6 (position angle of 64$^{\circ}$). 
	The source catalogues were generated using Python Blob Detector and Source Finder  \citep[PyBDSF\footnote{\url{https://github.com/lofar-astron/PyBDSF}};][]{Mohan15} on the mosaiced images of each band.
	It is important to note that for wide-band imaging, the effective observing frequency is affected by the frequency dependence of the primary beam and by the spectral index of the source. We derived accurate frequency for each source in band-3 and band-4 using the effective-frequency maps generated from the full-band data. Considering the effective frequency maps and steep spectral index ($\alpha$ $\geq$ -1) of our sources, we take 400 MHz as the central frequency of band-3 and 650 MHz for band-4.

	\begin{table*}
		\begin{minipage}{160mm}
			\centering
			\caption{Summary of deep multi$-$frequency radio observations.}
			\label{tab:RadioData}
			\begin{tabular}{ccccccccc} 
				\hline
				Frequency & Telescope & Area  &  {\em u-v} coverage  & 5$\sigma$ (median) & Beam$-$size & PA & No. of total  & Reference \\
				(MHz)   &           & (deg$^2$) & ($k \lambda$) & (mJy~beam$^{-1}$) & ($\prime\prime \times \prime\prime$) & (deg) & sources  &     \\
				\hline
				144  & LOFAR & 27.0 & 0.15$-$43 & 1.4$-$1.97 & 7.5 $\times$ 8.5 &  106.0    &  3044  & 1  \\
				323  & GMRT & 12.5 &  0.1$-$27 & 0.75 & 10.2 $\times$ 7.9 & 74.6     &  3739  & 2 \\
				400$^{a}$ & uGMRT &  6.87 &  0.08$-$32 & 0.16  & 6.9 $\times$ 4.6 & 64.0 &  10931  &   3  \\
				650$^{a}$ & uGMRT & 5.02 &  0.09$-$38 &  0.04  & 5.0 $\times$ 4.7 & 52.0 &  16284 &  3   \\
				1284 & MeerKAT & 14.4 &  0.09$-$34 &  0.02 & 8.9 $\times$ 8.9 & 0.0  &  97684 &  4   \\
				1500  & JVLA & 5.0 &  0.1$-$30 &  0.08  & 4.5 $\times$ 4.5 & 0.0   &  5760 & 5 \\
				\hline
			\end{tabular}
			\\
			Notes : $^{a}$ - Parameters correspond to the image generated using robust parameter $-$0.4. References - 1 : \citet{Hale19}; 2 :  \citet{Singh14}; 3 : \citet{Lal25}; 4 : \citet{Hale25};
			5: \citet{Heywood20}.
			
		\end{minipage}
	\end{table*}
	\subsection{1.284 GHz ($L-$band) MeerKAT Observations}
	The MeerKAT International GHz Tiered Extragalactic Explorations \citep[MIGHTEE;][]{Jarvis16} project includes radio continuum, spectro-polarimetry, and spectral line observations from the MeerKAT telescope in four different deep fields {\ie} XMM-LSS, COSMOS, E-CDFS and ELAIS-S1 at a central frequency of 1284 MHz. The MIGHTEE Data Release 1 \citep[MIGHTEE-DR1;][] {Hale25} contains radio continuum imaging of nearly 14.4 deg$^{2}$ with a 45-pointing mosaic in the XMM-LSS (see footprints in Figure~\ref{fig:Footprints}). The total on-target time was 297.9 hours in this field. 
	The Briggs weighting of 0.0 provides improved sensitivity with low median noise-rms of 3.2 $\mu$Jy, making the angular resolution to 8\farcs9. The Briggs weighting of -1.2 gives images with improved resolution of 5\farcs0 but with compromised median noise-rms of 5.1 $\mu$Jy beam$^{-1}$. In this work, we use the lower resolution image which has increased sensitivity to extended emission.
	The source catalogue is generated using PyBDSF. A total of 97684 radio sources are detected in the low-resolution image of 14.4 deg$^{2}$.
	%
	
	
	%
	\subsection{Auxiliary Multi-frequency Radio Observations}
	In addition to the recent deep uGMRT and MeerKAT observations from the superMIGHTEE project, the XMM-LSS field possesses wealth of multi-frequency radio continuum data. The details of the multi-frequency radio observations are given in Table~\ref{tab:RadioData}. In the following subsections, we briefly describe 144 MHz LOFAR, 323 MHz GMRT and 1.5 GHz JVLA observations that we used for our study. 
	
	\subsubsection{144 MHz LOFAR Observations}
	\cite{Hale19} performed 144 MHz High Band Antenna (HBA; 110$-$240 MHz) LOFAR observations in the XMM-LSS field. With three pointings, each observed for four hours, these observations cover 27 deg$^{2}$ sky area (see footprints in Figure~\ref{fig:Footprints}). The LOFAR observations cover an elliptical-shaped region aligned in the north$-$south direction and centered at RA = 02$^{\rm h}$ 20$^{\rm m}$ 00$^{\rm s}$ and DEC = -04$^{\circ}$ 30$^{\prime}$ 00$^{\prime\prime}$ (J2000) (see Figure~\ref{fig:Footprints}). The final mosaiced map has an angular resolution of 7\farcs5 $\times$ 8\farcs5 and a median noise$-$rms of 0.40 mJy beam$^{-1}$, while noise$-$rms reaches down to 0.28 mJy beam$^{-1}$ in the central region. The LOFAR observations detect a total of 3044 individual radio sources.
	\subsubsection{323 MHz GMRT Observations}
	Nearly 12.5 deg$^2$ sky-area centered at RA = 02$^{\rm h}$ 21$^{\rm m}$ 00$^{\rm s}$ and DEC = -04$^{\circ}$ 30$^{\prime}$ 00$^{\prime\prime}$ (J2000) in the XMM-LSS field has been surveyed with 323 MHz GMRT observations (see footprints in Figure~\ref{fig:Footprints}). These observations were performed with the 32 MHz narrow band receiver in the legacy GMRT.
	To optimize uv-coverage, these observations were performed in snap-shot mode with a total of 2.5 hours time per pointings. The final mosaiced image from 16 contiguous pointings has an average noise-rms of 150 $\mu$Jy beam$^{-1}$ which reaches down to 120~$\mu$Jy. The synthesized beam-size is 10\farcs2 $\times$ 7\farcs9. There are a total of 3739 individual radio sources detected at the flux limit of 5$\sigma$ in the deeper regions (noise-rms $<$200 $\mu$Jy beam$^{-1}$) and 6$\sigma$ in the relatively shallow regions (noise-rms $>$200 $\mu$Jy beam$^{-1}$). More details on these observations can be found in \cite{Singh14}.
	\subsubsection{1.5 GHz JVLA Observations}
	\cite{Heywood20} performed 1.5 GHz wide$-$band (0.994$-$2.018 GHz) JVLA radio observations of the near-IR VIDEO survey region in the XMM$-$LSS field. With a total of 32 pointings these observations cover 5.0 deg$^{2}$ sky-area (see footprints in Figure~\ref{fig:Footprints}). Using upgraded VLA in B-configuration and a total time of 67.5 minutes per pointing they achieved nearly uniform noise-rms with a median value of 16 $\mu$Jy beam$^{-1}$ and angular resolution of 4\farcs5. These are a total of 5760 radio sources detected above a $5\sigma$ detection threshold.
	%
	%
	\section{The Sample}
	\label{sec:sample}
	Our sample consists of 14 remnant candidates which are gleaned from our previous studies \citep{Singh21,Dutta23}. In our previous works we have searched and characterized small-size ($<$ 30$^{\prime\prime}$) as well as large-size ($\geq$ 30$^{\prime\prime}$) remnant candidates in the XMM-LSS field. The large-size remnant candidates were selected using morphological criterion ({\ie} absent core), strong spectral curvature (${\alpha}_{\rm 144~MHz}^{\rm 323~MHz}$ - ${\alpha}_{\rm 323~MHz}^{\rm 1.5~GHz}$ $\geq$ 0.5) and ultra-steep radio spectral criteria \citep[see][]{Dutta23}, while small-size sources with only slightly resolved structures are selected from mainly spectral curvature criterion (${\alpha}_{\rm 144~MHz}^{\rm 323~MHz}$ - ${\alpha}_{\rm 323~MHz}^{\rm 1.5~GHz}$ $\geq$ 0.5) with an undetected radio core in 1.5~GHz JVLA images at 3$\sigma$ $\sim$ 0.24~mJy~beam$^{-1}$ \citep[see][]{Singh21}. We also examined the 3.0 GHz Very Large Array Sky Survey (VLASS) images for all the sources in our sample but found no detections. The absence of detections at 3.0 GHz VLASS, especially for relatively bright sources at low frequencies, can be explained if these sources exhibit relic emission with a low surface-brightness and a steep spectral index. For the relatively faint remnant candidates, the non-detection is likely due to the limited sensitivity of VLASS, with a 5$\sigma$ threshold of approximately 0.6 mJy~beam$^{-1}$.
	In Table~\ref{tab:table1} we list our sample sources, their redshifts, radio luminosities at 144 MHz and flux densities in different bands. 
	\par
	Identifying host galaxies of RRGs is challenging due to the absence of a radio core, which complicates precise astrometric matching and redshift determination. In our previous studies \citep{Singh21,Dutta23}, we have identified potential hosts of our remnant sources by overlying radio contours onto the deep $i-$band optical images from the Hyper Suprime-Cam Subaru Strategic Program \citep[HSC-SSP;][]{Aihara18}. The optical source matching with the flux density weighted barycenter position lying on the intersection of two lobes axes is considered as the potential host galaxy. 
	The redshifts of potential hosts are found by matching their positions with sources in the recent photometric redshift catalogue \citep{Nyland17,Nyland23}. We also cross-matched the photometric redshifts from \cite{Adams23} and found them to be consistent with the redshifts reported by \cite{Nyland23}.
	In addition, we obtained spectroscopic redshift (spec-$z$) estimates for 6 out of the 14 sources in our sample from various spec-$z$ catalogues available in the XMM–LSS field  \citep[e.g.][]{Cool13,Garilli14,Vaccari15,Vaccari23,Vaccari26}.
	\par
	\cite{Singh21} and \cite{Dutta23} have also examined the large-scale environments of a sample of remnant candidates. 
	In our previous work, we investigated the large-scale environments by assessing the association of the remnant candidates with clusters identified in the XMM-LSS region. We employed an optically-selected cluster catalog from the HSC-SSP survey alongside an X-ray selected cluster catalog based on deep observations from the {\em XMM-Newton} survey. Our findings indicate that, with the exception of J021528-044045, remaining 13 remnant candidates in our sample are situated in non-cluster (less dense) environments \citep[see][]{Dutta23}. 
	In our earlier work \citep{Dutta22}, we derived spectral ages for J022318–044522 and J022338–045415 through spectral modeling based on integrated flux densities measured at 144 MHz, 323 MHz, 400 MHz, and 1.5 GHz. With the addition of new uGMRT band-3 and band-4 data, together with 1.284 GHz MeerKAT observations, we now revisit the spectral ageing analysis of both sources to obtain more reliable and robust age estimates.

	\begin{table*}
		\begin{minipage}{180mm}
			\centering
			\caption{The remnant sample and full band radio flux densities at different frequencies.}
			\label{tab:table1}
			\begin{tabular}{ccccccccc}
				\hline
				Source &    Redshift     &  $S_{\rm 144~MHz}$ & $S_{\rm 323~MHz}$ & $S_{\rm 400~MHz}$ & $S_{\rm 650~MHz}$  & $S_{\rm 1284~MHz}$ & $S_{\rm 1.5~GHz}$ & log $L_{\rm 144~MHz}$ \\
				Name   &     ($z$)     & (mJy)         & (mJy)         & (mJy)         & (mJy)          & (mJy)           & (mJy) & (W/Hz) \\
				\hline
				J021528-044045 & 0.35$^{\rm +0.003}_{\rm -0.003}$ (s) & 29.96$\pm$0.96 & 17.63$\pm$0.54 & 15.54$\pm$0.39 & 10.09$\pm$0.25 & 4.96$\pm$0.08 & 4.09$\pm$0.05 & 25.21 \\
				J021536-045220 & 0.67$^{\rm +0.07}_{\rm -0.11}$ (ph) & 3.06$\pm$0.96 & 1.91$\pm$0.19 & 1.35$\pm$0.10 & 0.68$\pm$0.13 & 0.27$\pm$0.01 & 0.24$\pm$0.03 & 24.79 \\
				J021646-051004 & 2.85$^{\rm +0.20}_{\rm -0.16}$ (ph) & 2.93$\pm$0.49 & 1.95$\pm$0.09 & 1.69$\pm$0.05 & 1.23$\pm$0.03 & 0.73$\pm$0.02 & 0.62$\pm$0.02 & 26.32 \\
				J021659-044918 & 1.32$^{\rm +0.001}_{\rm -0.001}$ (s) & 428.17$\pm$1.80 & 181.96$\pm$4.55 & 147.64$\pm$2.09 & 66.14$\pm$0.84 & 14.97$\pm$0.15 & 9.52$\pm$0.02 & 27.60 \\
				J021917-042654 & 1.52$^{\rm +0.04}_{\rm -0.14}$ (ph) & 9.72$\pm$0.45  & 3.62$\pm$0.22 & 2.86$\pm$0.11 & 1.24$\pm$0.05 & 0.25$\pm$0.01 & 0.16$\pm$0.01 & 26.17 \\
				J021926-051535 & 1.47$^{\rm +0.03}_{\rm -0.03}$ (ph) & 20.45$\pm$0.67 & 11.71$\pm$0.91 & 9.53$\pm$0.40 & 5.69$\pm$0.20 & 2.21$\pm$0.07   & 1.84$\pm$0.07 & 26.46 \\
				J022106-043928 & 0.99$^{\rm +0.03}_{\rm -0.03}$ (s) & 3.55$\pm$0.31  & 2.11$\pm$0.08 & 1.75$\pm$0.06 & 1.17$\pm$0.05 & 0.54$\pm$0.02 & 0.44$\pm$0.02 & 25.78 \\
				J022231-042757 & 2.22$^{\rm +0.31}_{\rm -0.07}$ (ph) & 48.74$\pm$0.56 & 25.70$\pm$1.05 & 19.51$\pm$0.97 & 10.70$\pm$0.62 & 3.65$\pm$0.13 & 2.67$\pm$0.11 & 27.28 \\
				J022302-042849 & 1.18$^{\rm +0.001}_{\rm -0.001}$ (s) & 12.33$\pm$0.55 & 6.24$\pm$0.28 & 5.21$\pm$0.15 & 2.90$\pm$0.05 & 1.34$\pm$0.01 & 1.08$\pm$0.04 & 26.12 \\
				J022305-041232 & 1.05$^{\rm +0.007}_{\rm -0.007}$ (s) & 13.78$\pm$0.62 & 8.06$\pm$0.81 & 6.39$\pm$0.45 & 3.85$\pm$0.23 & 1.89$\pm$0.07 & 1.40$\pm$0.07 & 25.88 \\
				J022318-044522 & 1.26$^{\rm +0.01}_{\rm -0.02}$ (ph) & 33.98$\pm$0.97 & 19.11$\pm$1.85 & 15.12$\pm$0.69 & 9.31$\pm$0.36 & 4.18$\pm$0.06 & 3.26$\pm$0.06 & 26.51 \\
				J022338-045415 & 0.81$^{\rm +0.001}_{\rm -0.001}$ (s) & 7.05$\pm$0.32 & 3.98$\pm$0.18 & 3.31$\pm$0.14 & 2.10$\pm$0.10 & 0.92$\pm$0.05 & 0.71$\pm$0.04 & 25.36 \\
				J022413-044643 & 1.24$^{\rm +0.02}_{\rm -0.02}$ (ph) & 7.97$\pm$0.65  & 3.49$\pm$0.28 & 2.65$\pm$0.20 & 1.36$\pm$0.07 & 0.37$\pm$0.03 & 0.23$\pm$0.02 & 25.86 \\
				J022433-043709 & 2.64$^{\rm +0.18}_{\rm -0.05}$ (ph) & 4.08$\pm$0.52 & 1.69$\pm$0.11 & 1.39$\pm$0.07 & 0.80$\pm$0.04 & 0.38$\pm$0.02 & 0.32$\pm$0.02 & 26.38 \\
				\hline
			\end{tabular}
		\end{minipage}
		Note - s: spectroscopic redshift ($z_{\rm spec}$), ph: photometric redshift ($z_{\rm phot}$)
	\end{table*}

	\section{Spectral Ageing of Radio Sources}
	\label{sec:modelling}

	\subsection{Spectral Ages Estimation from Radio SED Modeling}
	\label{subsec:spec-aging}
	%
	Active radio galaxies generally show a power law radio spectrum 
	which can be explained as the synchrotron emission from relativistic particles having a power law energy distribution (N(E) $\propto$ E$^{\rm -p}$). The power law radio spectral index ($\alpha$) is related to the energy distribution index p as : $\alpha$ = (p$-$1)/2.
	The power law radio spectrum begins to develop a break due to continuous radiative losses suffered by the lobe's plasma. Since high energy relativistic electrons lose their energy faster than the low energy electrons the break in the spectrum occurs towards high frequencies. The radio spectrum above break frequency (${\nu}_{\rm b}$) steepens compared to the initial value of spectral index (${\alpha}_{\rm inj}$).
	\par
	%
	In the remnant phase, there is no supply of plasma to the lobes. Due to lack of replenishment of new plasma and ongoing radiative losses, the radio spectrum develops another break frequency (${\nu}_{\rm b,~high}$) beyond which the spectrum falls off exponentially. As the radio source evolves both break frequencies shift towards lower frequencies. The lower break frequency, $\nu_{\rm b, low}$, is the synchrotron break frequency associated with the oldest electron population in the lobes, set by the total radiative age of the source since the onset of particle injection.
	The ratio of $\nu_{\rm b, low}$ and $\nu_{\rm b, high}$ depends on the duration of remnant phase (t$_{\rm OFF}$) {\em w.r.t.} the total source age and can be expressed by the following formula.
	\begin{equation} 
		\label{eq:1}
		\centering
		\frac{t_{\rm OFF}}{t_{\rm s}} = \left(\frac{\nu_{\rm b,~low}}{\nu_{\rm b,~high}}\right)^{0.5},
	\end{equation}
	where t$_{\rm s}$ (= t$_{\rm ON}$ + t$_{\rm OFF}$) is the total source age. This ratio is sensitive to the assumed magnetic field strength and spectral index, and may vary across different environments and source morphologies.
	For a sufficiently old remnant $\nu_{\rm b, high}$ approaches close to $\nu_{\rm b, low}$.
	%
	%
	The relation between the spectral age of the source, t$_{\rm s}$ and the break frequency ($\nu_{\rm b}$) is given by the formula (see \citet{Komissarov94,Slee01,Parma07})
	\begin{equation} 
		\label{eq:2}
		\centering
		t_{\rm s} = 1590 \left[\frac{B_{\rm eq}^{0.5}}{(B_{\rm eq}^2 + B_{\rm CMB}^2) \sqrt{\nu_{\rm b} (1 + z)}}\right]~{\rm Myr,}
	\end{equation}
	where B$_{\rm eq}$ is the equipartition magnetic field, B$_{\rm CMB}$ = 3.25(1+$z$)$^2$ is the inverse Compton equivalent magnetic field in the unit of $\mu$G and $\nu_{\rm b}$ is in GHz. These models consider only radiative losses without accounting for the losses due to expansion.

	\subsection{Estimation of Magnetic Field Strength}
	\label{subsec:Magnetic Fields}
	The estimation of spectral ages also requires the computation of magnetic field strength (see Equation~\ref{eq:2}). We used the equipartition magnetic field values estimated using the {\sc PySynch}\footnote{\url{https://github.com/mhardcastle/pysynch}} code \citep{Hardcastle98}, which computes synchrotron and inverse-Compton emission properties under the assumption of energy equipartition between relativistic particles and magnetic fields. 
	It is important to note that \cite{Croston05} examined X-ray emission from the lobes of quasars and FR-II radio galaxies, reporting that the ratio of observed to predicted emission from synchrotron-emitting electrons at equipartition follows a distribution that aligns with magnetic field strengths within 35\% of the equipartition value \citep{Ineson17,Pinjarkar23,Charlton25}. Accordingly, we estimated the magnetic field for our sample as 0.4 times the equipartition magnetic field. To achieve this, we supplied source parameters to the {\sc PySynch} code, including shape (assumed to be spherical, cylindrical, or ellipsoidal), dimensions (height, width, and depth), redshift ($z$), and injection index ($\alpha_{\rm inj}$). Various geometries are utilized for the sources due to their distinct shapes, as a single geometry cannot effectively accommodate all types of shapes. Additionally, the geometrical properties of all sources in our sample are examined using the highly sensitive 1284 MHz MeerKAT observations, a frequency particularly well-suited for tracing diffuse emission that is also evident in the lower-frequency uGMRT band-3 data. We caution that {\sc PySynch} code does not provide uncertainties on the derived parameters, formal error estimates for the magnetic field strengths cannot be quoted, and the reported values should be interpreted as representative.
	
		\begin{table*}
		\begin{minipage}{180mm}
			\centering
			\caption{Estimates of the source parameters obtained by using {\sc PySynch} code.} 
		\label{tab:table2}
		\begin{tabular}{ccc*{4}{c}}
			\hline
			\multirow{1}{*}{Source} & \multirow{1}{*}{Assumed} & \multirow{1}{*}{Source} & \multirow{1}{*}{Volume of} & \multicolumn{2}{c}{Magnetic Field Strength} \\\cline{5-6}
			Name & Shape & Size & the Source & Minimum Energy & Equipartition \\
			&  & ($^{\prime\prime}\times^{\prime\prime}$) & ($\times$10$^{64}$ m$^3$) & ($\mu$G) & ($\mu$G) \\
			\hline
			J021528-044045 & Ellipsoidal & 23.3$\times$19.1 & 1.6 & 3.5 & 4.6 \\
			J021536-045220 & Spherical & 12.0 & 7.7 & 3.3 & 4.2 \\
			J021646-051004 & Spherical & 12.5 & 12.2 & 4.9 & 5.7 \\
			J021659-044918 & Cylindrical & 151.5$\times$27.7 & 634.5 & 2.8 & 3.9 \\
			J021917-042654 & Spherical & 9.0 & 5.7 & 6.2 & 8.1 \\
			J021926-051535 & Ellipsoidal & 21.3$\times$14.7 & 4.3 & 6.6 & 8.6 \\
			J022106-043928 & Ellipsoidal & 19.7$\times$12.0 & 2.6 & 9.2 & 12.0 \\
			J022231-042757 & Ellipsoidal & 33.7$\times$15.7 & 7.2 & 7.1 & 9.2 \\
			J022302-042849 & Spherical & 10.9 & 10.1 & 4.7 & 6.1 \\		
			J022305-041232 & Ellipsoidal & 28.3$\times$15.2 & 5.2 & 4.9 & 6.3 \\
			J022318-044522 & Cylindrical & 94.3$\times$29.0 & 426.3 & 3.2 & 4.3 \\
			J022338-045415 & Spherical & 22.5 & 64.1 & 1.9 & 2.4 \\
			J022413-044643 & Spherical & 10.7 & 9.1 & 4.7 & 6.0 \\
			J022433-043709 & Spherical & 11.4 & 10.0 & 7.2 & 9.2 \\  
			\hline
		\end{tabular}
	\end{minipage}
	Note: For cylindrical and ellipsoidal geometries, the size components correspond to the source height and width, whereas for the spherical geometry they refer to the diameter. {\sc PySynch} does not return uncertainties; the quoted values are representative.
	\end{table*}
	
	The dimensions of each source were determined using Cube Analysis and Rendering Tool for Astronomy ({\sc CARTA\footnote{\url{https://cartavis.github.io}}}) software, with measurements taken from the 1284 MHz MeerKAT image (emission up to 3$\sigma$ level). The height and width correspond to the end-to-end linear extent along the axes of maximum and minimum elongation, respectively, while the cross-sectional diameter is considered to represent the source depth. The physical size of the radio source is derived from its observed angular extent by converting the angular size to a linear scale using the angular diameter distance at the corresponding redshift. The resulting equipartition magnetic field strength in our sample varies between 2.37 and 11.99 $\mu$G. The equipartition magnetic field strengths derived for our sample are relatively higher than those reported for some classical, extended remnants. However, the majority of our sources are likely young remnants in which the jets have only recently switched off. As a result, the lobes have not yet undergone substantial adiabatic expansion, and the magnetic fields are expected to remain comparatively strong. Such values are consistent with previous studies that report B$_{\rm eq} \sim$ 10 $\mu$G for remnant or evolved radio galaxies \citep[e.g.][]{Koekemoer98,Murgia11,Randriamanakoto20}. The parameter estimates of {\sc PySynch} are given in Table~\ref{tab:table2}. 

\subsection{Injection Index Estimation}
\label{subsec:Injection Index}
The injection index ($\alpha_{\rm inj}$) is a key parameter in determining the spectral age of radio sources, representing the initial slope of the power-law energy distribution of relativistic electrons injected into the source. 
To estimate the injection index, we utilized the {\it findinject} command in {\sc BRATS} (see the {\em BRATS cookbook\footnote{\url{http://www.askanastronomer.co.uk/brats/downloads/bratscookbook.pdf}}}). This command treats $\alpha_{\rm inj}$ as a pseudo-free variable, allowing the software to adjust its value within predefined limits to achieve the best fit with the observed spectral data. This approach enables an estimation of the injection index based on the spectral properties of the source. 
The $\alpha_{\rm inj}$ values obtained for our sample range from -0.53 to -0.68 (see second column of Table~\ref{tab:table3}), consistent with values typically observed in remnant, restarted, and active radio galaxies \citep[see][]{Murgia11}. These values fall well within the theoretically expected range of -0.5 $>$ $\alpha_{\rm inj}$ $>$ -0.8 \citep{Bell78,Carilli91,Harwood13}. 

\subsection{Radio SED Modeling of Our Remnant Candidates}
\label{subsec:Details}
For modeling the radio spectrum of an active radio galaxy, the continuous injection model \citep[CI$_{\rm ON}$;][]{Kardashev62,Jaffe73} that assumes continuous supply of plasma to the radio lobes at a constant rate for the duration of t$_{\rm ON}$, has been commonly used in the literature. To model the SEDs of our remnant candidates, we use continuous injection-off model \citep[CI$_{\rm OFF}$ or KGJP;][]{Komissarov94} which considers the remnant phase duration t$_{\rm OFF}$ after cessation of an active phase that lasted for a period of t$_{\rm ON}$. We also explored the CI$_{\rm ON}$ model, and found that the SEDs of two sources in our sample are better fitted by this model. 
For spectral modeling with CI$_{\rm ON}$ and CI$_{\rm OFF}$ in BRATS, we adopted input parameters including the magnetic field (0.4$\times$B$_{\rm eq}$) and the injection spectral index ($\alpha_{\rm inj}$). The fits yield key output parameters such as the low- and high-frequency spectral breaks ($\nu_{\rm b,low}$ and $\nu_{\rm b,high}$), the durations of the active (t$_{\rm ON}$) and remnant (t$_{\rm OFF}$) phases, the total source age (t$_{\rm s}$), and the reduced $\chi^2$ statistic as a measure of goodness-of-fit. The 
best-fitting parameters are summarized in Table~\ref{tab:table3}. The remnants exhibit $\nu_{\rm b,low}$ values in the range 390-1090 MHz, above which their spectra steepen significantly. These values are consistent with the strong curvature observed between 144 MHz and 1.5 GHz. For all sources, the $\nu_{\rm b,high}$ $-$ above which the spectrum steepens exponentially $-$ lies beyond our observing band (1.74-5.88 GHz; Table~\ref{tab:table3}), and is therefore constrained through SED modeling. The determination of these spectral break frequencies enables robust estimates of radiative lifetimes. From the best-fitting models, we find total spectral ages (t$_{\rm s}$) ranging from 8.06 to 41.97 Myr, with active phases lasting 1.48-28.96 Myr and subsequent remnant phases spanning 0.60-16.81 Myr.

\begin{table*}
	\begin{minipage}{180mm}
		\centering
		\caption{Spectral ageing and source parameters obtained by using CI$_{\rm ON}$ and CI$_{\rm OFF}$ model.}
		\label{tab:table3}
		\begin{tabular}{cccccccccc}
			\hline
			Source & $\alpha_{\rm inj}$ & B$_{\rm eq}$ & $\nu_{\rm b,low}$ & $\nu_{\rm b,high}$ & t$_{\rm s}$ & t$_{\rm ON}$ & t$_{\rm OFF}$ & t$_{\rm OFF}$/t$_{\rm s}$ & $\chi_{\rm red}^2$ \\
			Name & & ($\mu$G) & (GHz) & (GHz) & (Myr) & (Myr) & (Myr) & & \\
			\hline
			J021528-044045 & -0.58 & 4.57 & 0.76 & 4.14 & 41.97$^{+1.46}_{-1.36}$ & 28.96$^{+1.32}_{-1.32}$ & 13.01$^{+0.63}_{-0.34}$ & 0.30$^{+0.02}_{-0.01}$ & 1.51 \\
			J021536-045220 & -0.60 & 4.23 & 0.67 & 2.71 & 26.77$^{+0.38}_{-0.52}$ & 9.96$^{+0.01}_{-0.01}$ & 16.81$^{+0.38}_{-0.52}$ & 0.63$^{+0.02}_{-0.02}$ & 1.02 \\
			J021646-051004* & -0.54 & 5.66 & 0.67 & 2.71 & 0.53$^{+0.01}_{-0.02}$ & 0.53$^{+0.01}_{-0.02}$ & - & - & 1.20 \\
			J021659-044918 & -0.54 & 3.85 & 0.84 & 1.93 & 9.04$^{+0.01}_{-0.01}$ & 2.03$^{+0.01}_{-0.01}$ & 7.01$^{+0.00}_{-0.01}$ & 0.77$^{+0.01}_{-0.01}$ & 1.22 \\
			J021917-042654 & -0.56 & 8.05 & 0.85 & 1.74 & 8.47$^{+0.10}_{-0.12}$ & 1.48$^{+0.10}_{-0.10}$ & 6.99$^{+0.03}_{-0.06}$ & 0.83$^{+0.01}_{-0.02}$ & 1.26 \\
			J021926-051535 & -0.54 & 8.58 & 1.09 & 5.88 & 8.06$^{+0.13}_{-0.13}$ & 5.05$^{+0.13}_{-0.11}$ & 3.01$^{+0.01}_{-0.07}$ & 0.37$^{+0.01}_{-0.01}$ & 1.15 \\
			J022106-043928 & -0.62 & 11.99 & 0.88 & 6.22 & 11.07$^{+1.10}_{-0.53}$ & 8.06$^{+1.07}_{-0.45}$ & 3.01$^{+0.27}_{-0.27}$ & 0.27$^{+0.04}_{-0.03}$ & 1.47 \\
			J022231-042757 & -0.54 & 9.20 & 0.79 & 3.48 & 13.35$^{+0.07}_{-0.07}$ & 6.96$^{+0.01}_{-0.01}$ & 6.39$^{+0.07}_{-0.07}$ & 0.48$^{+0.01}_{-0.01}$ & 1.07 \\	
			J022302-042849 & -0.60 & 6.06 & 0.39 & 4.65 & 15.56$^{+0.57}_{-0.22}$ & 14.96$^{+0.56}_{-0.12}$ & 0.60$^{+0.11}_{-0.18}$ & 0.04$^{+0.10}_{-0.16}$ & 1.02 \\
			J022305-041232 & -0.57 & 6.29 & 1.03 & 5.51 & 12.78$^{+0.59}_{-1.06}$ & 7.99$^{+0.55}_{-0.98}$ & 4.79$^{+0.23}_{-0.40}$ & 0.38$^{+0.03}_{-0.04}$ & 1.14 \\
			J022318-044522 & -0.53 & 4.30 & 0.82 & 6.31 & 12.07$^{+0.24}_{-0.20}$ & 9.06$^{+0.24}_{-0.19}$ & 3.01$^{+0.01}_{-0.07}$ & 0.25$^{+0.01}_{-0.01}$ & 1.01 \\
			J022338-045415 & -0.65 & 2.37 & 0.98 & 3.77 & 13.95$^{+0.73}_{-1.32}$ & 8.96$^{+0.71}_{-1.22}$ & 4.99$^{+0.18}_{-0.50}$ & 0.36$^{+0.02}_{-0.05}$ & 1.50 \\
			J022413-044643 & -0.68 & 6.02 & 0.87 & 2.60 & 11.75$^{+1.09}_{-0.71}$ & 4.96$^{+1.07}_{-0.68}$ & 6.79$^{+0.24}_{-0.19}$ & 0.58$^{+0.06}_{-0.04}$ & 0.77 \\
			J022433-043709*  & -0.52 & 9.23 & 0.67 & 2.71 & 4.02$^{+0.00}_{-0.15}$ & 4.02$^{+0.00}_{-0.15}$ & - & - & 0.95 \\
			
			\hline
		\end{tabular}
	\end{minipage}
	Note - `*' represents the sources modeled with CI$_{\rm ON}$ model.
\end{table*}

\section{Spectral Age Maps}
\label{sec:spec-age-maps}
Spectral age maps offer a unique window into the lifecycles of radio galaxies, jet dynamics, and environmental interactions. These maps provide spatially resolved estimates of the radiative ages of relativistic electrons in synchrotron-emitting plasma.
\par
Due to the variations in beam sizes and pixel scales across different frequencies, we initially applied smoothing and regridding to all images using the {\it imsmooth} and {\it imregrid} commands in CASA. This process ensured consistency and comparability among the datasets. After this exercise, we obtained a circular PSF of 15$^{\prime\prime}\times15^{\prime\prime}$
and a pixel size of 2$^{\prime\prime}$. This beam size is optimal as it strikes a balance between angular resolution and sensitivity. We then employed {\sc BRATS} to load the images at each frequency. We clarify that no u–v tapering has been applied in this study. Instead, we note that smoothing in the image plane is mathematically equivalent to applying a taper in the u–v plane. This equivalence follows from the convolution theorem, whereby multiplying the visibilities by a u–v weighting (taper) function corresponds to convolving the image with its Fourier transform. Consequently, u–v tapering suppresses high spatial frequencies, resulting in a smoothed image in the image plane.
For the initial source detection, we established a threshold of 3$\sigma$ based on the rms values of images at each frequency. 
\par
{\sc BRATS} provides pixel-by-pixel analysis for both spectral index and spectral age maps. We generated spectral age maps using radio images at six different frequencies: 144 MHz, 323 MHz, 390 MHz, 650 MHz, 1.284 GHz, and 1.5 GHz. We applied astrometric corrections to the individual maps in CASA by shifting them according to the measured mean offsets, and subsequently generated the spectral age maps. For sources comparable in size to the beam, residual alignment uncertainties and limited resolution can affect pixel-based spectral fitting, potentially introducing or suppressing gradients. In contrast, gradients that persist after correction are likely intrinsic and reflect genuine synchrotron ageing variations. We set a uniform 10\% flux calibration uncertainty to all images \citep[see][]{Hale19,Heywood22} by specifying the {\em fluxcalerror} parameter, ensuring consistency across the dataset and minimizing systematic biases in flux density measurements. The {\sc BRATS} package offers several spectral ageing models, including the JP model \citep{Jaffe73}, the KP model \citep{Komissarov94} and the Tribble model \citep{Tribble93}. The KP and JP models both describe the evolution of synchrotron spectrum produced by an impulsively injected population of electrons, where all electrons are injected at $t=0$ and subsequently experience radiative losses. 
The KP and JP models assume that the magnetic field strength remains constant and uniform across the entire source. In contrast, the Tribble model takes into consideration a more complex scenario, where the magnetic field strength varies locally within the source, allowing for spatial inhomogeneities in its distribution. This difference in approach reflects a more nuanced understanding of magnetic field behavior and accurate estimation of spectral age of the sources. 
We specifically focused on the JP Tribble model for our sample sources because it incorporates a more realistic treatment of the magnetic field structure in radio lobes and the resulting electron energy losses. Before producing the spectral age maps, we ensured data consistency and addressed any potential issues related to the alignment of the radio images by performing spectral index fitting using the {\it specindex} command in {\sc BRATS} .
\par   
Prior to fitting, the spectral‐age parameter ({\em myears}) was bounded to the range 0-200 Myr, which defines the domain over which the spectral‐ageing models were evaluated. Further, we generate spectral age maps by using {\it specagemap} command. {\sc BRATS} also provides the goodness-of-fit of the model using $\chi^2$ minimization test as

\begin{equation}
\chi^2 = \sum_{\nu = 1}^{N}  \left(\frac{S_{i, \nu} - S_{model, \nu}}{\Delta S_{i, \nu}}\right)^2,
\end{equation}

where $S_{i, \nu}$ is the flux density at a given frequency $\nu$ in the $i^{\rm th}$ region. The average reduced $\chi^2$ and the minimum and maximum spectral age estimates for our sample, based on the JP Tribble models, are provided in Table~\ref{tab:table4}. 
\par
\begin{figure*}
	\centering
	\includegraphics[scale=0.215]{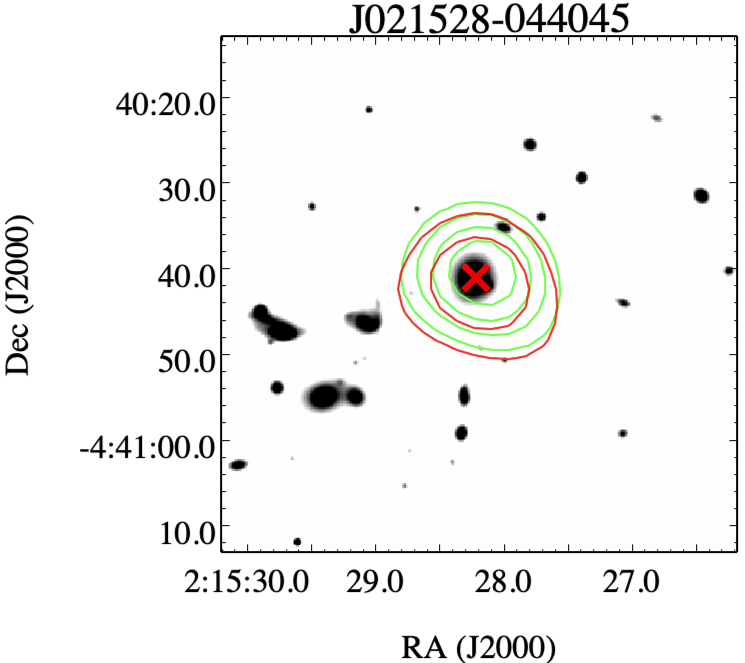}
	\includegraphics[scale=0.26]{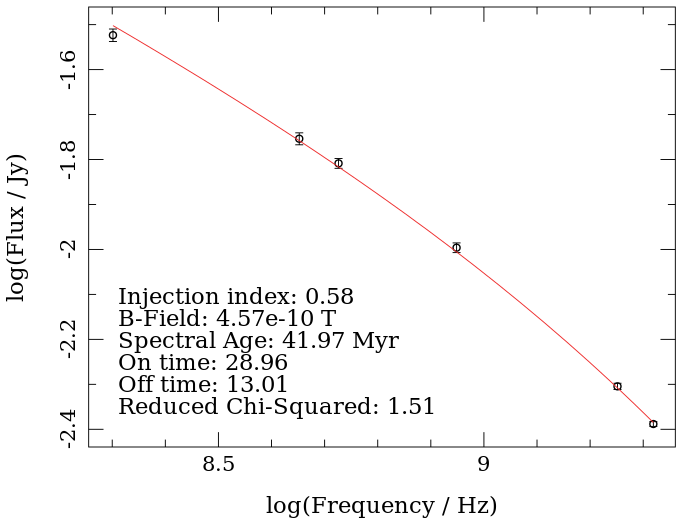}
	\includegraphics[scale=0.029]{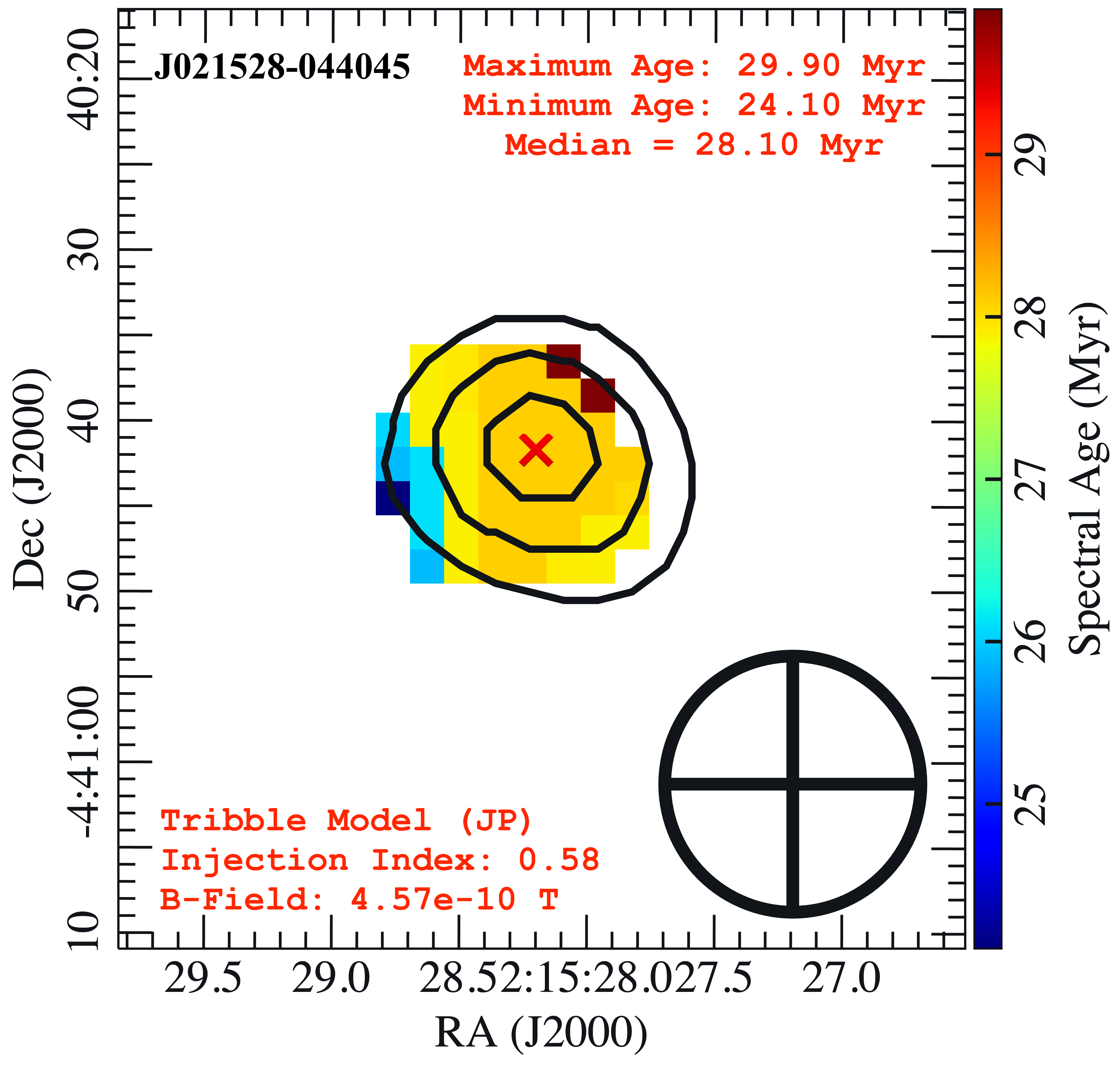}
	\includegraphics[scale=0.215]{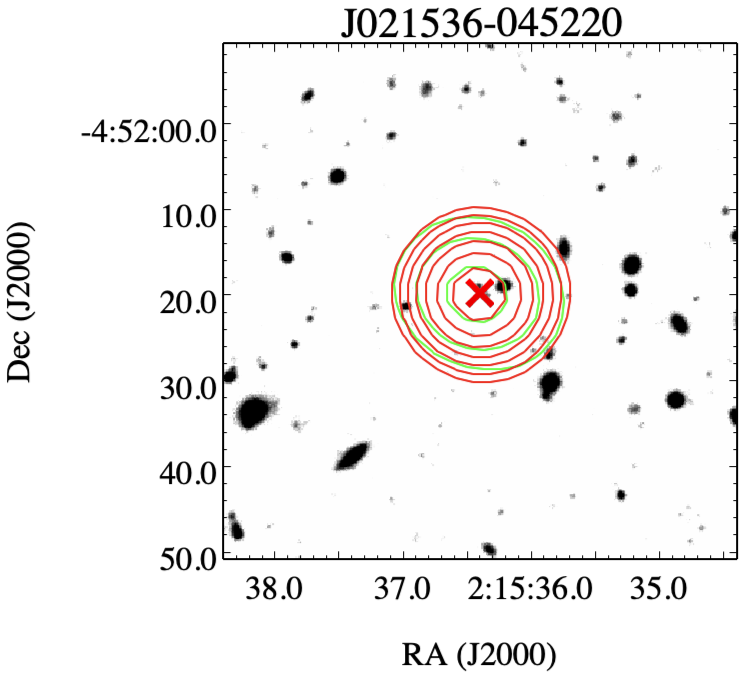}
	\includegraphics[scale=0.26]{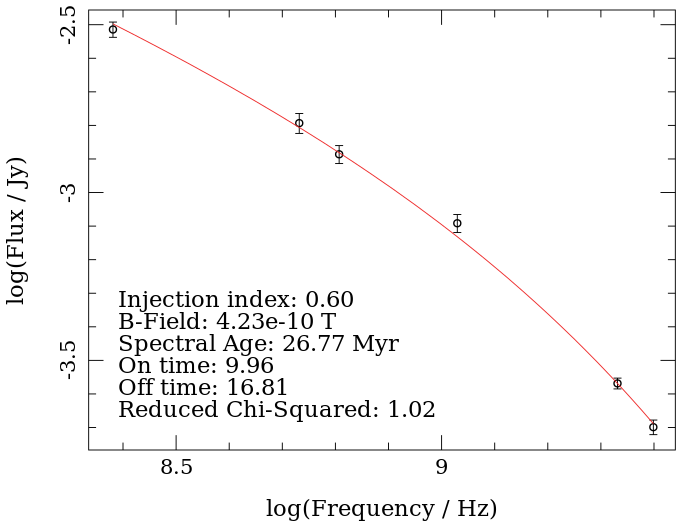}
	\includegraphics[scale=0.0524]{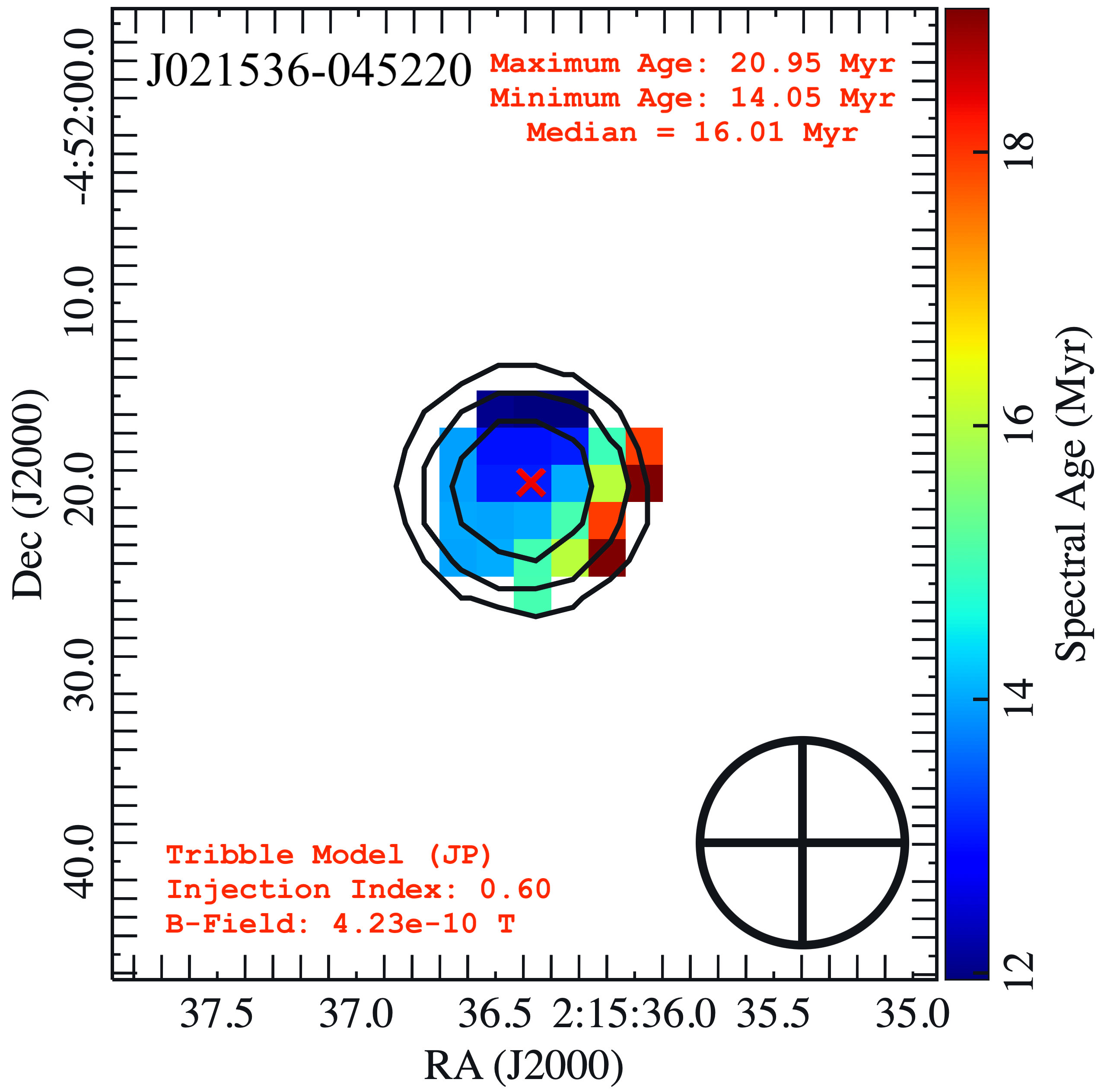}
	\includegraphics[scale=0.183]{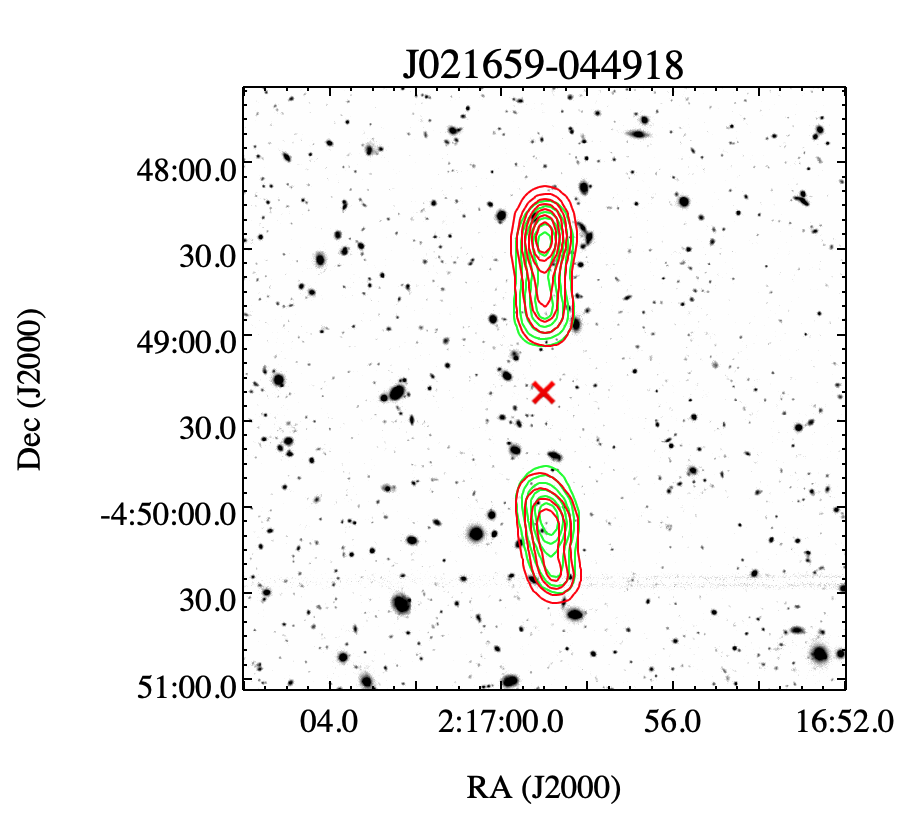}
	\includegraphics[scale=0.26]{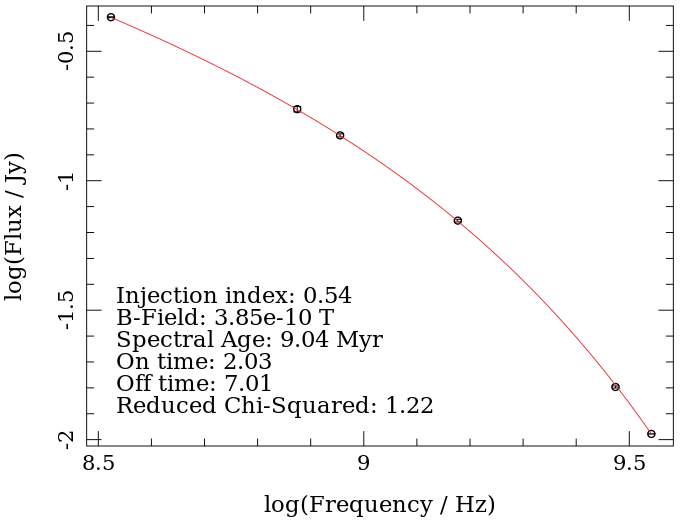}
	\includegraphics[scale=0.0524]{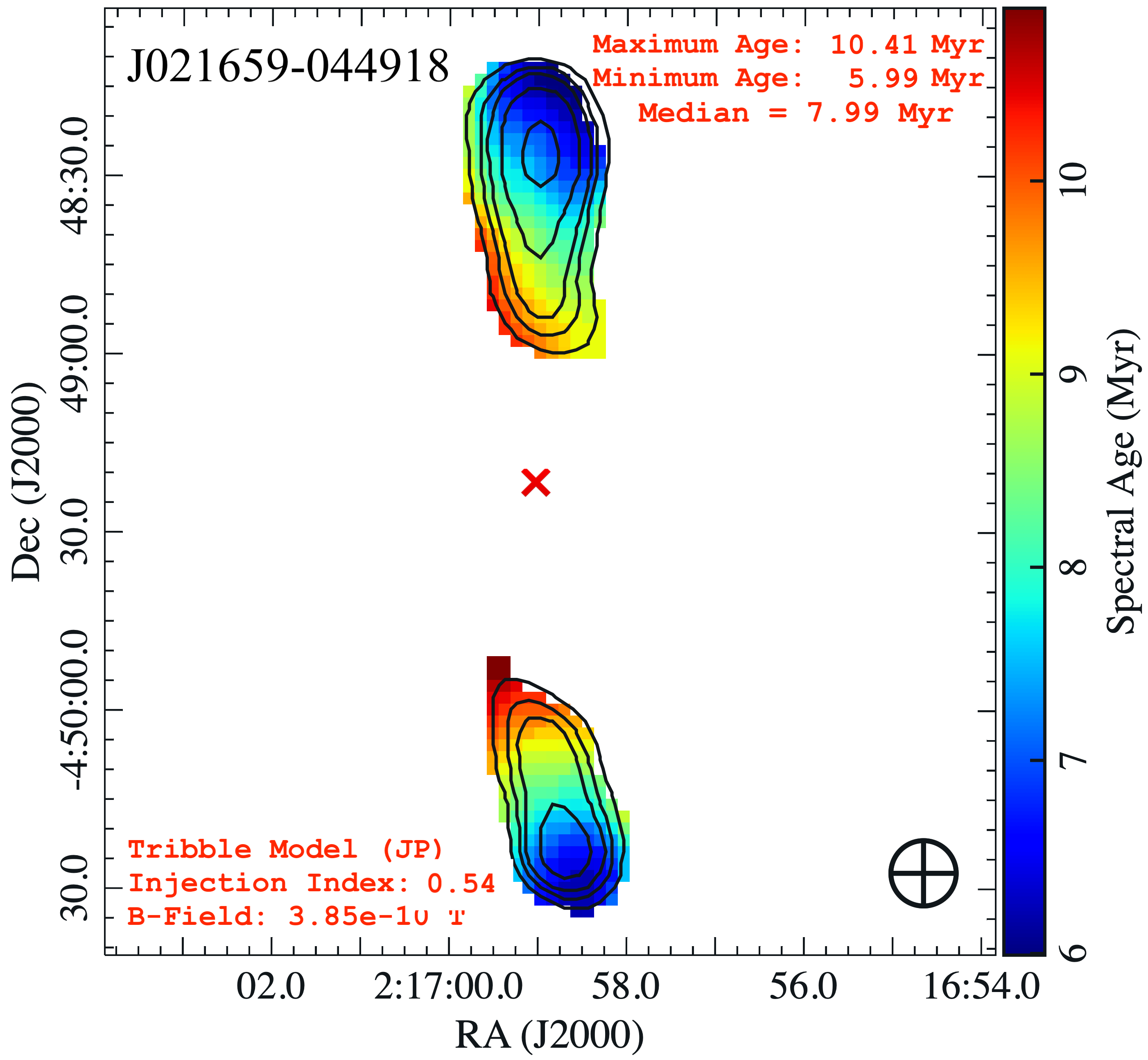}
	\includegraphics[scale=0.178]{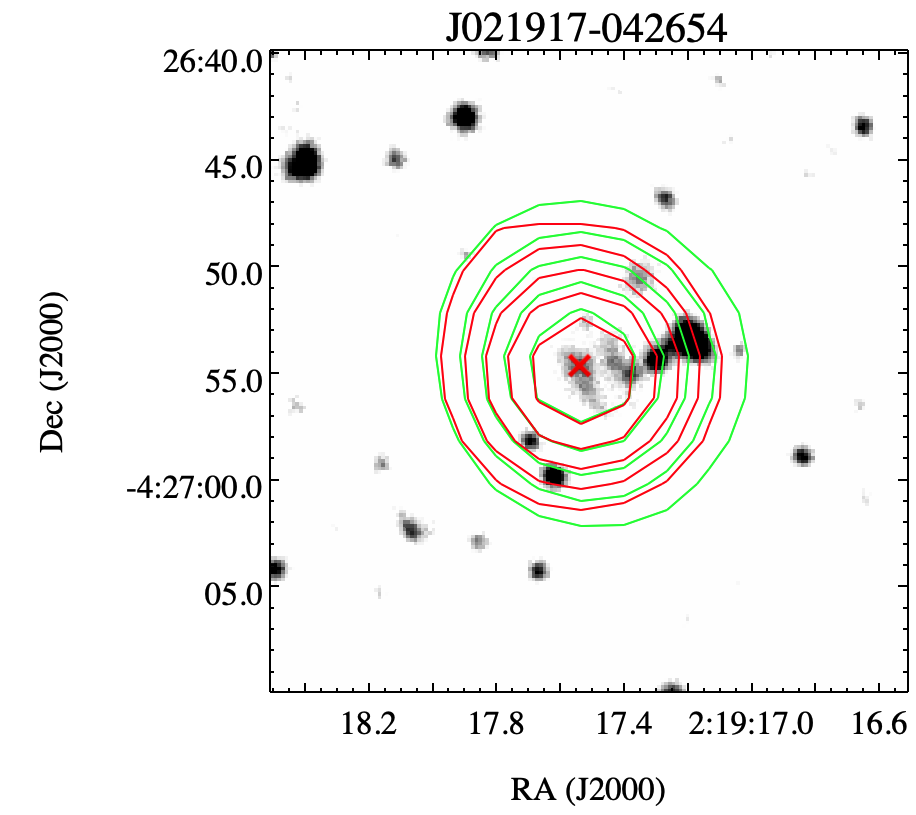}
	\includegraphics[scale=0.26]{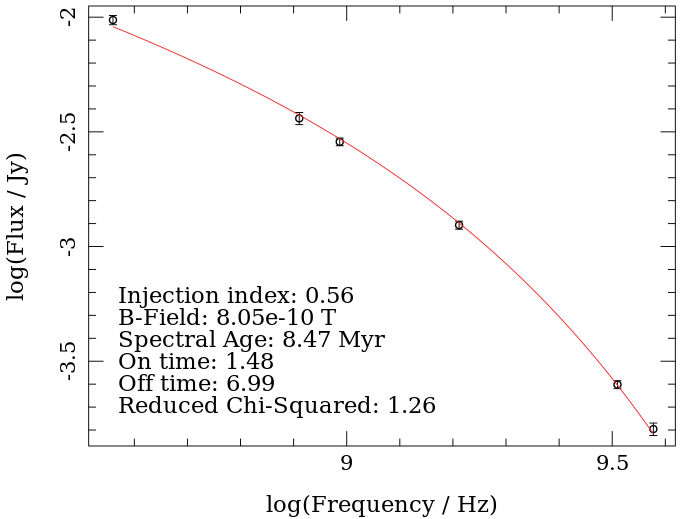}
	\includegraphics[scale=0.0523]{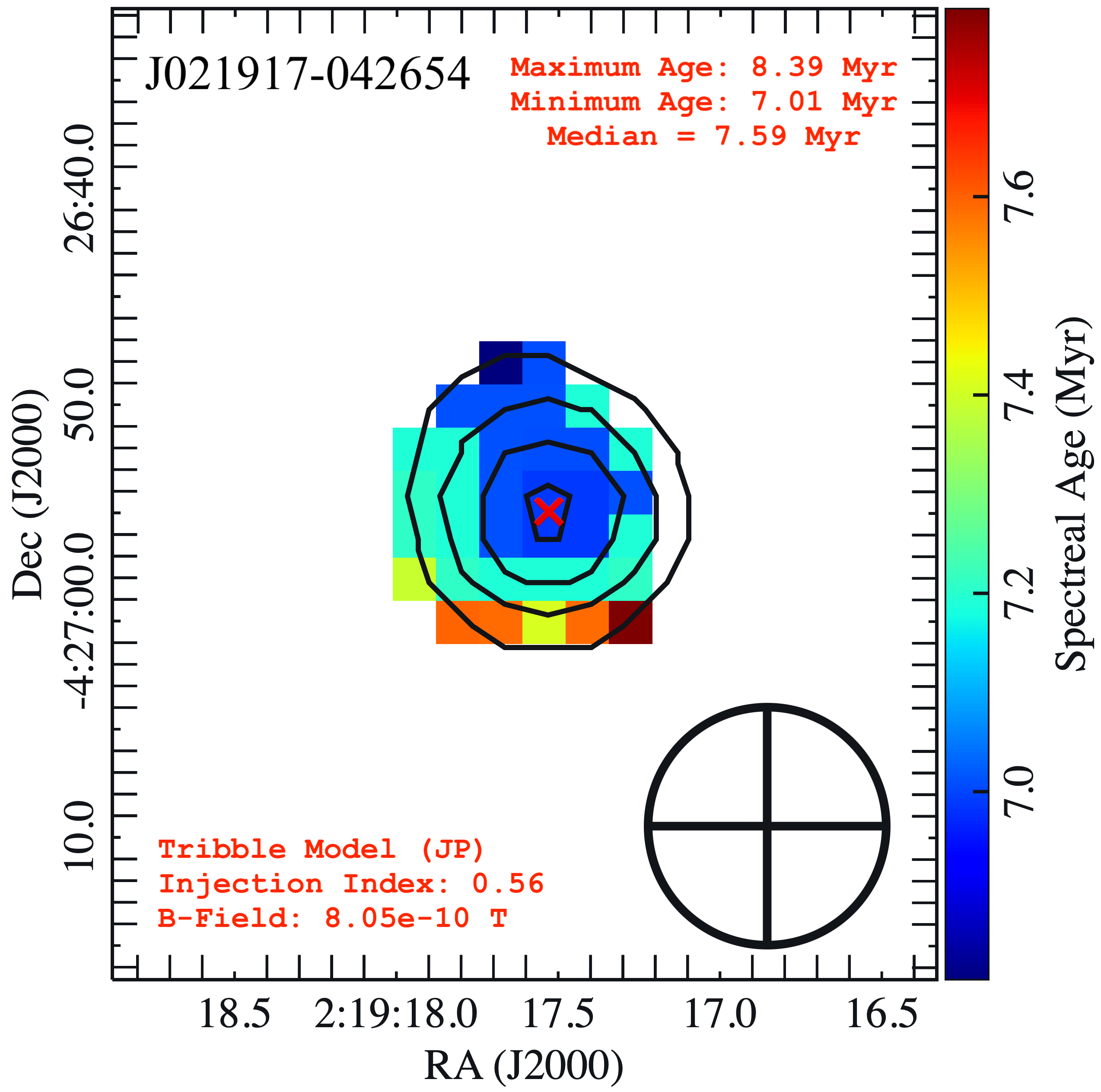}
	\caption{ {\it Left Panel} : Band-3 (in green) uGMRT and 1.284 GHz (in red) MeerKAT radio contours overplotted onto the grey-scale HSC-SSP $i-$band image. The radio contour levels are at 3$\sigma$ $\times$ (1, 2, 4, 8, 16 ...) and the corresponding optical image is logarithmically scaled. The magenta cross is the position of the potential host galaxy. Only 12/14 genuine remnant sources are shown in this figure. {\it Middle panel} : Best fit radio SEDs of the corresponding remnant sources. The solid red curve represents model fitted to the data points. {\it Right panel} : Spectral age maps derived by JP Tribble model of the corresponding remnant sources. The contours overlaid on the maps are 1.284 GHz MeerKAT at 8\farcs2 resolution and the levels are at 3$\sigma$ $\times$ (1, 2, 4, 8, 16 ...). The position of the host galaxy is marked with red cross. The open circle with a cross represents a circular PSF beam of 15$^{\prime\prime}$. }
	\label{fig:SED1}
\end{figure*}

Spectral age maps for our remnant sources, modelled using the JP Tribble model, are presented in the right panel of Figure~\ref{fig:SED1}, with 1284 MHz MeerKAT radio contours overlaid to illustrate the spatial distribution of synchrotron-emitting regions and their spectral ageing characteristics. 

\begin{table*}
	\begin{minipage}{180mm}
		\centering
		\caption{Estimates of the parameters obtained after performing the spectral age analysis using the JP Tribble model.}
		\label{tab:table4}
		\begin{tabular}{ccccccccc}
			\hline
			Source & Redshift & Magnetic Field & Injection & Min. Age & Max. Age & Mean Age & Median Age & Average \\
			Name & ($z$) & ($\mu$G) & Index & (Myr) & (Myr) & (Myr) & (Myr) & $\chi_{\rm red}^2$ \\
			\hline 
			J021528-044045 & 0.35$^{\rm +0.003}_{\rm -0.003}$ & 4.57 & -0.58 & 16.44$^{+5.46}_{-8.12}$ & 43.43$^{+3.79}_{-4.99}$ & 30.94$^{+2.50}_{-3.17}$ & 31.46$^{+2.18}_{-2.62}$ & 1.23 \\  
			
			J021536-045220 & 0.67$^{\rm +0.07}_{\rm -0.11}$ & 4.23 & -0.60 & 14.05$^{+3.60}_{-6.01}$ & 20.95$^{+4.56}_{-7.40}$ & 16.28$^{+3.54}_{-5.11}$ & 16.01$^{+3.79}_{-4.90}$ & 0.73 \\ 
			
			J021646-051004 & 2.85$^{+0.20}_{-0.16}$ & 5.66 & -0.54 & 0.52$^{+0.06}_{-0.04}$ & 0.84$^{+0.06}_{-0.03}$ & 0.67$^{+0.06}_{-0.08}$ & 0.68$^{+0.06}_{-0.07}$ & 1.27 \\ 
			
			J021659-044918 & 1.32$^{\rm +0.001}_{\rm -0.001}$ & 3.85 & -0.54 & 5.99$^{+0.01}_{-0.03}$ & 10.41$^{+0.72}_{-0.92}$ & 7.93$^{+0.15}_{-0.17}$ & 7.99$^{+0.12}_{-0.13}$ & 1.95 \\
			
			J021917-042654 & 1.52$^{\rm +0.04}_{\rm -0.14}$ & 8.05 & -0.56 & 7.01$^{+0.38}_{-0.67}$ & 8.39$^{+0.73}_{-0.96}$ & 7.57$^{+0.63}_{-0.70}$ & 7.59$^{+0.61}_{-0.69}$ & 2.82 \\
			
			J021926-051535 & 1.47$^{\rm +0.03}_{\rm -0.03}$ & 8.58 & -0.54 & 2.21$^{+0.10}_{-0.16}$ & 5.81$^{+1.38}_{-2.25}$ & 4.46$^{+0.44}_{-0.58}$ & 4.59$^{+0.34}_{-0.43}$ & 0.94 \\ 
			
			J022106-043925 & 0.99$^{\rm +0.03}_{\rm -0.03}$ & 11.99 & -0.62 & 0.00$^{+3.14}_{-0.00}$ & 8.09$^{+1.01}_{-1.27}$ & 1.64$^{+1.54}_{-1.64}$ & 1.50$^{+1.27}_{-1.50}$ & 1.87 \\
			
			J022231-042757 & 2.22$^{\rm +0.31}_{\rm -0.07}$ & 9.20 & -0.54 & 1.05$^{+0.01}_{-0.03}$ & 2.87$^{+0.48}_{-0.78}$ & 1.76$^{+0.12}_{-0.15}$ & 1.68$^{+0.10}_{-0.11}$ & 1.88 \\
			
			J022302-042850 & 1.18$^{\rm +0.001}_{\rm -0.001}$ & 6.06 & -0.60 & 2.11$^{+1.52}_{-2.11}$ & 8.68$^{+0.78}_{-1.10}$ & 5.41$^{+0.50}_{-0.62}$ & 5.39$^{+0.39}_{-0.45}$ & 1.81 \\
			
			J022305-041232 & 1.05$^{\rm +0.007}_{\rm -0.007}$ & 6.29 & -0.57 & 0.24$^{+0.46}_{-0.24}$ & 11.48$^{+4.59}_{-5.53}$ & 5.58$^{+1.74}_{-2.44}$ & 5.08$^{+1.67}_{-2.23}$ & 1.26 \\
			
			J022318-044522 & 1.26$^{\rm +0.01}_{\rm -0.02}$ & 4.30 & -0.53 & 3.99$^{+0.03}_{-0.07}$ & 8.51$^{+1.31}_{-2.23}$ & 6.31$^{+0.43}_{-0.50}$ & 6.24$^{+0.38}_{-0.41}$ & 1.42 \\
			
			J022338-045418 & 0.81$^{\rm +0.001}_{\rm -0.001}$ & 2.37 & -0.65 & 0.51$^{+0.80}_{-0.51}$ & 11.48$^{+3.46}_{-5.02}$ & 5.02$^{+1.92}_{-2.19}$ & 5.48$^{+1.92}_{-2.24}$ & 1.61 \\
			
			J022413-044643 & 1.24$^{\rm +0.02}_{\rm -0.02}$ & 6.02 & -0.68 & 5.39$^{+2.50}_{-4.56}$ & 8.38$^{+1.26}_{-1.46}$ & 7.63$^{+1.00}_{-1.16}$ & 7.79$^{+0.81}_{-0.84}$ & 1.53 \\
			
			J022433-043709 & 2.64$^{+0.18}_{-0.05}$ & 9.23 & -0.52 & 0.60$^{+0.02}_{-0.03}$ & 0.90$^{+0.03}_{-0.02}$ & 0.78$^{+0.05}_{-0.03}$ & 0.80$^{+0.04}_{-0.03}$ & 1.24 \\
			\hline
		\end{tabular}
	\end{minipage}
\end{table*}

\addtocounter{figure}{-1}
\begin{figure*}
	\centering
	\includegraphics[scale=0.179]{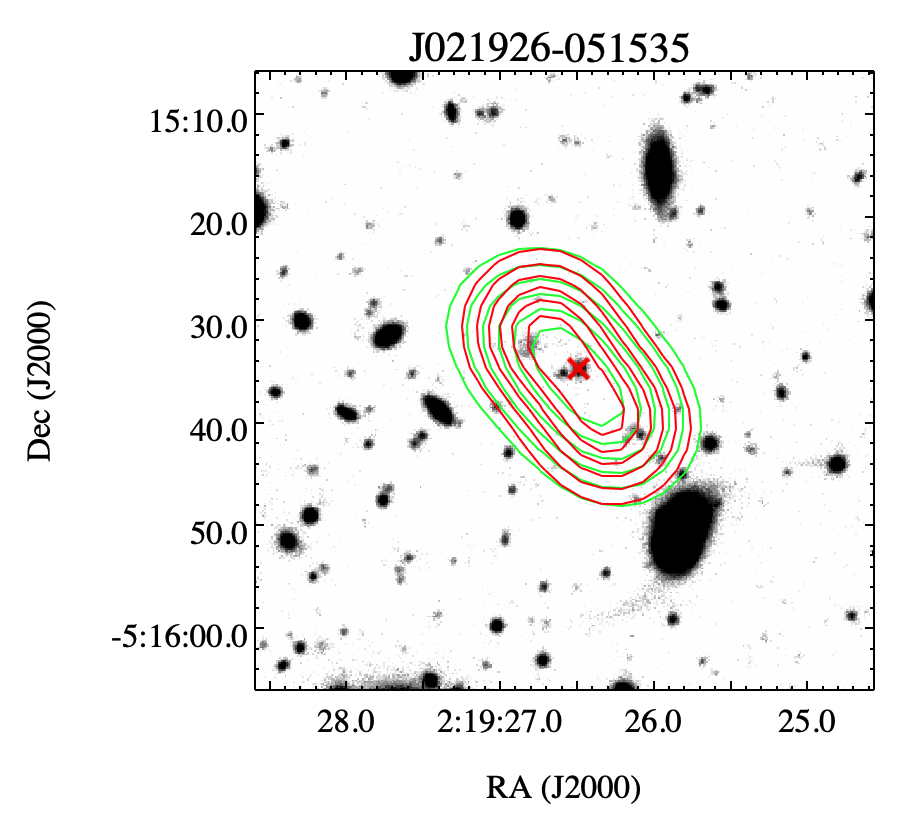}
	\includegraphics[scale=0.26]{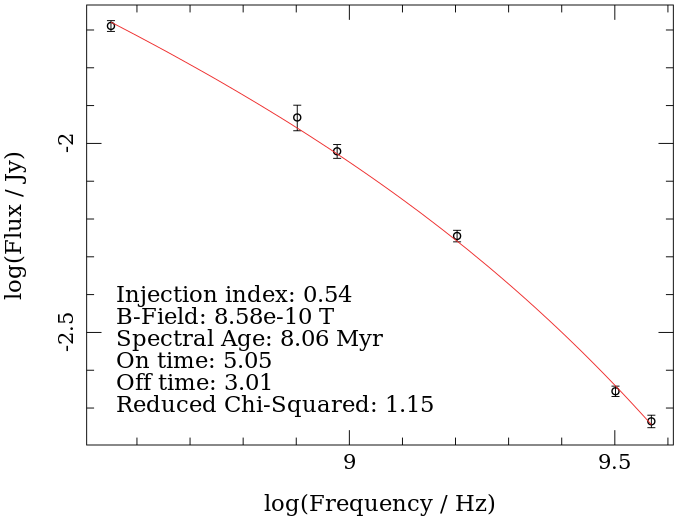}
	\includegraphics[scale=0.0523]{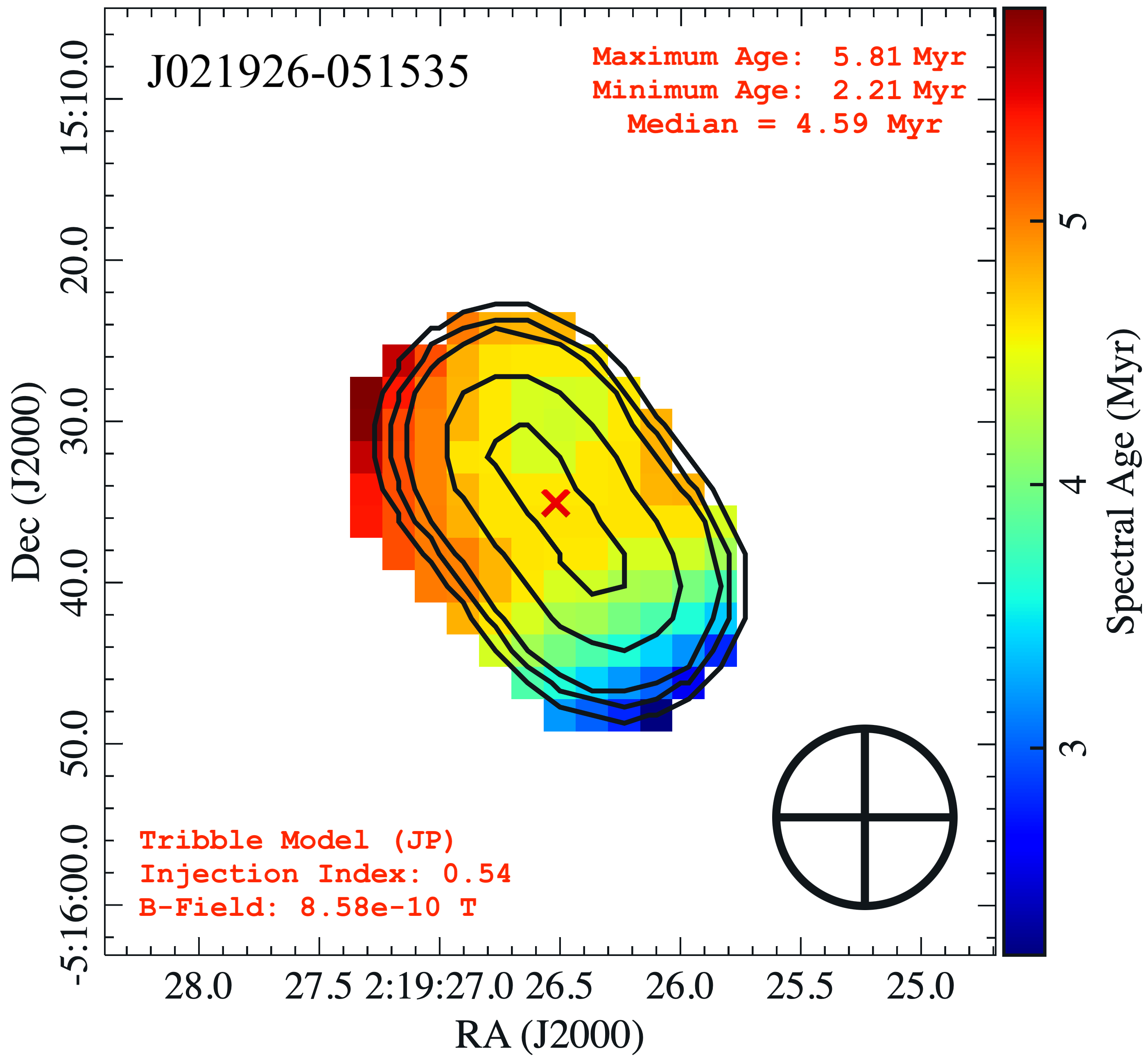}
	\includegraphics[scale=0.22]{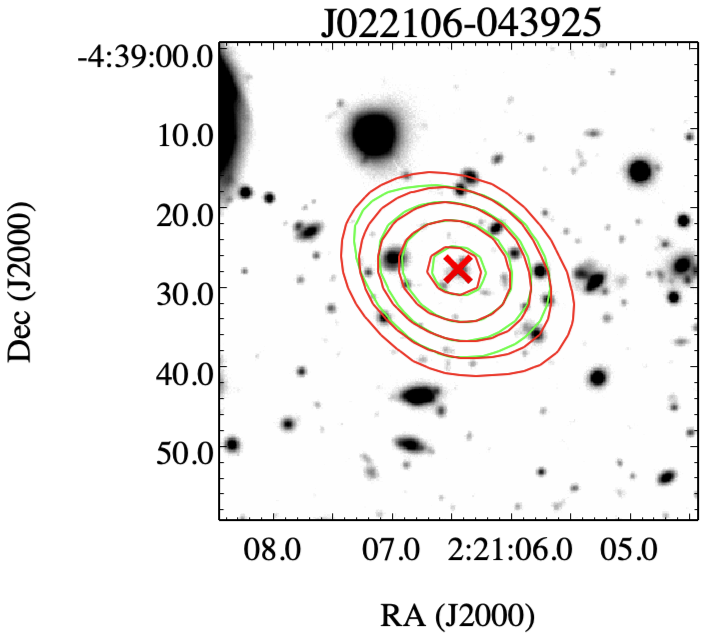}
	\includegraphics[scale=0.26]{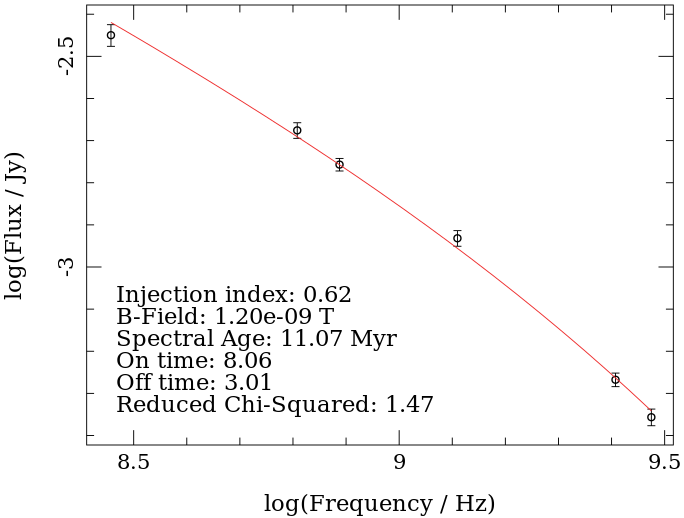}
	\includegraphics[scale=0.0524]{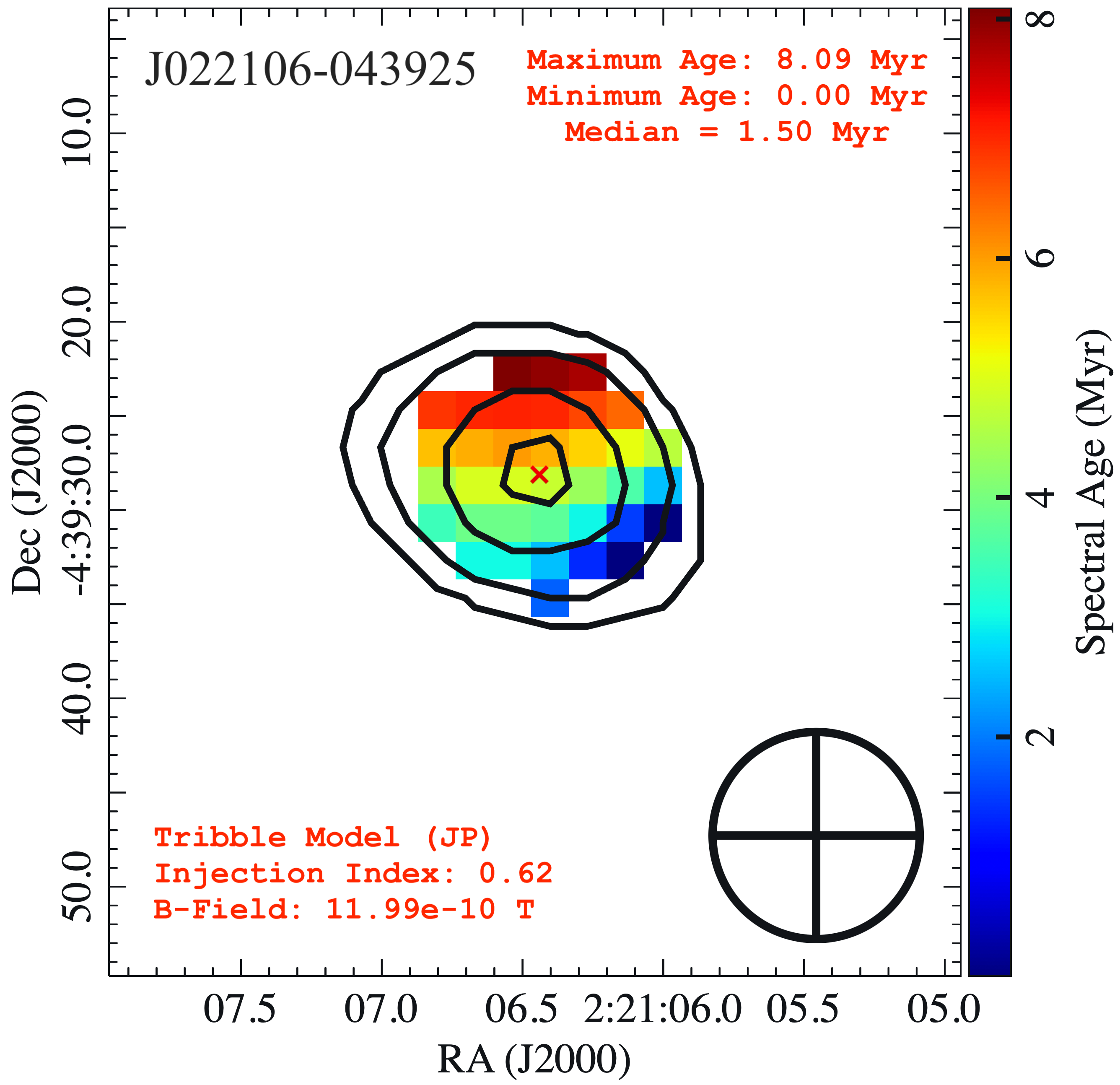}
	\includegraphics[scale=0.18]{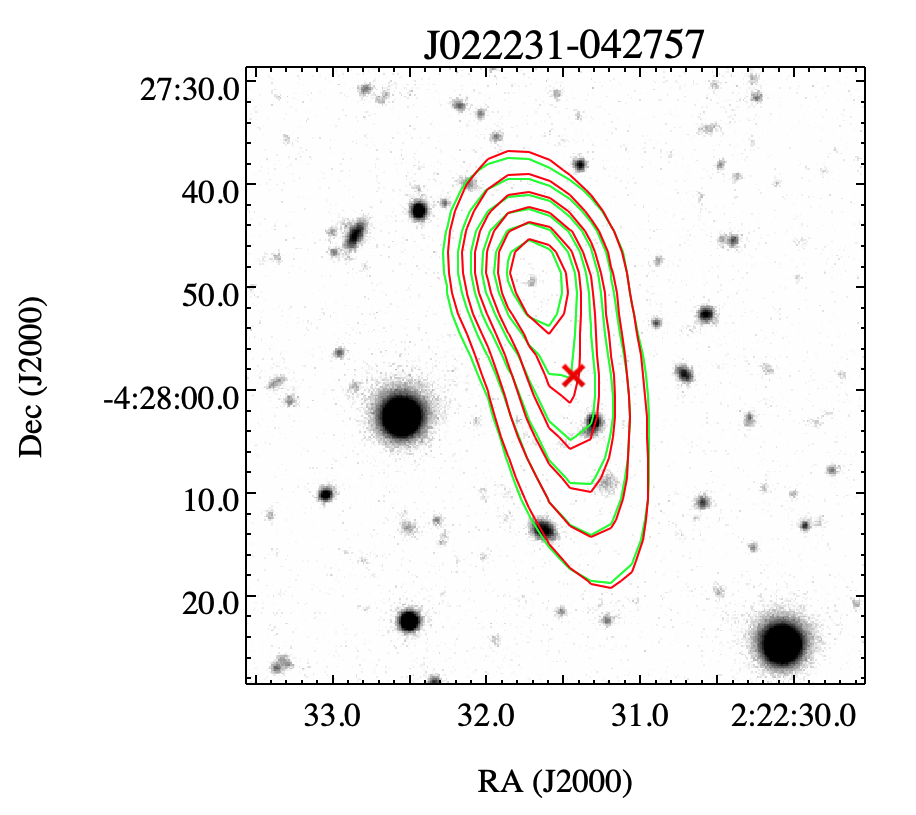}
	\includegraphics[scale=0.26]{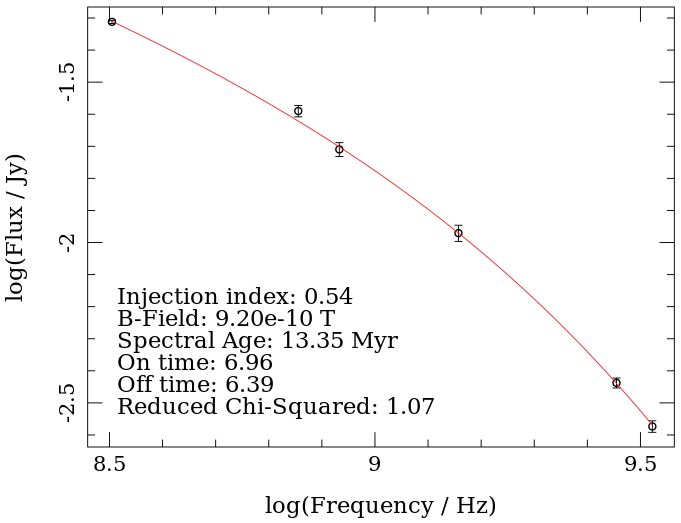}
	\includegraphics[scale=0.0523]{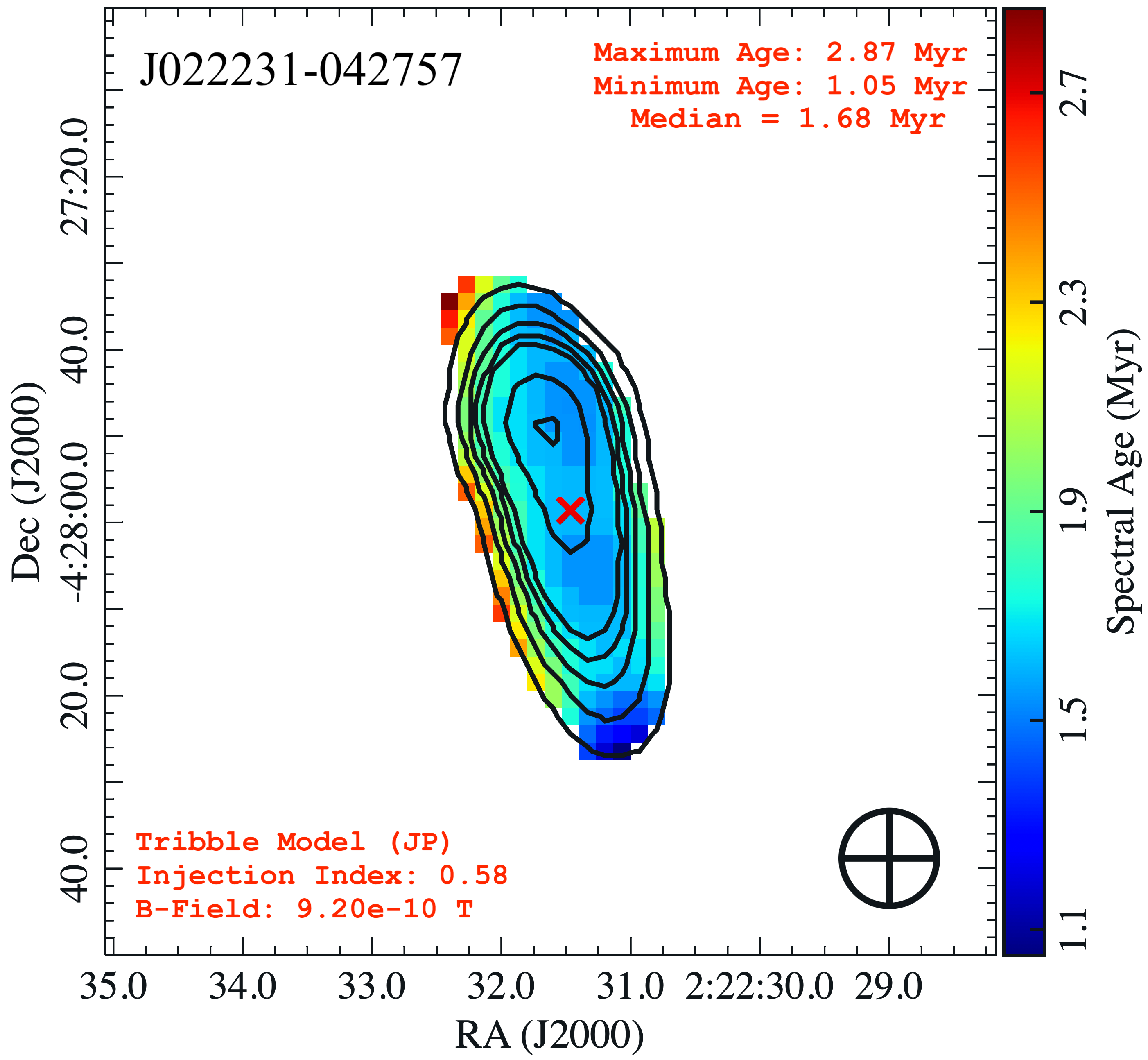}
	\includegraphics[scale=0.18]{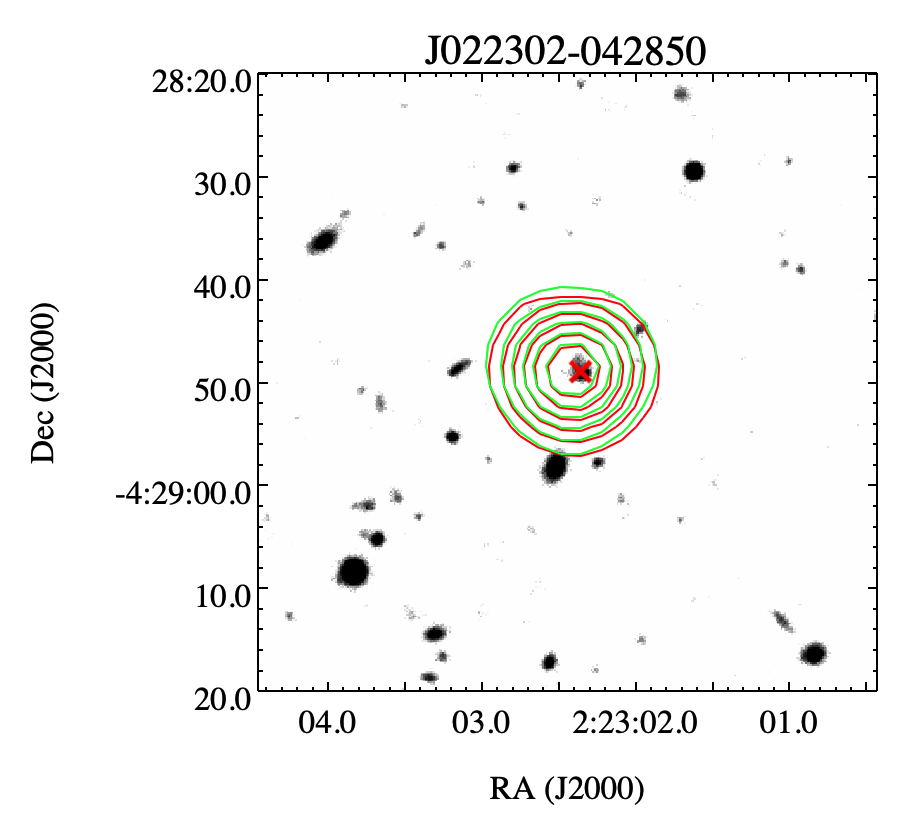}
	\includegraphics[scale=0.26]{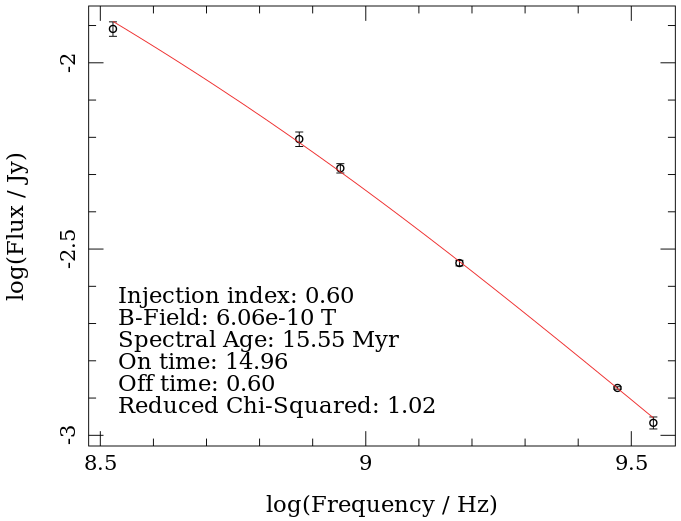}
	\includegraphics[scale=0.0524]{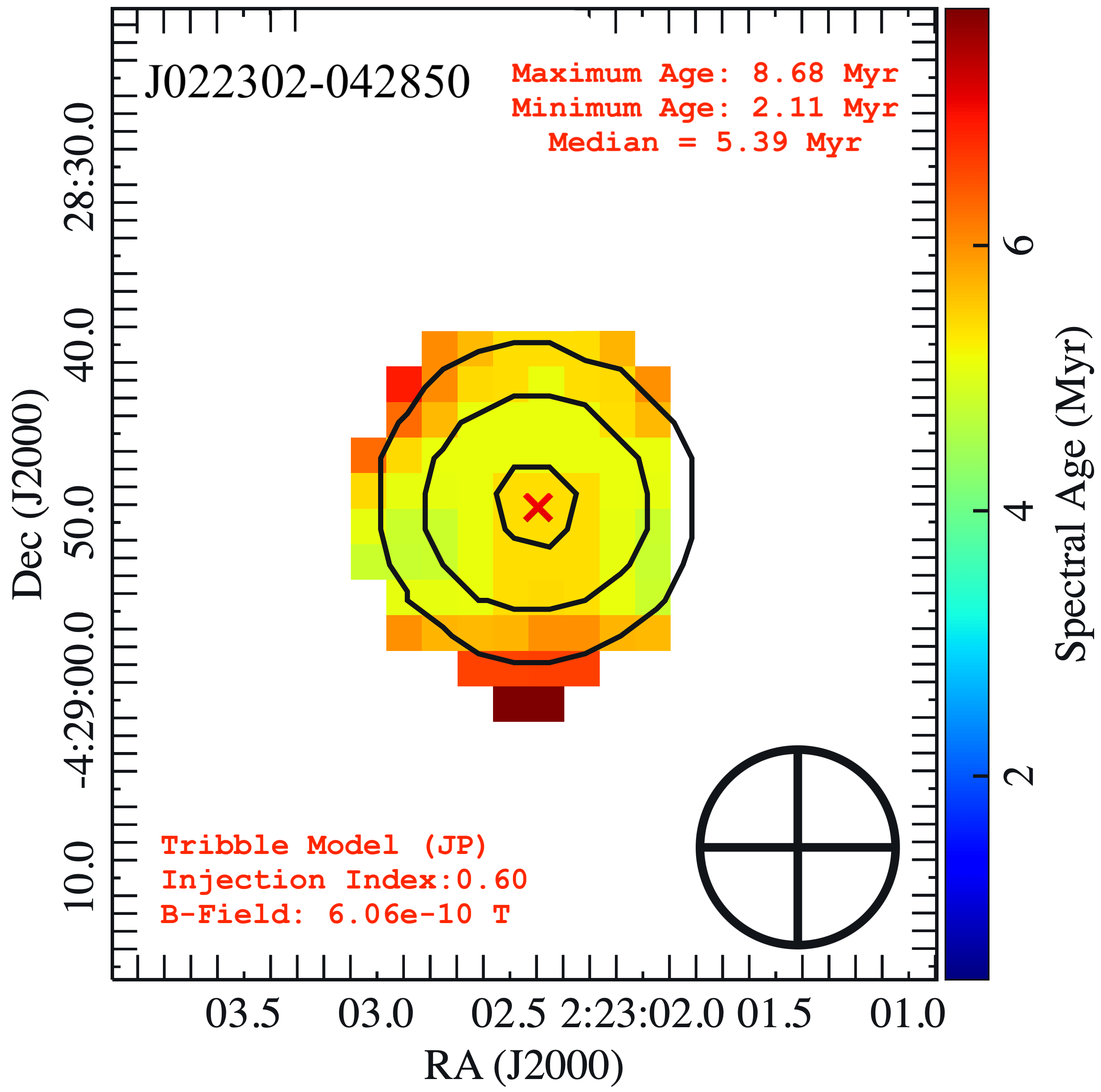}
	\label{fig:SED2}
\end{figure*}

\addtocounter{figure}{-1}

\begin{figure*}
	\centering
	\includegraphics[scale=0.195]{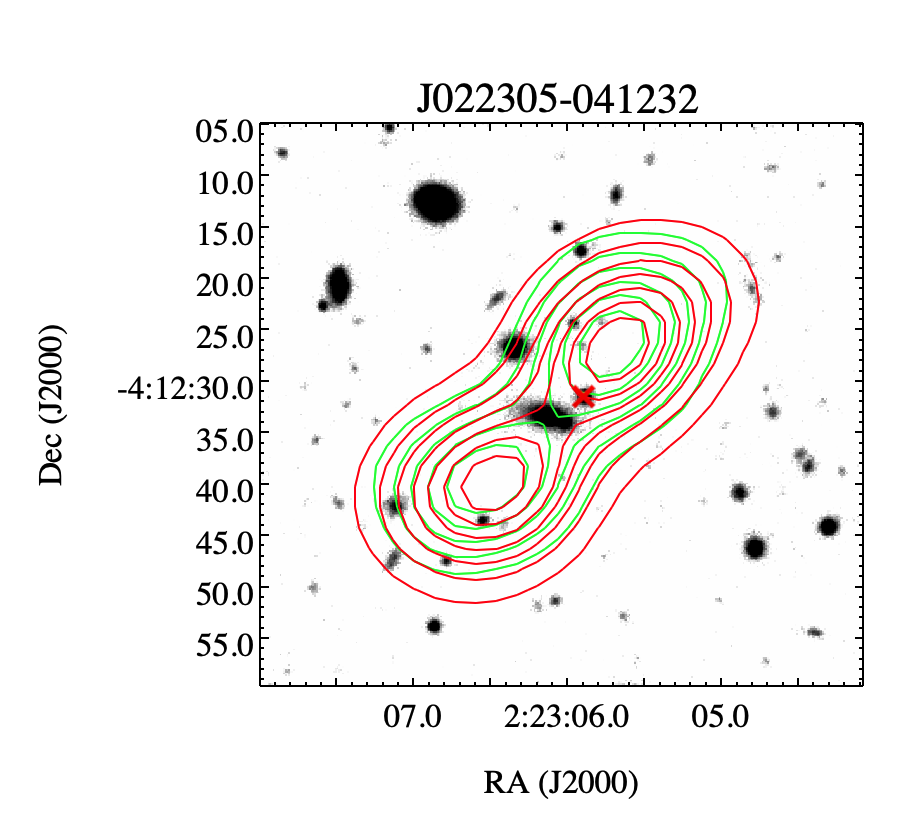}
	\includegraphics[scale=0.26]{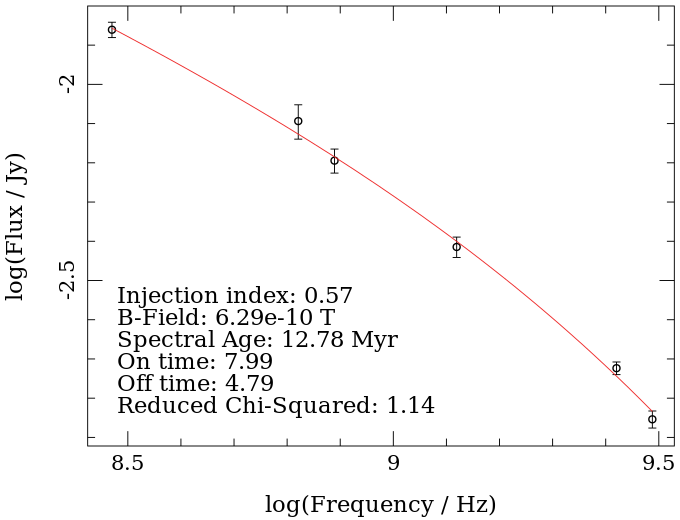}
	\includegraphics[scale=0.0509]{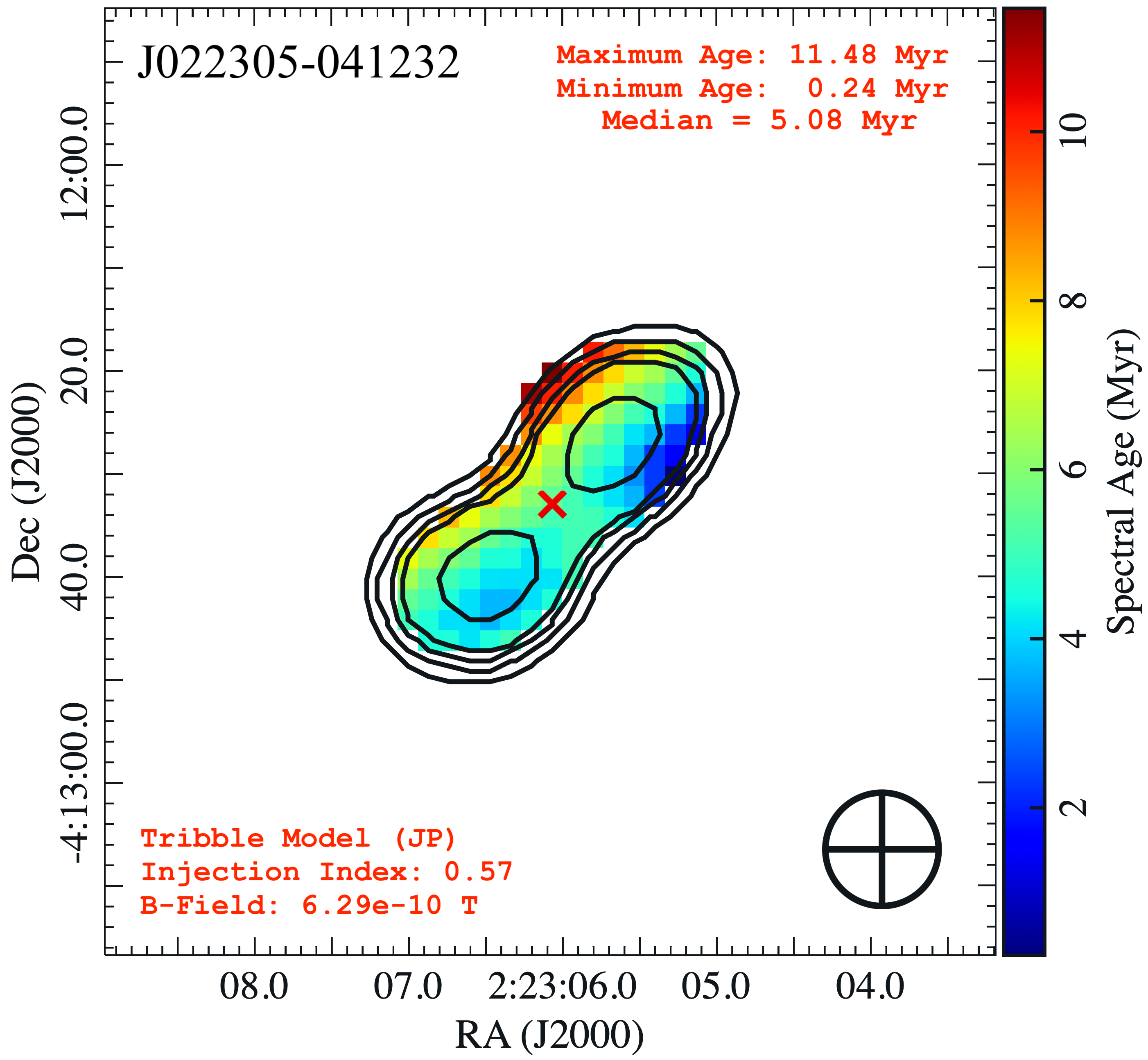}
	\includegraphics[scale=0.184]{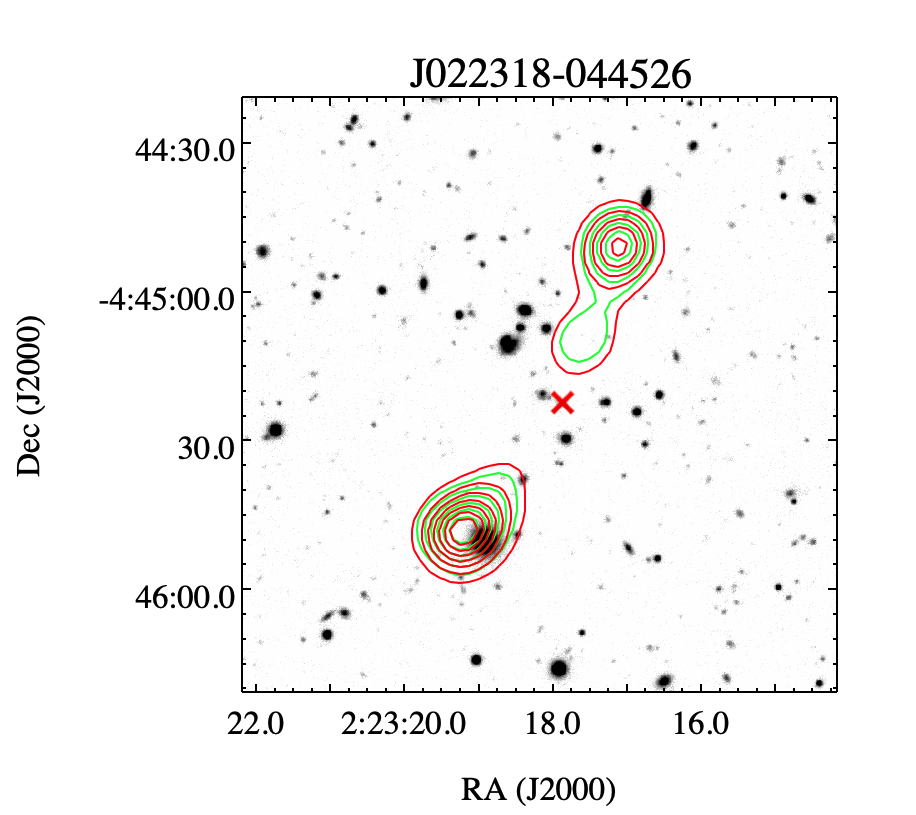}
	\includegraphics[scale=0.26]{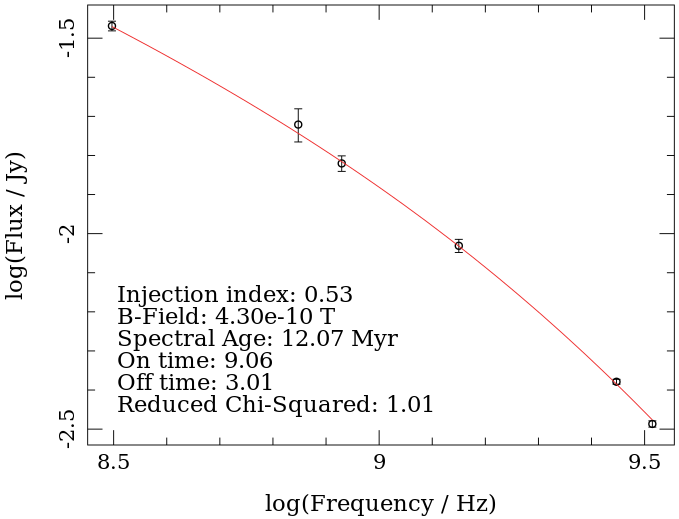}
	\includegraphics[scale=0.0524]{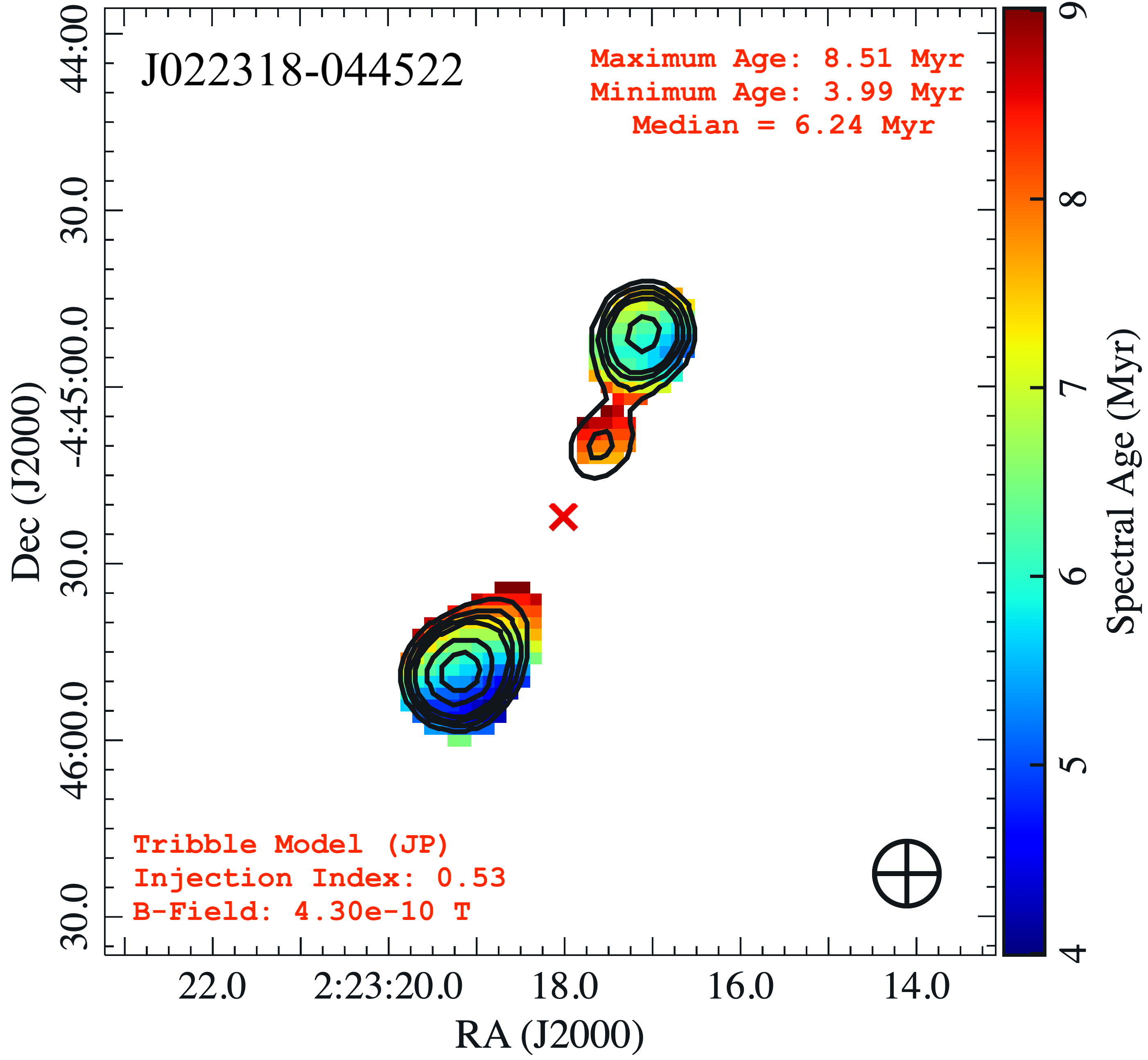}
	\includegraphics[scale=0.178]{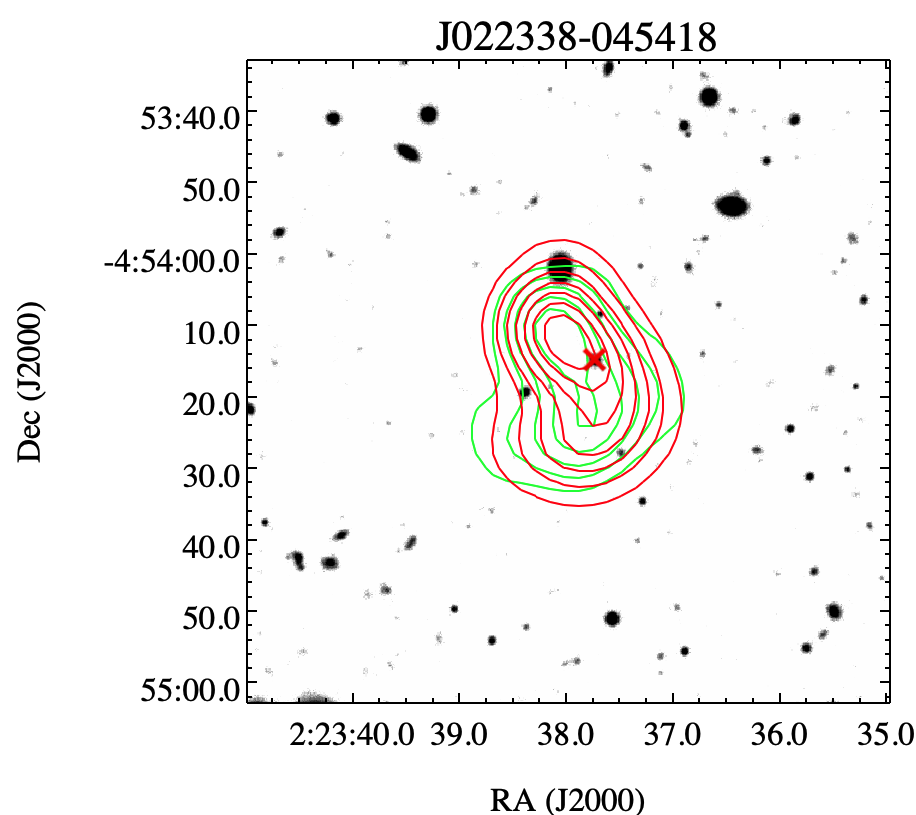}
	\includegraphics[scale=0.26]{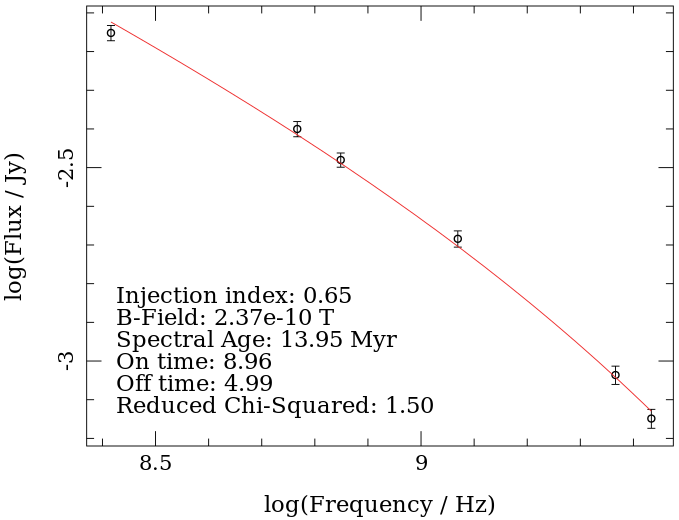}
	\includegraphics[scale=0.0524]{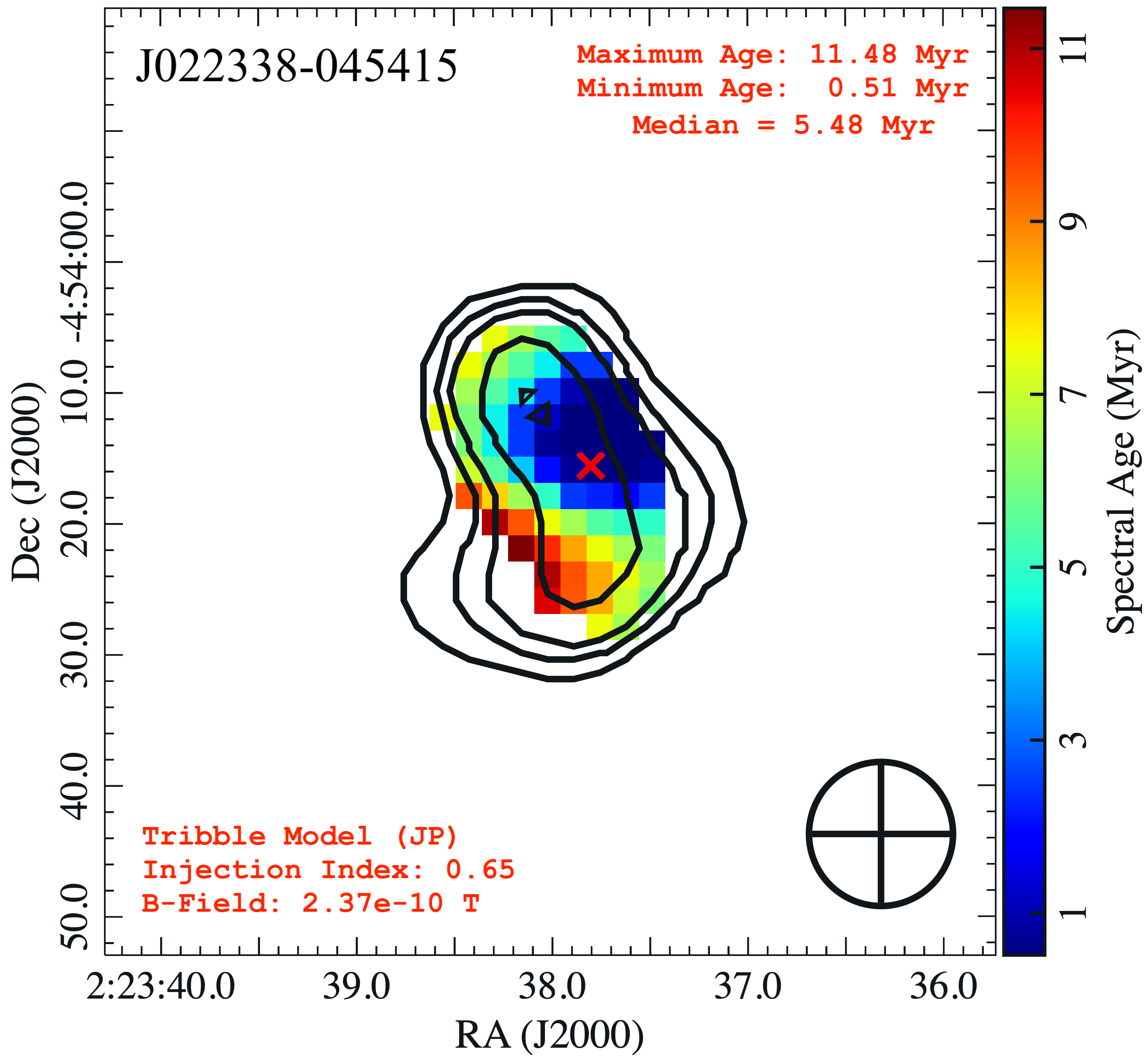}
	\includegraphics[scale=0.18]{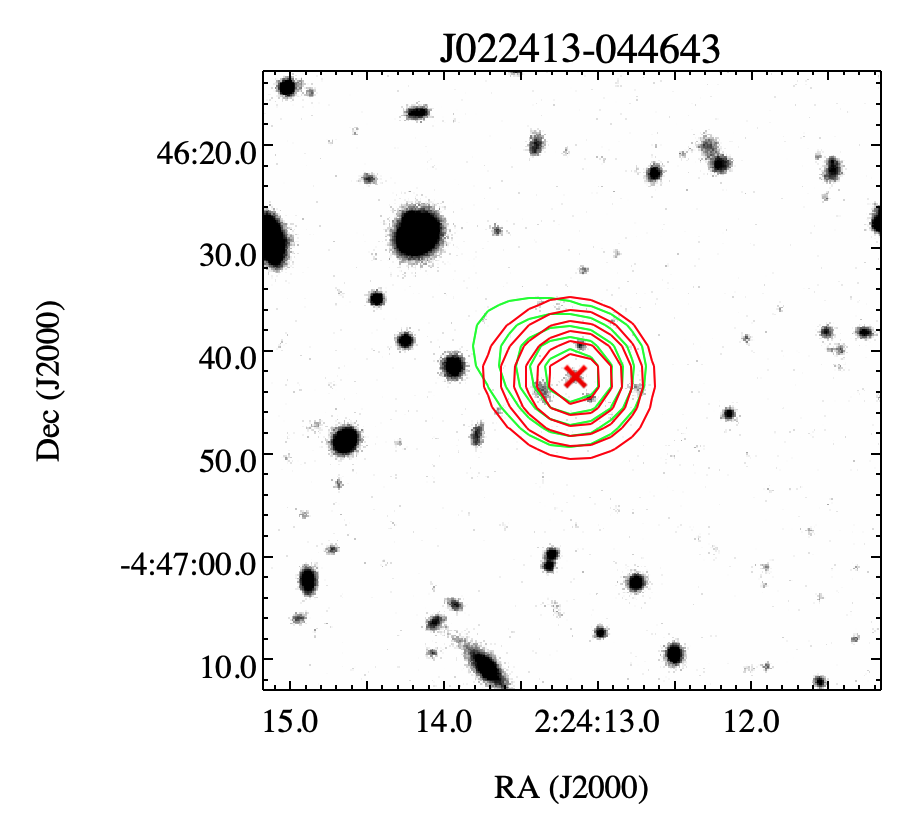}
	\includegraphics[scale=0.26]{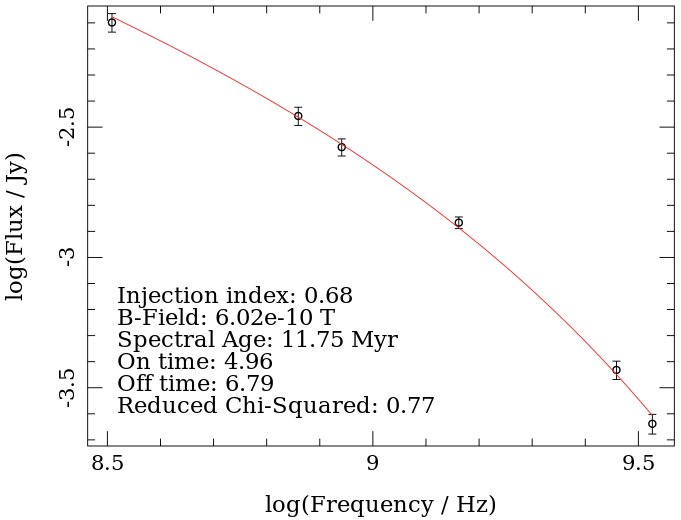}
	\includegraphics[scale=0.0522]{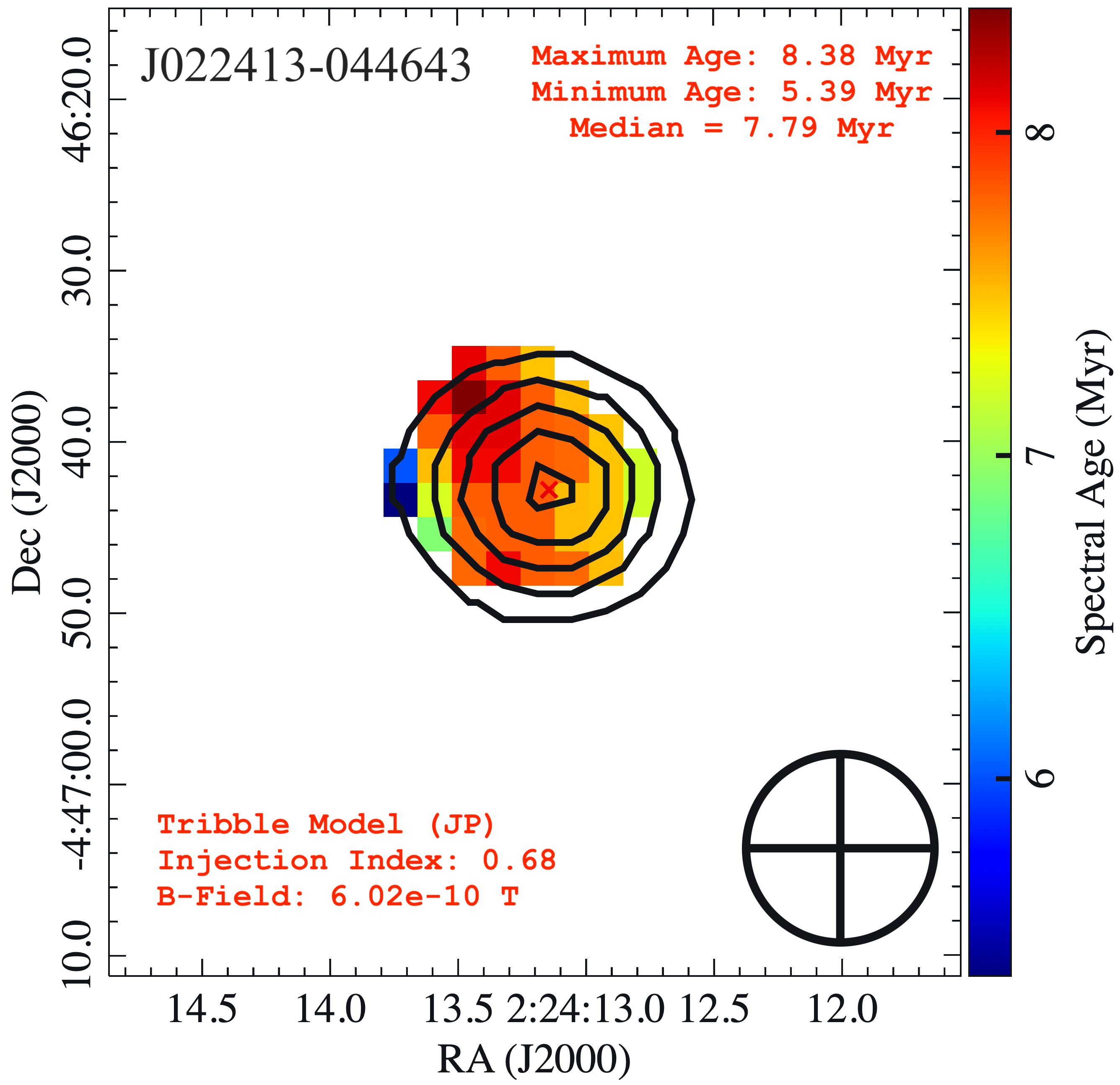}
	\label{fig:SED3}
\end{figure*}

Our analysis revealed that the well-resolved sources (J021659-044918 and J022318-044522) display steeper spectral indices in the central regions of the lobes and flatter indices toward the core and peripheral regions. This is consistent with the ``back-flow" model of plasma movement within the lobes \citep[see][]{Leahy84,Leahy89,Capetti02,Hodges11,Rossi17}.

\section{Results}
\label{sec:results}
Of the 14 sources analyzed, 12 are best fit with the CI$_{\rm OFF}$ model and are consistent with being remnant radio galaxies. Two sources (J021646–051004 and J022433–043709) are best fitted with CI$_{\rm ON}$ model, {\em i. e.,} require ongoing particle injection and are therefore classified as active (see Figure~\ref{fig:SED4}). This confirms that the majority of our sample consists of genuine remnants, demonstrating the efficiency of our selection criteria. These two sources were originally classified as remnant candidates by \cite{Singh21} based on radio SEDs with only three  (150 MHz, 325 MHz and 1.4 GHz) data points. Their re-classification here as active systems underscores the critical role of sensitive, broad-band, multi-frequency radio observations in reliably identifying the genuine remnants. Additionally, the reduced $\chi^2$ values for most of the sources in our sample are close to 1, indicating that the SED modeling is statistically robust and provides a good fit to the data. We note that the typical uncertainties in spectral age estimates are below 10\%. However, these quoted uncertainties arise from model fitting, while the true errors may be larger due to various underlying assumptions in spectral age calculations \citep{Harwood17}. Therefore, we emphasize that spectral age estimates should be interpreted as indicative timescales subject to systematic uncertainties arising from assumptions on magnetic field strength, source geometry, and spectral model limitations, and should be treated with appropriate caution.

%
\addtocounter{figure}{+2}
\begin{figure*}
	\centering
	\includegraphics[scale=0.21]{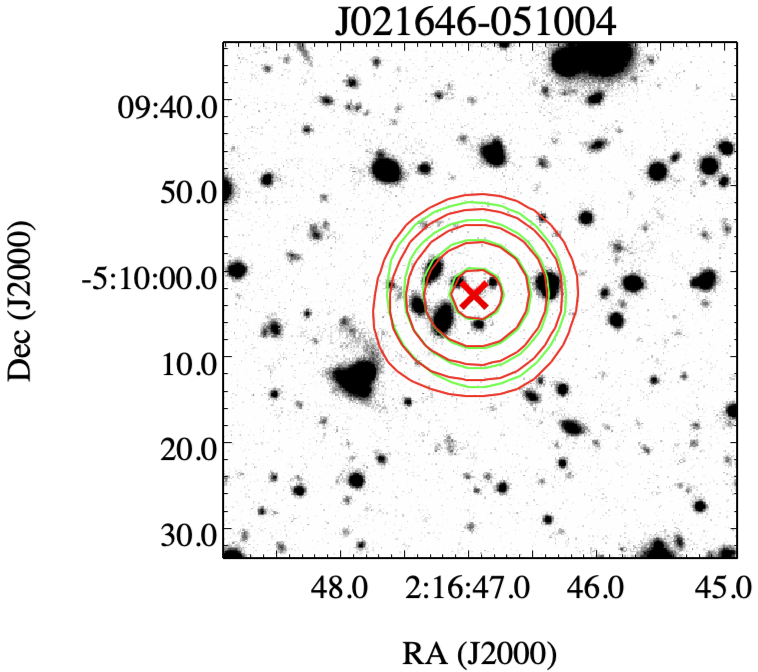}
	\includegraphics[scale=0.26]{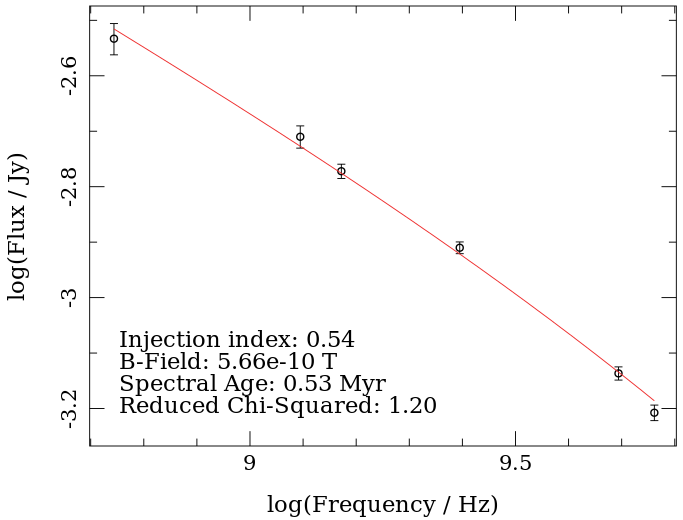}
	\includegraphics[scale=0.029]{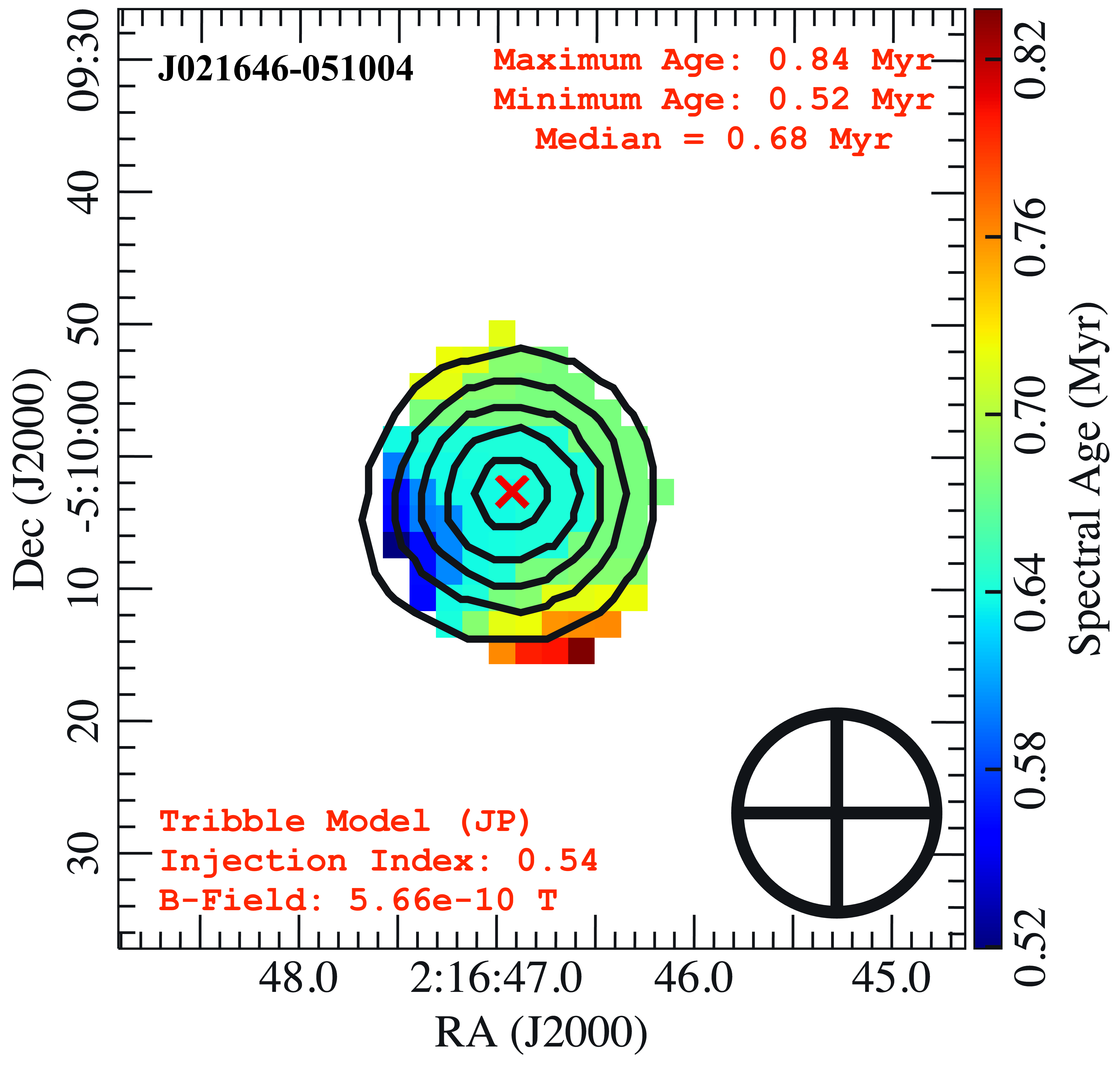}
	\includegraphics[scale=0.21]{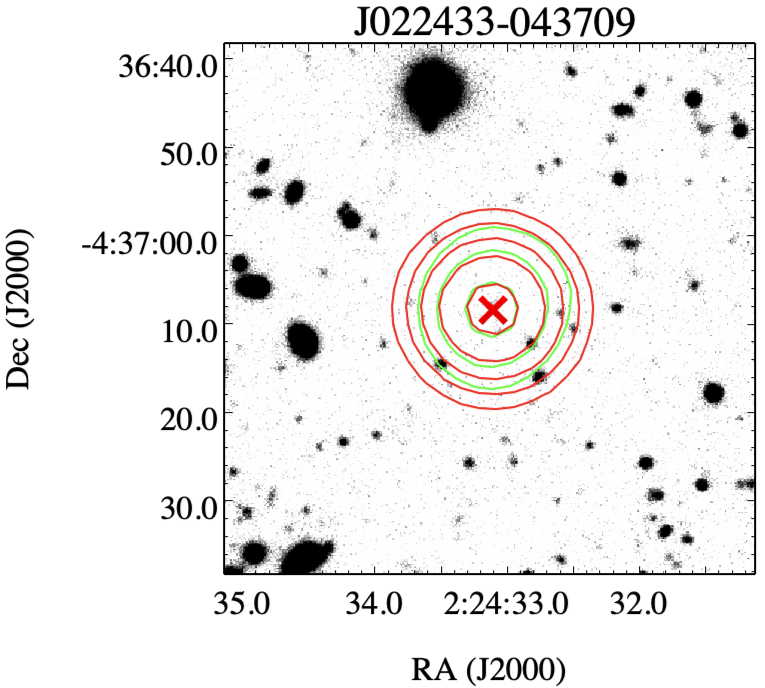}
	\includegraphics[scale=0.26]{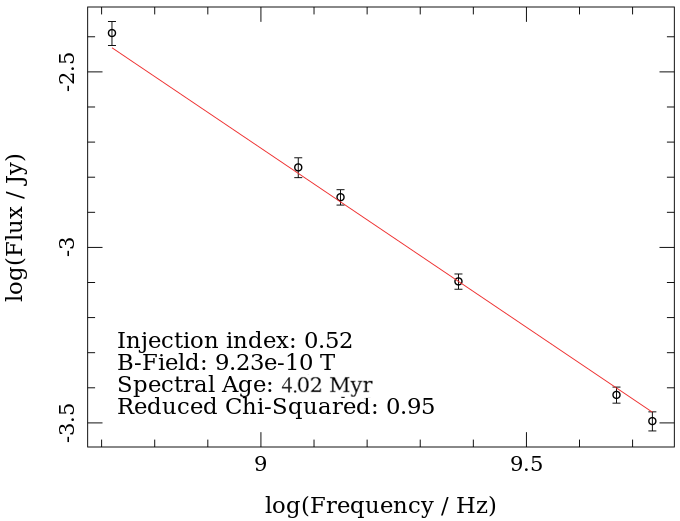}
	\includegraphics[scale=0.029]{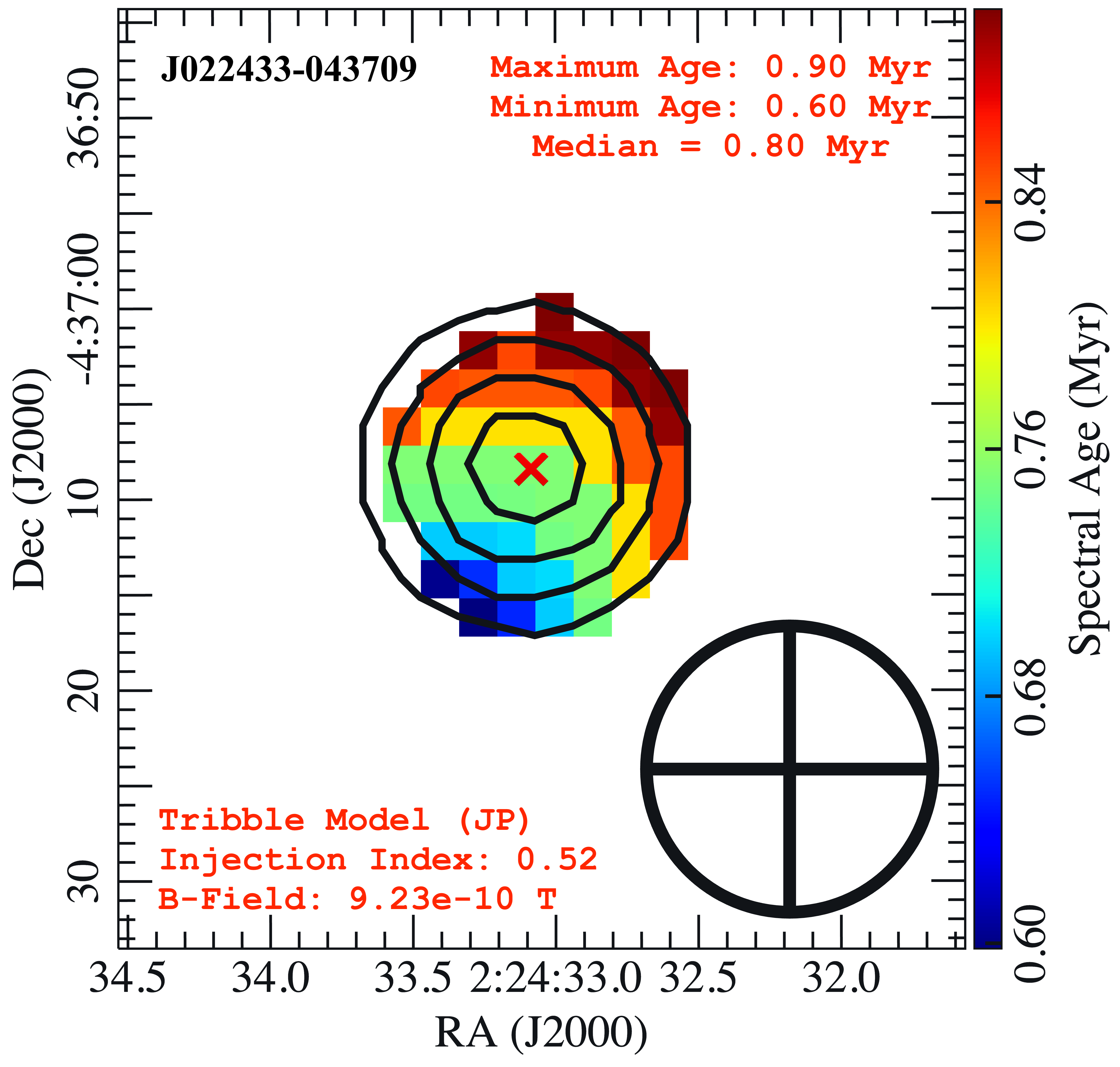}
	\caption{ {\it Left Panel} : Band-3 (in green) uGMRT and 1.284 GHz (in red) MeerKAT radio contours overplotted onto the grey-scale HSC-SSP $i-$band image. The radio contour levels are at 3$\sigma$ $\times$ (1, 2, 4, 8, 16 ...) and the corresponding optical image is logarithmically scaled. The magenta cross is the position of the potential host galaxy. Only 02/14 active sources are shown in this figure. {\it Middle panel} : Best fit radio SEDs obtained by using CI$_{\rm ON}$ of the corresponding active sources. The solid red curve represents model fitted to the data points. {\it Right panel} : Spectral age maps derived by JP Tribble model of the corresponding active sources. The contours overlaid on the maps are 1.284 GHz MeerKAT at 8\farcs2 resolution and the levels are at 3$\sigma$ $\times$ (1, 2, 4, 8, 16 ...). The position of the host galaxy is marked with red cross. The open circle with a cross represents a circular PSF beam of 15$^{\prime\prime}$.}
	\label{fig:SED4}
\end{figure*}

\par
Modeling with the CI$_{\rm OFF}$ (or KGJP) indicates that the oldest source in our sample exhibits a spectral age of 41.97 Myr, while the youngest source has a spectral age of 8.06 Myr (median t$_{\rm s}$ is 12.43 Myr). However, the CI$_{\rm ON}$ model estimates the total source ages (t$_{\rm s}$) for J021646-051004 and J022433-043709 to be 0.53 Myr and 20.02 Myr, respectively. These ages are systematically younger than those of classical remnants reported in literature. This indicates that our sample represent a new population of short-lived remnants with brief active and quiescent phases, consistent with rapidly evolving AGN duty cycles.  
\par
In addition, assuming the jet axis is oriented at 90$^{\circ}$ relative to the line-of-sight ({\em los}), the average lobe speed along the jet axis can be determined by dividing the mean lobe separation by the active phase timescale, t$_{\rm ON}$. We found that our remnant candidates exhibit average lobe speeds in the range of 0.01$c$ to 0.12$c$, which aligns well with those typically observed in powerful radio galaxies \citep[see][]{Machalski07,Odea09,Dutta22}. In other words, if dynamical ages are estimated using a representative average lobe speed, they are expected to be consistent with the spectral ages. We caution that, for compact sources, it is not possible to define a meaningful lobe separation or jet axis. Such sources are therefore treated as single components, and beam-deconvolved size estimates are used to place an upper limit on the characteristic expansion scale. Consequently, the derived velocity should be interpreted as an upper limit on the average jet-head advance speed, rather than a direct measurement.
\par
For small-size sources in our sample, JP Tribble spectral age maps yield maximum pixel-based ages of 0.87$-$43.43 Myr with median values of 0.70$-$31.46 Myr (see columns 6 and 8 of Table~\ref{tab:table4}), whereas for the extended sources the corresponding ranges are 2.87$-$11.48 Myr and 1.68$-$7.99 Myr, respectively (see Table~\ref{tab:table4}). This broader spread in compact remnants likely reflects a combination of projection effects, limited spatial resolution, and intrinsically diverse evolutionary pathways, leading to less uniform spectral ageing compared to the more extended sources. Spectral age maps derived using the JP Tribble model yield results that are broadly consistent with those obtained from the CI$_{\rm OFF}$ spectral fitting. This agreement indicates that both models capture the key features of the synchrotron ageing process, despite their differing physical assumptions. The consistency suggests that our derived source ages are not strongly dependent on the adopted model, thereby reinforcing the reliability and robustness of the spectral ageing analysis. 
\par
Furthermore, the source J021659-044918 has been the subject of extensive investigation in the literature. \cite{Tamhane15} reported a magnetic field strength of 3.3 $\mu$G for this source from synchrotron ageing analysis. Using the {\sc PySynch} package, we derive an equipartition field of 3.85 $\mu$G, marginally higher than the 3.10 $\mu$G estimated by \citet{Pinjarkar23} with the same approach. This discrepancy may be attributed to differences in input parameters, such as source geometry, spectral index, or break frequencies, between the two studies. Our analysis yields mean and median spectral ages of 7.93$_{-0.17}^{+0.15}$ Myr and 7.99$_{-0.13}^{+0.12}$ Myr, respectively, which are consistent with the $\sim$8 Myr estimate of \cite{Tamhane15}. The maximum spectral age derived from the JP model, 8.59$_{-0.91}^{+0.81}$ Myr, closely matches the value reported by \cite{Pinjarkar23}. These results provide independent validation for the previously reported ages and further support the robustness of the models used in our analysis. 
\par
While the integrated CI$_{\rm OFF}$ model provides reliable global lifetimes, the JP Tribble maps add an important spatial dimension, revealing how spectral  ages vary across the lobes. In the extended sources, the spectral age maps exhibit relatively smooth age distributions/gradients ({\em e.g.,} J021659–044918, J021926–051535, and J022318–044522), consistent with spatially homogeneous radiative losses, whereas in the compact sources, no clear age gradients are observed. The close agreement between the methods highlights the robustness of our age estimates and establishes combined spectral modeling as a powerful tool for constraining AGN duty cycles and unveiling the mechanisms governing remnant evolution.
\par
We also note that the extended sources in our sample ({\em e.g.} J021659-044918 and J022318-044522) exhibit coherent spectral-age gradients along their lobes, indicative of backflow-driven plasma transport and spatially ordered radiative losses. The observed gradients, with younger plasma at the lobe extremities and older material toward the core, are consistent with FR II-type evolution, where backflowing synchrotron plasma from the jet termini undergoes radiative ageing during inward propagation \citep[for instance 3C 315, 3C 52, and 3C 452;][]{Leahy84, Harwood17}. The smoothness of these gradients suggests a stable magnetic field structure and limited re-acceleration during the post-jet phase. 
The JP Tribble model, which accounts for local magnetic field variations, reproduces these spatial differences effectively and indicates that post-activity lobes remain dynamically and radiatively active long after jet cessation.

\section{Discussion: Evolutionary Scenario of Remnant Sources}
\label{sec:Discussion}

\begin{figure}
	\centering
	\includegraphics[scale=0.45]{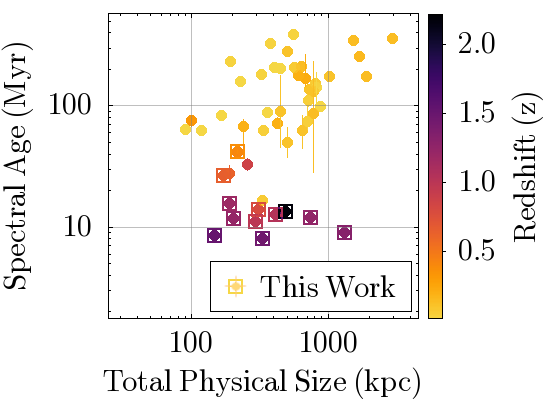}
	\caption{A diagnostic plot of total physical size at 1.284 GHz versus spectral age of our sources using CI$_{\rm OFF}$ model. We compare the spectral ages of our sample with those of remnant sources reported in the literature. The scatter plot is colour coded by median redshift values.}
	\label{fig:AgevsSize}
\end{figure}

The spectral age–radio size plot for remnants links radiative and physical growth histories, enabling constraints on their evolutionary stages and environmental influences. In Figure~\ref{fig:AgevsSize}, we compare the spectral ages and physical sizes of our remnant candidates, determined using the CI$_{\rm OFF}$ model, with those of remnants reported in the literature \citep[see][]{Parma07,Murgia11,Brienza16,Shulevski17,Duchesne19,Randriamanakoto20,Dutta22,Quici25}. The scatter plot (Figure~\ref{fig:AgevsSize}) shows that 10 out of 12 sources in our sample have linear sizes smaller than 500 kpc, while 5 of the 12 sources exhibit small-size radio sizes below 200 kpc ({\em i.e.}, LAS$<$30$^{\prime\prime}$). 
\par
It is also worth noting that most of our sample remnants display smaller radio sizes compared to those in previous studies. 
The smaller sizes may reflect confinement by a relatively dense local environment or a less efficient expansion mechanism. However, our sample sources reside in lower-density environments, {\em i.e.,} they are located in a non-clustered ambience, and show a wide range of spectral ages (8.06$-$41.97 Myr).   
\par
Contrary to the typical trend observed, 
we do not find a clear correlation between the spectral age of a source and its physical size in our sample. 
One would expect older sources to exhibit larger linear sizes due to the expansion of radio lobes over time \citep[see][]{Parma07,Murgia11,Brienza16,Brienza17,Quici21}; however, this trend is not apparent in our data. The lack of a size-age correlation may reflect a range of factors affecting the source age, such as varying ambient medium densities, differing expansion rates, or enhanced radiative losses by inverse-Compton effect at higher redshifts, which could complicate the interpretation of age-size relationships in these sources. Most of the small-size ($<$200 kpc) sources in our sample exhibit relatively lower spectral ages. This trend indicates that remnant lobes at higher redshifts undergo rapid fading, resulting in lower measured spectral ages.
\par
\begin{figure}
	\centering
	\includegraphics[scale=0.45]{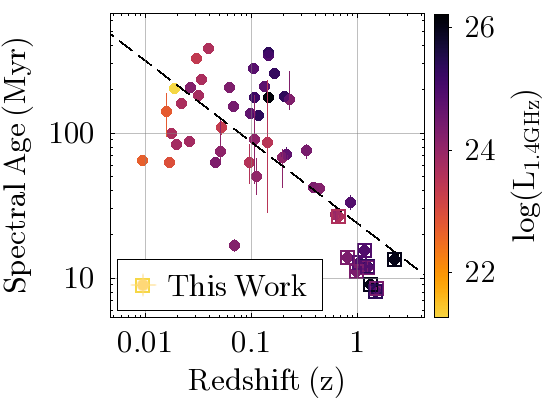}
	\caption{Diagnostic plot showing spectral age from the CI$_{\rm OFF}$ model as a function of redshift for our sources, with remnant sources from the literature included. A dashed black line indicates the best-fit linear trend (linear correlation = -0.72), and points are colour-coded by 1.4 GHz radio luminosity to illustrate the relationship between source age, redshift, and luminosity.}
	\label{fig:AgevsRedshift}
\end{figure}

Furthermore, at high redshifts, inverse Compton losses become the dominant energy loss mechanism due to the increased density of CMB photons. The inverse Compton equivalent magnetic field (B$_{\rm CMB}$) scales as (1+$z$)$^2$ (see Equation \ref{eq:2}), leading to enhanced energy losses for relativistic electrons. Consequently, radio sources at higher redshifts are expected to experience more rapid fading than those at lower redshifts, significantly influencing their estimated spectral ages. This is also evident in Figure~\ref{fig:AgevsRedshift}, where we compare the spectral ages and redshifts of our remnant sources with those reported in literature.
\par
In this context, we carried out partial and pairwise Pearson correlation analyses for both our sample and the remnant sources available in the literature to quantitatively assess and compare the underlying relationships among the key parameters (see Table~\ref{tab:partial_corr_comparison}). The pairwise Pearson correlations show a significant positive trend between radio luminosity and redshift (r = 0.47, p = 0.0004) and a strong negative correlation between redshift and spectral age (r = -0.56, p $<$ 0.001), while no significant relation is found between radio luminosity and spectral age. Partial correlation analysis confirms that the luminosity–redshift link remains robust (partial r = 0.53, p = 0.0001) after accounting for spectral age, indicating an intrinsic dependence influenced by observational selection. When controlling for redshift, a weak but significant positive relation emerges between luminosity and spectral age (partial r = 0.28, p = 0.05), whereas the negative redshift–spectral age correlation persists (partial r = -0.60, p $<$ 0.001), further suggesting younger sources dominate at higher redshifts.
\par
The redshifts of our remnant sources, ranging from 0.67 to 2.85, with the exception of J021528-044045 at a redshift of 0.35, distinguish them from lower-redshift (typically between 0.01$-$0.30) samples in previous studies. As a result, we expect to obtain lower spectral age estimates compared to those for remnant sources reported in the previous studies, which is consistent with the trends observed in Figure~\ref{fig:AgevsSize}. 

\begin{table*}
	\centering
	\caption{Comparison of pairwise and partial correlation coefficients among radio luminosity, redshift, and spectral age for our sample and remnant sources from the literature. }
	\label{tab:partial_corr_comparison}
	\begin{tabular}{ccccccc}
		\hline
		\multirow{2}{*}{\textbf{Variables}} & \multirow{2}{*}{\textbf{Correlation Type}} &
		\multicolumn{2}{c}{\textbf{Our Sample}} & & 
		\multicolumn{2}{c}{\textbf{Literature Remnants}} \\
		\cline{3-4} \cline{6-7}
		&  & \textbf{r} & \textbf{p-value} & & \textbf{r} & \textbf{p-value} \\
		\hline
		\multicolumn{7}{l}{\textbf{Pairwise Pearson Correlations}} \\
		\hline
		radio luminosity -- redshift & Pearson & 0.323 & 0.0418 & & 0.700 & 0.0113 \\
		radio luminosity -- spectral age & Pearson & 0.173 & 0.2867 & & -0.441 & 0.1514 \\
		redshift -- spectral age & Pearson & -0.298 & 0.0615 & & -0.666 & 0.0180 \\
		\hline
		\multicolumn{7}{l}{\textbf{Partial Correlations (controlling for third variable)}} \\
		\hline
		radio luminosity -- redshift \,|\, spectral age & Partial & 0.399 & 0.0119 & & 0.607 & 0.0478 \\
		radio luminosity -- spectral age \,|\, redshift & Partial & 0.298 & 0.0654 & & 0.048 & 0.8896 \\
		redshift -- radio luminosity \,|\, spectral age  & Partial & 0.607 & 0.0478 & & 0.399  &  0.0119 \\
		redshift -- spectral\_age \,|\, radio\_luminosity & Partial & -0.380 & 0.0170 & & -0.558 & 0.0745 \\
		spectral age -- radio luminosity \,|\, redshift &Partial & 0.048 & 0.8896 & & 0.298 & 0.0654 \\
		spectral age -- redshift \,|\, radio luminosity & Partial & -0.558 & 0.0745 & & -0.380 & 0.0170 \\
		\hline
	\end{tabular}
	
	\vspace{2mm}
	\small {\em Note.} The Pearson coefficient ($r$) measures the strength of linear correlation between variables, while the partial correlation ($r_{\rm{partial}}$) quantifies their relationship after accounting for the third variable. p-values indicate statistical significance; values $<0.05$ are considered significant.
\end{table*}

\begin{figure}
	\centering
	\includegraphics[scale=0.3]{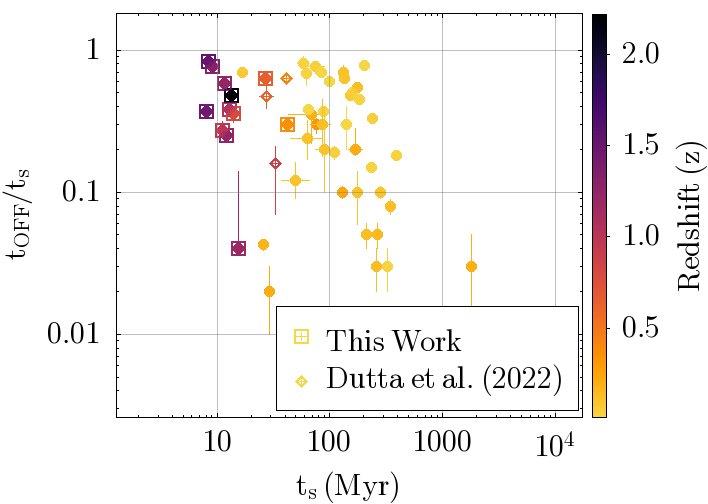}
	\caption{Plot of total source age t$_{\rm s}$ versus fractional remnant time-scale (t$_{\rm OFF}$/t$_{\rm s}$). The vertical color bar indicates the redshifts of remnant hosts. }
	\label{fig:comparison}
\end{figure} 

\begin{figure}
	\centering
	\includegraphics[scale=0.4]{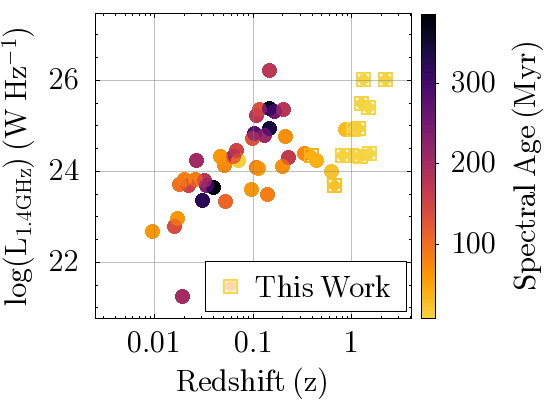}
	\caption{1.4 GHz rest-frame radio luminosity versus redshift plot. We compare the radio luminosities of our sample with those of remnant sources reported in the literature. The scatter plot is colour coded by median spectral age values obtained by CI$_{\rm OFF}$ model.}
	\label{fig:PowervsRedshift}
\end{figure}

\begin{figure}
	\centering
	\includegraphics[scale=0.45]{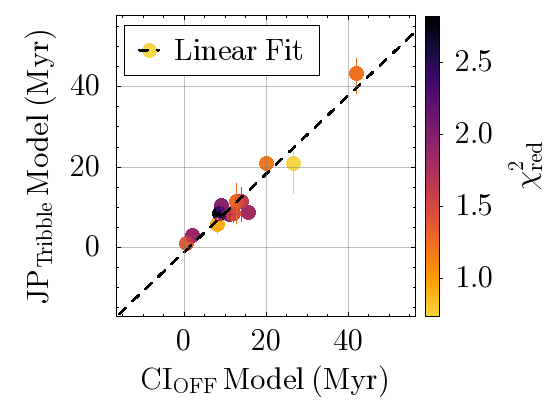}
	\caption{Comparison of two spectral ageing models applied to the sample. The dashed line shows the best-fitting linear relation (linear correlation = 0.96) used to assess the consistency between the models. The data points are colour coded according to their reduced $\chi^2$ values.}
	\label{fig:comparemodel}
\end{figure}

\par
In Figure~\ref{fig:comparison}, we present a plot of total source age (t$_{\rm s}$) versus the fractional remnant duration (t$_{\rm OFF}$/t$_{\rm s}$), which provides insight into the evolutionary stage of these radio galaxies. 
The t$_{\rm OFF}$/t$_{\rm s}$ ratios for our sample range from 0.04 to 0.83 (see Table~\ref{tab:table3}), indicating that the sources span a continuum of evolutionary stages—from recently switched-off, young remnants to systems that have remained inactive for a substantial fraction of their lifetimes. For instance, in the case of J021917-042654, the ratio t$_{\rm OFF}$/t$_{\rm s}$ = 0.83 indicates that the source has spent nearly 83\% of its total lifetime in the remnant phase, suggesting a long inactive period following the cessation of jet activity. In contrast, J022302-042849 exhibits a ratio of 0.04, implying that only about 4\% of its lifetime has been spent in the remnant stage, consistent with it being a recently switched-off or young remnant source. 
\par
Moreover, in comparison with remnants reported in the literature, 
our sample exhibits a more diverse population, spanning a broad range of t$_{\rm OFF}$/t$_{\rm s}$ ratios, physical sizes, and redshifts. This diversity indicates that the remnants in our study occupy different evolutionary stages within the radio galaxy life cycle.
\par
Considering all the sources shown in Figure~\ref{fig:comparison}, including the remnants compiled from the literature, the results further confirm the diversity observed among remnant populations. This diversity may be driven by variations in AGN duty cycles, where the length and frequency of active and inactive phases differ from source to source. 
\par
In Figure~\ref{fig:PowervsRedshift}, a correlation is observed between radio luminosity (L$_{\rm 1.4~GHz}$) and the redshift of our sources, indicating that our remnant sources are both relatively distant and luminous. This trend likely reflects the influence of the Malmquist bias, wherein only the brighter and more powerful sources remain detectable at higher redshifts, while intrinsically fainter systems fall below the survey sensitivity limit. 
\par
As discussed in Section~\ref{sec:results}, the JP Tribble and CI$_{\rm OFF}$ models yield comparable spectral age distributions, confirming that both adequately describe synchrotron ageing despite their differing assumptions. The strong concordance evident in Figure~\ref{fig:comparemodel} validates the robustness of the analysis and the credibility of the derived ages.

\section{Conclusions}
\label{sec:conclusions}
In this paper, we present new band-3 ($\nu_{\rm central}$ = 400 MHz) and band-4 ($\nu_{\rm central}$ = 650 MHz) superMIGHTEE observations of 14 remnant candidates. The spectral ageing analysis was performed by modeling the radio SEDs using spectral fitting (CI$_{\rm OFF}$) model across 144 MHz$-$1.5 GHz. We also applied JP Tribble model to generate spectral age maps for all sources, enabling spatially resolved estimates of their radiative lifetimes. 
The key conclusions of our study are summarized below.

\begin{itemize}
	\item 
	Out of these 14 candidate remnants, 12 sources are confirmed as remnants based on CI$_{\rm OFF}$ modeling and the presence of strong curvature, while two (J021646-051004 and J022433-043709) are re-classified as active. This highlights that sparse radio SEDs can lead to misclassification, and that reliable remnant identification requires dense, multi-frequency sampling. 
	
	\item CI$_{\rm OFF}$ fits yield total spectral ages of 8.06-41.97 Myr (median $\approx$ 12 Myr), systematically younger than many previously studied remnants. This suggests the presence of a hitherto unexplored population of short-lived remnants with comparatively brief active and quiescent phases, unlike classical long-lived systems. \\
	
	 \item JP Tribble spectral age maps provide pixel-based maximum ages of 0.87-43.43 Myr for compact sources and 2.87-11.48 Myr for extended ones, broadly consistent with integrated CI$_{\rm OFF}$ estimates but revealing localized age gradients. The agreement between models reinforces the robustness of our results. \\
	 
	 \item The ratio t$_{\rm OFF}$/t$_{\rm s}$ spans 0.04–0.83, indicating remnants at a wide range of evolutionary stages. This confirms that our sample do not represent a uniform class but instead span a continuum of evolutionary stages, from freshly quenched systems to long-lived relics, reflecting the diversity of AGN duty cycles.  \\
	
	\item The extended sources ({\em e.g.} J021659-044918 and J022318-044522) exhibit systematic spectral-age gradients consistent with backflow-driven plasma transport and ordered radiative losses, whereas compact remnants show irregular ageing patterns likely shaped by environmental asymmetries and magnetic field inhomogeneities. These results highlight that post-jet lobes can remain dynamically and radiatively active well beyond the cessation of AGN activity, underscoring the complex interplay between magnetic fields, plasma flows, and ambient conditions in governing remnant evolution. \\
	
	\item Unlike most previous studies, our work presents spectral age estimates for faint, high-redshift remnant radio galaxies. The unusually short spectral ages we find for these new faint remnants provide an important stepping stone toward the search and characterization of the yet-unexplored faint, high-z remnant population that will be uncovered by deep SKA radio continuum surveys reaching sub-$\mu$Jy sensitivities.  
	
\end{itemize}

\section*{Acknowledgments}
SD and RT acknowledge financial support from the Inter-University Institute for Data Intensive Astronomy (IDIA). VS acknowledges the support from Physical Research Laboratory, Ahmedabad, funded by the Department of Space, Government of India. ICCH and YW acknowledge support from the Department of Atomic Energy under project 12-R\&D-TFR-5.02-0700. MV and LM acknowledge financial support from the IDIA, a partnership of the University of Cape Town, the University of Pretoria and the University of the Western Cape, and from the South African Department of Science and Innovation’s National Research Foundation under the ISARP RADIOMAP Joint Research Scheme (DSI-NRF Grant Number 150551) and the CPRR HIPPO Project (DSI-NRF Grant Number SRUG22031677) and the CPRR Project (DSI-NRF Grant Number SRUG2204254729). CLH acknowledges support from the Oxford Hintze Centre for Astrophysical Surveys which is funded through generous support from the Hintze Family Charitable Foundation and support from the Science and Technology Facilities Council (STFC) through grant ST/Y000951/1. ZR acknowledges support from the South African Astronomical Observatory, which is a facility of the National Research Foundation, an agency of the Department of Science, Technology and Innovation. We thank the staff of GMRT who have made these observations possible. GMRT is run by the National Centre for Radio Astrophysics of the Tata Institute of Fundamental Research.
%
The MeerKAT telescope is operated by the South African Radio Astronomy Observatory, which is a facility of the National Research Foundation, an agency of the Department of Science and Innovation. We acknowledge use of the IDIA data intensive research cloud for data processing. IDIA is a South African university partnership involving the University of Cape Town, the University of Pretoria and the University of the Western Cape. This work made use of the CARTA (Cube Analysis and Rendering Tool for Astronomy) software (DOI: 10.5281/zenodo.3377984 – https://cartavis.github.io).
%
\section*{Facilities}

uGMRT, MeerKAT, LOFAR, JVLA.


\section*{Data Availability}
The raw data from uGMRT observations are available from the NCRA achieve proposal system http://naps.ncra.tifr.res.in. The raw MeerKAT visibilities for which any proprietary period has expired can be obtained from the SARAO archive at https://archive.sarao.ac.za. The MIGHTEE continuum DR1 images can be obtained from https://doi.org/10.48479/7msw-r692 or \cite{Hale25}.
%



\bibliographystyle{mnras}
\bibliography{RRGSEDpaper} 

\begin{thebibliography}{}
\makeatletter
\relax
\def\mn@urlcharsother{\let\do\@makeother \do\$\do\&\do\#\do\^\do\_\do\%\do\~}
\def\mn@doi{\begingroup\mn@urlcharsother \@ifnextchar [ {\mn@doi@}
  {\mn@doi@[]}}
\def\mn@doi@[#1]#2{\def\@tempa{#1}\ifx\@tempa\@empty \href
  {http://dx.doi.org/#2} {doi:#2}\else \href {http://dx.doi.org/#2} {#1}\fi
  \endgroup}
\def\mn@eprint#1#2{\mn@eprint@#1:#2::\@nil}
\def\mn@eprint@arXiv#1{\href {http://arxiv.org/abs/#1} {{\tt arXiv:#1}}}
\def\mn@eprint@dblp#1{\href {http://dblp.uni-trier.de/rec/bibtex/#1.xml}
  {dblp:#1}}
\def\mn@eprint@#1:#2:#3:#4\@nil{\def\@tempa {#1}\def\@tempb {#2}\def\@tempc
  {#3}\ifx \@tempc \@empty \let \@tempc \@tempb \let \@tempb \@tempa \fi \ifx
  \@tempb \@empty \def\@tempb {arXiv}\fi \@ifundefined
  {mn@eprint@\@tempb}{\@tempb:\@tempc}{\expandafter \expandafter \csname
  mn@eprint@\@tempb\endcsname \expandafter{\@tempc}}}

\bibitem[\protect\citeauthoryear{{Adams}, {Bowler}, {Jarvis}, {Varadaraj}  \&
  {H{\"a}u{\ss}ler}}{{Adams} et~al.}{2023}]{Adams23}
{Adams} N.~J.,  {Bowler} R.~A.~A.,  {Jarvis} M.~J.,  {Varadaraj} R.~G.,
  {H{\"a}u{\ss}ler} B.,  2023, \mn@doi [\mnras] {10.1093/mnras/stad1333}, \href
  {https://ui.adsabs.harvard.edu/abs/2023MNRAS.523..327A} {523, 327}

\bibitem[\protect\citeauthoryear{{Aihara} et~al.,}{{Aihara}
  et~al.}{2018}]{Aihara18}
{Aihara} H.,  et~al., 2018, \mn@doi [\pasj] {10.1093/pasj/psx066}, \href
  {https://ui.adsabs.harvard.edu/abs/2018PASJ...70S...4A} {70, S4}

\bibitem[\protect\citeauthoryear{{Andernach} \& {Br{\"u}ggen}}{{Andernach} \&
  {Br{\"u}ggen}}{2025}]{Andernach25}
{Andernach} H.,  {Br{\"u}ggen} M.,  2025, \mn@doi [\aap]
  {10.1051/0004-6361/202452961}, \href
  {https://ui.adsabs.harvard.edu/abs/2025A&A...699A.257A} {699, A257}

\bibitem[\protect\citeauthoryear{{Bell}}{{Bell}}{1978}]{Bell78}
{Bell} A.~R.,  1978, \mn@doi [\mnras] {10.1093/mnras/182.2.147}, \href
  {https://ui.adsabs.harvard.edu/abs/1978MNRAS.182..147B} {182, 147}

\bibitem[\protect\citeauthoryear{{Bennett}, {Larson}, {Weiland}  \&
  {Hinshaw}}{{Bennett} et~al.}{2014}]{Bennett14}
{Bennett} C.~L.,  {Larson} D.,  {Weiland} J.~L.,   {Hinshaw} G.,  2014, \mn@doi
  [\apj] {10.1088/0004-637X/794/2/135}, \href
  {https://ui.adsabs.harvard.edu/abs/2014ApJ...794..135B} {794, 135}

\bibitem[\protect\citeauthoryear{{Brienza}, {Morganti}, {Shulevski}, {Godfrey}
  \& {Vilchez}}{{Brienza} et~al.}{2016}]{Brienza16}
{Brienza} M.,  {Morganti} R.,  {Shulevski} A.,  {Godfrey} L.,   {Vilchez} N.,
  2016, \mn@doi [Astronomische Nachrichten] {10.1002/asna.201512260}, \href
  {https://ui.adsabs.harvard.edu/abs/2016AN....337...31B} {337, 31}

\bibitem[\protect\citeauthoryear{{Brienza} et~al.,}{{Brienza}
  et~al.}{2017}]{Brienza17}
{Brienza} M.,  et~al., 2017, \mn@doi [\aap] {10.1051/0004-6361/201730932},
  \href {https://ui.adsabs.harvard.edu/abs/2017A&A...606A..98B} {606, A98}

\bibitem[\protect\citeauthoryear{{Capetti}, {Zamfir}, {Rossi}, {Bodo}, {Zanni}
  \& {Massaglia}}{{Capetti} et~al.}{2002}]{Capetti02}
{Capetti} A.,  {Zamfir} S.,  {Rossi} P.,  {Bodo} G.,  {Zanni} C.,   {Massaglia}
  S.,  2002, \mn@doi [\aap] {10.1051/0004-6361:20021070}, \href
  {https://ui.adsabs.harvard.edu/abs/2002A&A...394...39C} {394, 39}

\bibitem[\protect\citeauthoryear{{Carilli}, {Perley}, {Dreher}  \&
  {Leahy}}{{Carilli} et~al.}{1991}]{Carilli91}
{Carilli} C.~L.,  {Perley} R.~A.,  {Dreher} J.~W.,   {Leahy} J.~P.,  1991,
  \mn@doi [\apj] {10.1086/170813}, \href
  {https://ui.adsabs.harvard.edu/abs/1991ApJ...383..554C} {383, 554}

\bibitem[\protect\citeauthoryear{{Charlton} et~al.,}{{Charlton}
  et~al.}{2025}]{Charlton25}
{Charlton} K.~K.~L.,  et~al., 2025, \mn@doi [\mnras] {10.1093/mnras/stae2543},
  \href {https://ui.adsabs.harvard.edu/abs/2025MNRAS.537..272C} {537, 272}

\bibitem[\protect\citeauthoryear{{Cool} et~al.,}{{Cool} et~al.}{2013}]{Cool13}
{Cool} R.~J.,  et~al., 2013, \mn@doi [\apj] {10.1088/0004-637X/767/2/118},
  \href {https://ui.adsabs.harvard.edu/abs/2013ApJ...767..118C} {767, 118}

\bibitem[\protect\citeauthoryear{{Cordey}}{{Cordey}}{1987}]{Cordey87}
{Cordey} R.~A.,  1987, \mn@doi [\mnras] {10.1093/mnras/227.3.695}, \href
  {https://ui.adsabs.harvard.edu/abs/1987MNRAS.227..695C} {227, 695}

\bibitem[\protect\citeauthoryear{{Croston}, {Hardcastle}, {Harris}, {Belsole},
  {Birkinshaw}  \& {Worrall}}{{Croston} et~al.}{2005}]{Croston05}
{Croston} J.~H.,  {Hardcastle} M.~J.,  {Harris} D.~E.,  {Belsole} E.,
  {Birkinshaw} M.,   {Worrall} D.~M.,  2005, \mn@doi [\apj] {10.1086/430170},
  \href {https://ui.adsabs.harvard.edu/abs/2005ApJ...626..733C} {626, 733}

\bibitem[\protect\citeauthoryear{{Duchesne} \& {Johnston-Hollitt}}{{Duchesne}
  \& {Johnston-Hollitt}}{2019}]{Duchesne19}
{Duchesne} S.~W.,  {Johnston-Hollitt} M.,  2019, \mn@doi [\pasa]
  {10.1017/pasa.2018.26}, \href
  {https://ui.adsabs.harvard.edu/abs/2019PASA...36...16D} {36, e016}

\bibitem[\protect\citeauthoryear{{Dutta}, {Singh}, {Chandra}, {Wadadekar}  \&
  {Kayal}}{{Dutta} et~al.}{2022}]{Dutta22}
{Dutta} S.,  {Singh} V.,  {Chandra} C.~H.~I.,  {Wadadekar} Y.,   {Kayal} A.,
  2022, \mn@doi [Journal of Astrophysics and Astronomy]
  {10.1007/s12036-022-09883-y}, \href
  {https://ui.adsabs.harvard.edu/abs/2022JApA...43...96D} {43, 96}

\bibitem[\protect\citeauthoryear{{Dutta}, {Singh}, {Chandra}, {Wadadekar},
  {Kayal}  \& {Heywood}}{{Dutta} et~al.}{2023}]{Dutta23}
{Dutta} S.,  {Singh} V.,  {Chandra} C.~H.~I.,  {Wadadekar} Y.,  {Kayal} A.,
  {Heywood} I.,  2023, \mn@doi [\apj] {10.3847/1538-4357/acaf01}, \href
  {https://ui.adsabs.harvard.edu/abs/2023ApJ...944..176D} {944, 176}

\bibitem[\protect\citeauthoryear{{Dwarakanath} \& {Kale}}{{Dwarakanath} \&
  {Kale}}{2009}]{DK09}
{Dwarakanath} K.~S.,  {Kale} R.,  2009, \mn@doi [\apjl]
  {10.1088/0004-637X/698/2/L163}, \href
  {https://ui.adsabs.harvard.edu/abs/2009ApJ...698L.163D} {698, L163}

\bibitem[\protect\citeauthoryear{{Eales} et~al.,}{{Eales}
  et~al.}{2010}]{Eales10}
{Eales} S.,  et~al., 2010, \mn@doi [\pasp] {10.1086/653086}, \href
  {https://ui.adsabs.harvard.edu/abs/2010PASP..122..499E} {122, 499}

\bibitem[\protect\citeauthoryear{{Garilli} et~al.,}{{Garilli}
  et~al.}{2014}]{Garilli14}
{Garilli} B.,  et~al., 2014, \mn@doi [\aap] {10.1051/0004-6361/201322790},
  \href {https://ui.adsabs.harvard.edu/abs/2014A&A...562A..23G} {562, A23}

\bibitem[\protect\citeauthoryear{{Gebhardt} et~al.,}{{Gebhardt}
  et~al.}{2021}]{Gebhardt21}
{Gebhardt} K.,  et~al., 2021, \mn@doi [\apj] {10.3847/1538-4357/ac2e03}, \href
  {https://ui.adsabs.harvard.edu/abs/2021ApJ...923..217G} {923, 217}

\bibitem[\protect\citeauthoryear{{Hale} et~al.,}{{Hale} et~al.}{2019}]{Hale19}
{Hale} C.~L.,  et~al., 2019, \mn@doi [\aap] {10.1051/0004-6361/201833906},
  \href {https://ui.adsabs.harvard.edu/abs/2019A&A...622A...4H} {622, A4}

\bibitem[\protect\citeauthoryear{{Hale} et~al.,}{{Hale} et~al.}{2025}]{Hale25}
{Hale} C.~L.,  et~al., 2025, \mn@doi [\mnras] {10.1093/mnras/stae2528}, \href
  {https://ui.adsabs.harvard.edu/abs/2025MNRAS.536.2187H} {536, 2187}

\bibitem[\protect\citeauthoryear{{Hardcastle} \& {Croston}}{{Hardcastle} \&
  {Croston}}{2020}]{Hardcastle20}
{Hardcastle} M.~J.,  {Croston} J.~H.,  2020, \mn@doi [\nar]
  {10.1016/j.newar.2020.101539}, \href
  {https://ui.adsabs.harvard.edu/abs/2020NewAR..8801539H} {88, 101539}

\bibitem[\protect\citeauthoryear{{Hardcastle}, {Birkinshaw}  \&
  {Worrall}}{{Hardcastle} et~al.}{1998}]{Hardcastle98}
{Hardcastle} M.~J.,  {Birkinshaw} M.,   {Worrall} D.~M.,  1998, \mn@doi
  [\mnras] {10.1111/j.1365-8711.1998.01159.x10.1046/j.1365-8711.1998.01159.x},
  \href {https://ui.adsabs.harvard.edu/abs/1998MNRAS.294..615H} {294, 615}

\bibitem[\protect\citeauthoryear{{Harwood}}{{Harwood}}{2017}]{Harwood17}
{Harwood} J.~J.,  2017, \mn@doi [\mnras] {10.1093/mnras/stw3318}, \href
  {https://ui.adsabs.harvard.edu/abs/2017MNRAS.466.2888H} {466, 2888}

\bibitem[\protect\citeauthoryear{{Harwood}, {Hardcastle}, {Croston}  \&
  {Goodger}}{{Harwood} et~al.}{2013}]{Harwood13}
{Harwood} J.~J.,  {Hardcastle} M.~J.,  {Croston} J.~H.,   {Goodger} J.~L.,
  2013, \mn@doi [\mnras] {10.1093/mnras/stt1526}, \href
  {https://ui.adsabs.harvard.edu/abs/2013MNRAS.435.3353H} {435, 3353}

\bibitem[\protect\citeauthoryear{{Harwood}, {Hardcastle}  \&
  {Croston}}{{Harwood} et~al.}{2015}]{Harwood15}
{Harwood} J.~J.,  {Hardcastle} M.~J.,   {Croston} J.~H.,  2015, \mn@doi
  [\mnras] {10.1093/mnras/stv2194}, \href
  {https://ui.adsabs.harvard.edu/abs/2015MNRAS.454.3403H} {454, 3403}

\bibitem[\protect\citeauthoryear{{Harwood} et~al.,}{{Harwood}
  et~al.}{2016}]{Harwood16}
{Harwood} J.~J.,  et~al., 2016, \mn@doi [\mnras] {10.1093/mnras/stw638}, \href
  {https://ui.adsabs.harvard.edu/abs/2016MNRAS.458.4443H} {458, 4443}

\bibitem[\protect\citeauthoryear{{Heeschen}}{{Heeschen}}{1975}]{Heeschen75}
{Heeschen} D.~S.,  1975, \skytel, \href
  {https://ui.adsabs.harvard.edu/abs/1975S&T....49..344H} {49, 344}

\bibitem[\protect\citeauthoryear{{Heywood}, {Hale}, {Jarvis}, {Makhathini},
  {Peters}, {Sebokolodi}  \& {Smirnov}}{{Heywood} et~al.}{2020}]{Heywood20}
{Heywood} I.,  {Hale} C.~L.,  {Jarvis} M.~J.,  {Makhathini} S.,  {Peters}
  J.~A.,  {Sebokolodi} M.~L.~L.,   {Smirnov} O.~M.,  2020, \mn@doi [\mnras]
  {10.1093/mnras/staa1770}, \href
  {https://ui.adsabs.harvard.edu/abs/2020MNRAS.496.3469H} {496, 3469}

\bibitem[\protect\citeauthoryear{{Heywood} et~al.,}{{Heywood}
  et~al.}{2022}]{Heywood22}
{Heywood} I.,  et~al., 2022, \mn@doi [\mnras] {10.1093/mnras/stab3021}, \href
  {https://ui.adsabs.harvard.edu/abs/2022MNRAS.509.2150H} {509, 2150}

\bibitem[\protect\citeauthoryear{{Hodges-Kluck} \& {Reynolds}}{{Hodges-Kluck}
  \& {Reynolds}}{2011}]{Hodges11}
{Hodges-Kluck} E.~J.,  {Reynolds} C.~S.,  2011, \mn@doi [\apj]
  {10.1088/0004-637X/733/1/58}, \href
  {https://ui.adsabs.harvard.edu/abs/2011ApJ...733...58H} {733, 58}

\bibitem[\protect\citeauthoryear{{Ichikawa}, {Ueda}, {Shidatsu}, {Kawamuro}  \&
  {Matsuoka}}{{Ichikawa} et~al.}{2016}]{Ichikawa16}
{Ichikawa} K.,  {Ueda} J.,  {Shidatsu} M.,  {Kawamuro} T.,   {Matsuoka} K.,
  2016, \mn@doi [\pasj] {10.1093/pasj/psv112}, \href
  {https://ui.adsabs.harvard.edu/abs/2016PASJ...68....9I} {68, 9}

\bibitem[\protect\citeauthoryear{{Ineson}, {Croston}, {Hardcastle}  \&
  {Mingo}}{{Ineson} et~al.}{2017}]{Ineson17}
{Ineson} J.,  {Croston} J.~H.,  {Hardcastle} M.~J.,   {Mingo} B.,  2017,
  \mn@doi [\mnras] {10.1093/mnras/stx189}, \href
  {https://ui.adsabs.harvard.edu/abs/2017MNRAS.467.1586I} {467, 1586}

\bibitem[\protect\citeauthoryear{{Jaffe} \& {Perola}}{{Jaffe} \&
  {Perola}}{1973}]{Jaffe73}
{Jaffe} W.~J.,  {Perola} G.~C.,  1973, \aap, \href
  {https://ui.adsabs.harvard.edu/abs/1973A&A....26..423J} {26, 423}

\bibitem[\protect\citeauthoryear{{Jamrozy}, {Klein}, {Mack}, {Gregorini}  \&
  {Parma}}{{Jamrozy} et~al.}{2004}]{Jamrozy04}
{Jamrozy} M.,  {Klein} U.,  {Mack} K.~H.,  {Gregorini} L.,   {Parma} P.,  2004,
  \mn@doi [\aap] {10.1051/0004-6361:20048056}, \href
  {https://ui.adsabs.harvard.edu/abs/2004A&A...427...79J} {427, 79}

\bibitem[\protect\citeauthoryear{{Jarvis} et~al.,}{{Jarvis}
  et~al.}{2016}]{Jarvis16}
{Jarvis} M.,  et~al., 2016, in MeerKAT Science: On the Pathway to the SKA. p.~6
  (\mn@eprint {arXiv} {1709.01901}), \mn@doi{10.22323/1.277.0006}

\bibitem[\protect\citeauthoryear{{Jonas}}{{Jonas}}{2009}]{Jonas09}
{Jonas} J.,  2009, in Panoramic Radio Astronomy: Wide-field 1-2 GHz Research on
  Galaxy Evolution. p.~4

\bibitem[\protect\citeauthoryear{{Jurlin} et~al.,}{{Jurlin}
  et~al.}{2020}]{Jurlin20}
{Jurlin} N.,  et~al., 2020, \mn@doi [\aap] {10.1051/0004-6361/201936955}, \href
  {https://ui.adsabs.harvard.edu/abs/2020A&A...638A..34J} {638, A34}

\bibitem[\protect\citeauthoryear{{Jurlin}, {Brienza}, {Morganti}, {Wadadekar},
  {Ishwara-Chandra}, {Maddox}  \& {Mahatma}}{{Jurlin} et~al.}{2021}]{Jurlin21}
{Jurlin} N.,  {Brienza} M.,  {Morganti} R.,  {Wadadekar} Y.,  {Ishwara-Chandra}
  C.~H.,  {Maddox} N.,   {Mahatma} V.,  2021, \mn@doi [\aap]
  {10.1051/0004-6361/202040102}, \href
  {https://ui.adsabs.harvard.edu/abs/2021A&A...653A.110J} {653, A110}

\bibitem[\protect\citeauthoryear{{Jurlin}, {Morganti}, {Sweijen}, {Morabito},
  {Brienza}, {Barthel}  \& {Miley}}{{Jurlin} et~al.}{2024}]{Jurlin24}
{Jurlin} N.,  {Morganti} R.,  {Sweijen} F.,  {Morabito} L.~K.,  {Brienza} M.,
  {Barthel} P.,   {Miley} G.~K.,  2024, \mn@doi [\aap]
  {10.1051/0004-6361/202245821}, \href
  {https://ui.adsabs.harvard.edu/abs/2024A&A...682A.118J} {682, A118}

\bibitem[\protect\citeauthoryear{{Kardashev}}{{Kardashev}}{1962}]{Kardashev62}
{Kardashev} N.~S.,  1962, \sovast, \href
  {https://ui.adsabs.harvard.edu/abs/1962SvA.....6..317K} {6, 317}

\bibitem[\protect\citeauthoryear{{Koekemoer} \& {Bicknell}}{{Koekemoer} \&
  {Bicknell}}{1998}]{Koekemoer98}
{Koekemoer} A.~M.,  {Bicknell} G.~V.,  1998, \mn@doi [\apj] {10.1086/305490},
  \href {https://ui.adsabs.harvard.edu/abs/1998ApJ...497..662K} {497, 662}

\bibitem[\protect\citeauthoryear{{Komissarov} \& {Gubanov}}{{Komissarov} \&
  {Gubanov}}{1994}]{Komissarov94}
{Komissarov} S.~S.,  {Gubanov} A.~G.,  1994, \aap, \href
  {https://ui.adsabs.harvard.edu/abs/1994A&A...285...27K} {285, 27}

\bibitem[\protect\citeauthoryear{{Ku{\'z}micz}, {Jamrozy},
  {Kozie{\l}-Wierzbowska}  \& {We{\.z}gowiec}}{{Ku{\'z}micz}
  et~al.}{2017}]{Kuzmicz17}
{Ku{\'z}micz} A.,  {Jamrozy} M.,  {Kozie{\l}-Wierzbowska} D.,   {We{\.z}gowiec}
  M.,  2017, \mn@doi [\mnras] {10.1093/mnras/stx1830}, \href
  {https://ui.adsabs.harvard.edu/abs/2017MNRAS.471.3806K} {471, 3806}

\bibitem[\protect\citeauthoryear{{Lal}}{{Lal}}{2021}]{Lal21}
{Lal} D.~V.,  2021, \mn@doi [\apj] {10.3847/1538-4357/ac042d}, \href
  {https://ui.adsabs.harvard.edu/abs/2021ApJ...915..126L} {915, 126}

\bibitem[\protect\citeauthoryear{Lal, Taylor, Sekhar, Ishwara-Chandra, Dutta
  \& Kolwa}{Lal et~al.}{2025}]{Lal25}
Lal D.~V.,  Taylor R.,  Sekhar S.,  Ishwara-Chandra C.,  Dutta S.,   Kolwa S.,
  2025, \mn@doi [The Astrophysical Journal] {10.3847/1538-4357/adf6dc}, 991, 9

\bibitem[\protect\citeauthoryear{{Lazio}, {Kassim}, {Weiler}  \&
  {Gross}}{{Lazio} et~al.}{1999}]{Lazio99}
{Lazio} T.~J.~W.,  {Kassim} N.~E.,  {Weiler} K.,   {Gross} C.~A.,  1999, in
  American Astronomical Society Meeting Abstracts. p. 83.06

\bibitem[\protect\citeauthoryear{{Leahy} \& {Williams}}{{Leahy} \&
  {Williams}}{1984}]{Leahy84}
{Leahy} J.~P.,  {Williams} A.~G.,  1984, \mn@doi [\mnras]
  {10.1093/mnras/210.4.929}, \href
  {https://ui.adsabs.harvard.edu/abs/1984MNRAS.210..929L} {210, 929}

\bibitem[\protect\citeauthoryear{{Leahy}, {Muxlow}  \& {Stephens}}{{Leahy}
  et~al.}{1989}]{Leahy89}
{Leahy} J.~P.,  {Muxlow} T.~W.~B.,   {Stephens} P.~W.,  1989, \mn@doi [\mnras]
  {10.1093/mnras/239.2.401}, \href
  {https://ui.adsabs.harvard.edu/abs/1989MNRAS.239..401L} {239, 401}

\bibitem[\protect\citeauthoryear{{Machalski}, {Chy{\.z}y}, {Stawarz}  \&
  {Kozie{\l}}}{{Machalski} et~al.}{2007}]{Machalski07}
{Machalski} J.,  {Chy{\.z}y} K.~T.,  {Stawarz} {\L}.,   {Kozie{\l}} D.,  2007,
  \mn@doi [\aap] {10.1051/0004-6361:20066121}, \href
  {https://ui.adsabs.harvard.edu/abs/2007A&A...462...43M} {462, 43}

\bibitem[\protect\citeauthoryear{{Mahatma}}{{Mahatma}}{2023}]{Mahatma23}
{Mahatma} V.~H.,  2023, \mn@doi [Galaxies] {10.3390/galaxies11030074}, \href
  {https://ui.adsabs.harvard.edu/abs/2023Galax..11...74M} {11, 74}

\bibitem[\protect\citeauthoryear{{Mahatma} et~al.,}{{Mahatma}
  et~al.}{2018}]{Mahatma18}
{Mahatma} V.~H.,  et~al., 2018, \mn@doi [\mnras] {10.1093/mnras/sty025}, \href
  {https://ui.adsabs.harvard.edu/abs/2018MNRAS.475.4557M} {475, 4557}

\bibitem[\protect\citeauthoryear{{Mahatma} et~al.,}{{Mahatma}
  et~al.}{2019}]{Mahatma19}
{Mahatma} V.~H.,  et~al., 2019, \mn@doi [\aap] {10.1051/0004-6361/201833973},
  \href {https://ui.adsabs.harvard.edu/abs/2019A&A...622A..13M} {622, A13}

\bibitem[\protect\citeauthoryear{{Mahatma}, {Hardcastle}, {Croston}, {Harwood},
  {Ineson}  \& {Moldon}}{{Mahatma} et~al.}{2020}]{Mahatma20}
{Mahatma} V.~H.,  {Hardcastle} M.~J.,  {Croston} J.~H.,  {Harwood} J.,
  {Ineson} J.,   {Moldon} J.,  2020, \mn@doi [\mnras] {10.1093/mnras/stz3396},
  \href {https://ui.adsabs.harvard.edu/abs/2020MNRAS.491.5015M} {491, 5015}

\bibitem[\protect\citeauthoryear{{Mohan} \& {Rafferty}}{{Mohan} \&
  {Rafferty}}{2015}]{Mohan15}
{Mohan} N.,  {Rafferty} D.,  2015, {PyBDSF: Python Blob Detection and Source
  Finder}, Astrophysics Source Code Library, record ascl:1502.007 (\mn@eprint
  {ascl} {1502.007})

\bibitem[\protect\citeauthoryear{{Morganti}}{{Morganti}}{2017}]{Morganti17}
{Morganti} R.,  2017, \mn@doi [Nature Astronomy] {10.1038/s41550-017-0223-0},
  \href {https://ui.adsabs.harvard.edu/abs/2017NatAs...1..596M} {1, 596}

\bibitem[\protect\citeauthoryear{{Morganti}}{{Morganti}}{2024}]{Morganti24}
{Morganti} R.,  2024, \mn@doi [Galaxies] {10.3390/galaxies12020011}, \href
  {https://ui.adsabs.harvard.edu/abs/2024Galax..12...11M} {12, 11}

\bibitem[\protect\citeauthoryear{{Mostert} et~al.,}{{Mostert}
  et~al.}{2023}]{Mostert23}
{Mostert} R. I.~J.,  et~al., 2023, \mn@doi [\aap]
  {10.1051/0004-6361/202346035}, \href
  {https://ui.adsabs.harvard.edu/abs/2023A&A...674A.208M} {674, A208}

\bibitem[\protect\citeauthoryear{{Murgia} et~al.,}{{Murgia}
  et~al.}{2011}]{Murgia11}
{Murgia} M.,  et~al., 2011, \mn@doi [\aap] {10.1051/0004-6361/201015302}, \href
  {https://ui.adsabs.harvard.edu/abs/2011A&A...526A.148M} {526, A148}

\bibitem[\protect\citeauthoryear{{Myers} \& {Spangler}}{{Myers} \&
  {Spangler}}{1985}]{Myers85}
{Myers} S.~T.,  {Spangler} S.~R.,  1985, \mn@doi [\apj] {10.1086/163040}, \href
  {https://ui.adsabs.harvard.edu/abs/1985ApJ...291...52M} {291, 52}

\bibitem[\protect\citeauthoryear{{Nandi} \& {Saikia}}{{Nandi} \&
  {Saikia}}{2012}]{Nandi12}
{Nandi} S.,  {Saikia} D.~J.,  2012, Bulletin of the Astronomical Society of
  India, \href {https://ui.adsabs.harvard.edu/abs/2012BASI...40..121N} {40,
  121}

\bibitem[\protect\citeauthoryear{{Nityananda}}{{Nityananda}}{2009}]{Nityananda09}
{Nityananda} R.,  2009, in {Saikia} D.~J.,  {Green} D.~A.,  {Gupta} Y.,
  {Venturi} T.,  eds,  Astronomical Society of the Pacific Conference Series
  Vol. 407, The Low-Frequency Radio Universe. p.~389

\bibitem[\protect\citeauthoryear{{Nyland} et~al.,}{{Nyland}
  et~al.}{2017}]{Nyland17}
{Nyland} K.,  et~al., 2017, \mn@doi [\apjs] {10.3847/1538-4365/aa6fed}, \href
  {https://ui.adsabs.harvard.edu/abs/2017ApJS..230....9N} {230, 9}

\bibitem[\protect\citeauthoryear{{Nyland}, {Lacy}, {Brandt}, {Yang}, {Ni},
  {Sajina}, {Zou}  \& {Vaccari}}{{Nyland} et~al.}{2023}]{Nyland23}
{Nyland} K.,  {Lacy} M.,  {Brandt} W.~N.,  {Yang} G.,  {Ni} Q.,  {Sajina} A.,
  {Zou} F.,   {Vaccari} M.,  2023, \mn@doi [Research Notes of the American
  Astronomical Society] {10.3847/2515-5172/acbc72}, \href
  {https://ui.adsabs.harvard.edu/abs/2023RNAAS...7...33N} {7, 33}

\bibitem[\protect\citeauthoryear{{O'Dea}, {Daly}, {Kharb}, {Freeman}  \&
  {Baum}}{{O'Dea} et~al.}{2009}]{Odea09}
{O'Dea} C.~P.,  {Daly} R.~A.,  {Kharb} P.,  {Freeman} K.~A.,   {Baum} S.~A.,
  2009, \mn@doi [\aap] {10.1051/0004-6361:200809416}, \href
  {https://ui.adsabs.harvard.edu/abs/2009A&A...494..471O} {494, 471}

\bibitem[\protect\citeauthoryear{{Oei} et~al.,}{{Oei} et~al.}{2022}]{Oei22}
{Oei} M. S.~S.~L.,  et~al., 2022, \mn@doi [\aap] {10.1051/0004-6361/202142778},
  \href {https://ui.adsabs.harvard.edu/abs/2022A&A...660A...2O} {660, A2}

\bibitem[\protect\citeauthoryear{{Parma}, {Murgia}, {de Ruiter}, {Fanti},
  {Mack}  \& {Govoni}}{{Parma} et~al.}{2007}]{Parma07}
{Parma} P.,  {Murgia} M.,  {de Ruiter} H.~R.,  {Fanti} R.,  {Mack} K.~H.,
  {Govoni} F.,  2007, \mn@doi [\aap] {10.1051/0004-6361:20077592}, \href
  {https://ui.adsabs.harvard.edu/abs/2007A&A...470..875P} {470, 875}

\bibitem[\protect\citeauthoryear{{Pinjarkar}, {Hardcastle}, {Harwood}, {Lal},
  {Hatfield}, {Jarvis}, {Randriamanakoto}  \& {Whittam}}{{Pinjarkar}
  et~al.}{2023}]{Pinjarkar23}
{Pinjarkar} S.,  {Hardcastle} M.~J.,  {Harwood} J.~J.,  {Lal} D.~V.,
  {Hatfield} P.~W.,  {Jarvis} M.~J.,  {Randriamanakoto} Z.,   {Whittam} I.~H.,
  2023, \mn@doi [\mnras] {10.1093/mnras/stad1432}, \href
  {https://ui.adsabs.harvard.edu/abs/2023MNRAS.523..620P} {523, 620}

\bibitem[\protect\citeauthoryear{{Quici} et~al.,}{{Quici}
  et~al.}{2021}]{Quici21}
{Quici} B.,  et~al., 2021, \mn@doi [\pasa] {10.1017/pasa.2020.49}, \href
  {https://ui.adsabs.harvard.edu/abs/2021PASA...38....8Q} {38, e008}

\bibitem[\protect\citeauthoryear{{Quici}, {Turner}, {Seymour}  \&
  {Hurley-Walker}}{{Quici} et~al.}{2025}]{Quici25}
{Quici} B.,  {Turner} R.~J.,  {Seymour} N.,   {Hurley-Walker} N.,  2025,
  \mn@doi [\mnras] {10.1093/mnras/staf024}, \href
  {https://ui.adsabs.harvard.edu/abs/2025MNRAS.537..343Q} {537, 343}

\bibitem[\protect\citeauthoryear{{Randriamanakoto}, {Ishwara-Chandra}  \&
  {Taylor}}{{Randriamanakoto} et~al.}{2020}]{Randriamanakoto20}
{Randriamanakoto} Z.,  {Ishwara-Chandra} C.~H.,   {Taylor} A.~R.,  2020,
  \mn@doi [\mnras] {10.1093/mnras/staa1782}, \href
  {https://ui.adsabs.harvard.edu/abs/2020MNRAS.496.3381R} {496, 3381}

\bibitem[\protect\citeauthoryear{{Rossi}, {Bodo}, {Capetti}  \&
  {Massaglia}}{{Rossi} et~al.}{2017}]{Rossi17}
{Rossi} P.,  {Bodo} G.,  {Capetti} A.,   {Massaglia} S.,  2017, \mn@doi [\aap]
  {10.1051/0004-6361/201730594}, \href
  {https://ui.adsabs.harvard.edu/abs/2017A&A...606A..57R} {606, A57}

\bibitem[\protect\citeauthoryear{{Saikia}}{{Saikia}}{2022}]{Saikia22}
{Saikia} D.~J.,  2022, \mn@doi [Journal of Astrophysics and Astronomy]
  {10.1007/s12036-022-09863-2}, \href
  {https://ui.adsabs.harvard.edu/abs/2022JApA...43...97S} {43, 97}

\bibitem[\protect\citeauthoryear{{Saikia} \& {Jamrozy}}{{Saikia} \&
  {Jamrozy}}{2009}]{Saikia09}
{Saikia} D.~J.,  {Jamrozy} M.,  2009, Bulletin of the Astronomical Society of
  India, \href {https://ui.adsabs.harvard.edu/abs/2009BASI...37...63S} {37, 63}

\bibitem[\protect\citeauthoryear{{Saripalli}}{{Saripalli}}{2012}]{Saripalli12}
{Saripalli} L.,  2012, \mn@doi [\aj] {10.1088/0004-6256/144/3/85}, \href
  {https://ui.adsabs.harvard.edu/abs/2012AJ....144...85S} {144, 85}

\bibitem[\protect\citeauthoryear{{Schoenmakers}, {de Bruyn}, {R{\"o}ttgering}
  \& {van der Laan}}{{Schoenmakers} et~al.}{2000}]{Schoenmakers00}
{Schoenmakers} A.~P.,  {de Bruyn} A.~G.,  {R{\"o}ttgering} H.~J.~A.,   {van der
  Laan} H.,  2000, \mn@doi [\mnras] {10.1046/j.1365-8711.2000.03432.x}, \href
  {https://ui.adsabs.harvard.edu/abs/2000MNRAS.315..395S} {315, 395}

\bibitem[\protect\citeauthoryear{{Sebastian}, {Ishwara-Chandra}, {Joshi}  \&
  {Wadadekar}}{{Sebastian} et~al.}{2018}]{Sebastian18}
{Sebastian} B.,  {Ishwara-Chandra} C.~H.,  {Joshi} R.,   {Wadadekar} Y.,  2018,
  \mn@doi [\mnras] {10.1093/mnras/stx2631}, \href
  {https://ui.adsabs.harvard.edu/abs/2018MNRAS.473.4926S} {473, 4926}

\bibitem[\protect\citeauthoryear{{Sethi}, {Jamrozy}  \& {Ku{\'z}micz}}{{Sethi}
  et~al.}{2022}]{Sethi22}
{Sethi} S.,  {Jamrozy} M.,   {Ku{\'z}micz} A.,  2022, in EAS2022, European
  Astronomical Society Annual Meeting. p.~2173

\bibitem[\protect\citeauthoryear{{Sethi}, {Ku{\'z}micz}, {Hunik}  \&
  {Jamrozy}}{{Sethi} et~al.}{2025}]{Sethi25}
{Sethi} S.,  {Ku{\'z}micz} A.,  {Hunik} D.,   {Jamrozy} M.,  2025, \mn@doi
  [\aap] {10.1051/0004-6361/202554987}, \href
  {https://ui.adsabs.harvard.edu/abs/2025A&A...699L...4S} {699, L4}

\bibitem[\protect\citeauthoryear{{Shulevski} et~al.,}{{Shulevski}
  et~al.}{2017}]{Shulevski17}
{Shulevski} A.,  et~al., 2017, \mn@doi [\aap] {10.1051/0004-6361/201630008},
  \href {https://ui.adsabs.harvard.edu/abs/2017A&A...600A..65S} {600, A65}

\bibitem[\protect\citeauthoryear{{Shulevski} et~al.,}{{Shulevski}
  et~al.}{2024}]{Shulevski24}
{Shulevski} A.,  et~al., 2024, \mn@doi [\aap] {10.1051/0004-6361/202346824},
  \href {https://ui.adsabs.harvard.edu/abs/2024A&A...682A.171S} {682, A171}

\bibitem[\protect\citeauthoryear{{Singh} et~al.,}{{Singh}
  et~al.}{2014}]{Singh14}
{Singh} V.,  et~al., 2014, \mn@doi [\aap] {10.1051/0004-6361/201423644}, \href
  {https://ui.adsabs.harvard.edu/abs/2014A&A...569A..52S} {569, A52}

\bibitem[\protect\citeauthoryear{{Singh}, {Dutta}, {Wadadekar}  \&
  {Ishwara-Chandra}}{{Singh} et~al.}{2021}]{Singh21}
{Singh} V.,  {Dutta} S.,  {Wadadekar} Y.,   {Ishwara-Chandra} C.~H.,  2021,
  \mn@doi [Galaxies] {10.3390/galaxies9040121}, \href
  {https://ui.adsabs.harvard.edu/abs/2021Galax...9..121S} {9, 121}

\bibitem[\protect\citeauthoryear{{Slee}, {Roy}, {Murgia}, {Andernach}  \&
  {Ehle}}{{Slee} et~al.}{2001}]{Slee01}
{Slee} O.~B.,  {Roy} A.~L.,  {Murgia} M.,  {Andernach} H.,   {Ehle} M.,  2001,
  \mn@doi [\aj] {10.1086/322105}, \href
  {https://ui.adsabs.harvard.edu/abs/2001AJ....122.1172S} {122, 1172}

\bibitem[\protect\citeauthoryear{{Stewart}, {Shabala}, {Turner}, {Yates-Jones},
  {Krause}, {Wong}, {Power}  \& {Hardcastle}}{{Stewart}
  et~al.}{2025}]{Stewart25}
{Stewart} G. S.~C.,  {Shabala} S.~S.,  {Turner} R.~J.,  {Yates-Jones} P.~M.,
  {Krause} M. G.~H.,  {Wong} O.~I.,  {Power} C.,   {Hardcastle} M.~J.,  2025,
  \mn@doi [arXiv e-prints] {10.48550/arXiv.2511.01193}, \href
  {https://ui.adsabs.harvard.edu/abs/2025arXiv251101193S} {p. arXiv:2511.01193}

\bibitem[\protect\citeauthoryear{{Swarup}}{{Swarup}}{1991}]{Swarup91}
{Swarup} G.,  1991, in {Cornwell} T.~J.,  {Perley} R.~A.,  eds,  Astronomical
  Society of the Pacific Conference Series Vol. 19, IAU Colloq. 131: Radio
  Interferometry. Theory, Techniques, and Applications. pp 376--380

\bibitem[\protect\citeauthoryear{{Tamhane}, {Wadadekar}, {Basu}, {Singh},
  {Ishwara-Chandra}, {Beelen}  \& {Sirothia}}{{Tamhane}
  et~al.}{2015}]{Tamhane15}
{Tamhane} P.,  {Wadadekar} Y.,  {Basu} A.,  {Singh} V.,  {Ishwara-Chandra}
  C.~H.,  {Beelen} A.,   {Sirothia} S.,  2015, \mn@doi [\mnras]
  {10.1093/mnras/stv1768}, \href
  {https://ui.adsabs.harvard.edu/abs/2015MNRAS.453.2438T} {453, 2438}

\bibitem[\protect\citeauthoryear{Taylor}{Taylor}{2019}]{Taylor19}
Taylor A.~R.,  2019, in 2019 URSI Asia-Pacific Radio Science Conference
  (AP-RASC). pp~1--1, \mn@doi{10.23919/URSIAP-RASC.2019.8738702}

\bibitem[\protect\citeauthoryear{{Tribble}}{{Tribble}}{1993}]{Tribble93}
{Tribble} P.~C.,  1993, \mn@doi [\mnras] {10.1093/mnras/261.1.57}, \href
  {https://ui.adsabs.harvard.edu/abs/1993MNRAS.261...57T} {261, 57}

\bibitem[\protect\citeauthoryear{{Turner}}{{Turner}}{2018}]{Turner18}
{Turner} R.~J.,  2018, \mn@doi [\mnras] {10.1093/mnras/sty433}, \href
  {https://ui.adsabs.harvard.edu/abs/2018MNRAS.476.2522T} {476, 2522}

\bibitem[\protect\citeauthoryear{{Vaccari}}{{Vaccari}}{2015}]{Vaccari15}
{Vaccari} M.,  2015, in The Many Facets of Extragalactic Radio Surveys: Towards
  New Scientific Challenges. p.~27 (\mn@eprint {arXiv} {1604.02353}),
  \mn@doi{https://pos.sissa.it/267/027}

\bibitem[\protect\citeauthoryear{Vaccari}{Vaccari}{2023}]{Vaccari23}
Vaccari M.,  2023, The Spitzer Data Fusion, \mn@doi{10.5281/zenodo.8192777},
  \url {https://doi.org/10.5281/zenodo.8192777}

\bibitem[\protect\citeauthoryear{{Vaccari}}{{Vaccari}}{2026}]{Vaccari26}
{Vaccari} M.,  2026, \mn@doi [Research Notes of the American Astronomical
  Society] {10.3847/2515-5172/ae6a9a}, \href
  {https://ui.adsabs.harvard.edu/abs/2026RNAAS..10..118V} {10, 118}

\bibitem[\protect\citeauthoryear{{Wolnik}, {Jurusik}  \& {Jamrozy}}{{Wolnik}
  et~al.}{2024}]{Wolnik24}
{Wolnik} K.,  {Jurusik} W.,   {Jamrozy} M.,  2024, \mn@doi [\aap]
  {10.1051/0004-6361/202450897}, \href
  {https://ui.adsabs.harvard.edu/abs/2024A&A...691A..76W} {691, A76}

\makeatother
\end{thebibliography}




 \appendix
 
\section{Spectral Index Maps of Sample Sources} 
 
\begin{figure*}
	\centering
	\includegraphics[scale=0.035]{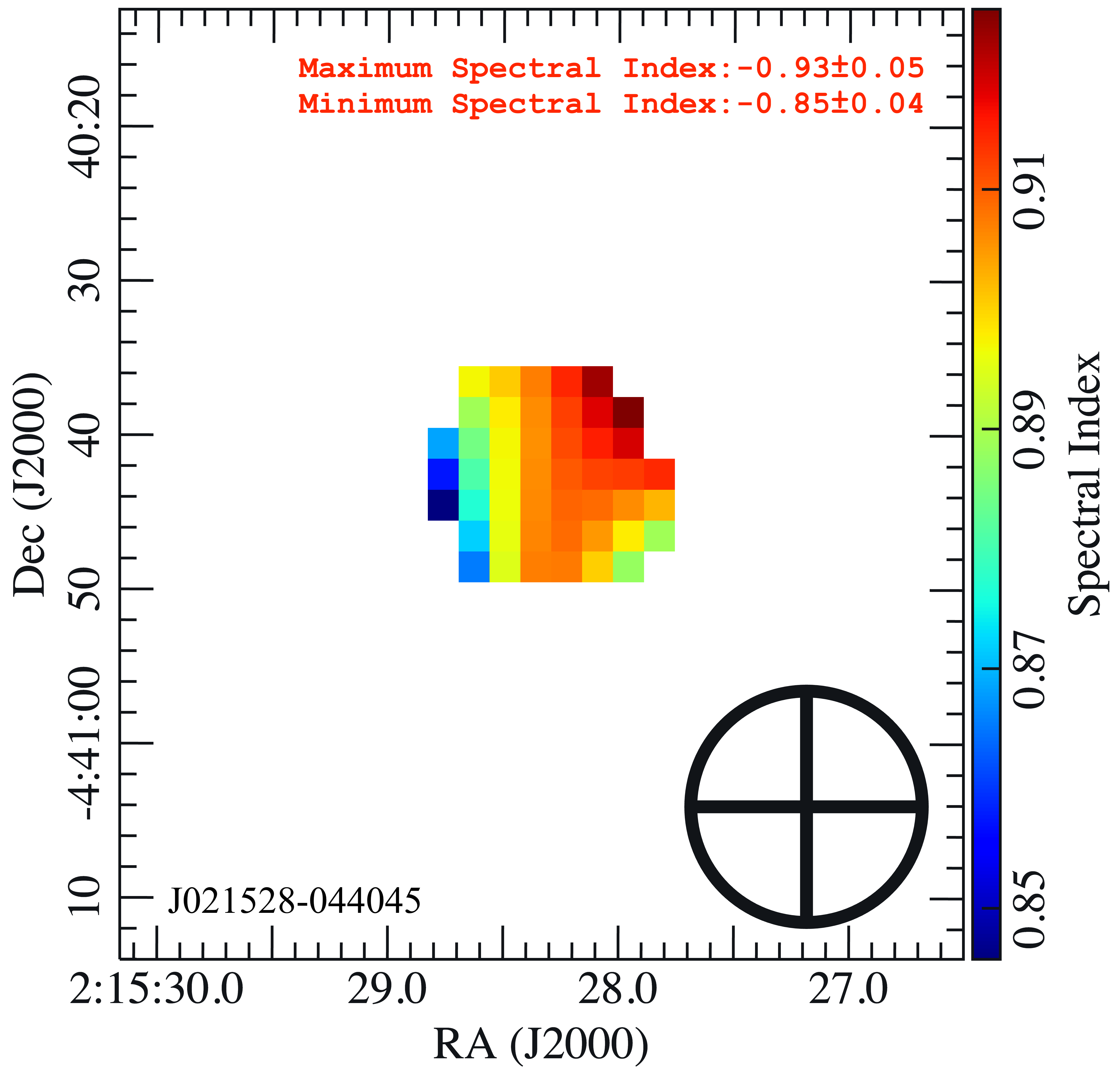}
	\includegraphics[scale=0.035]{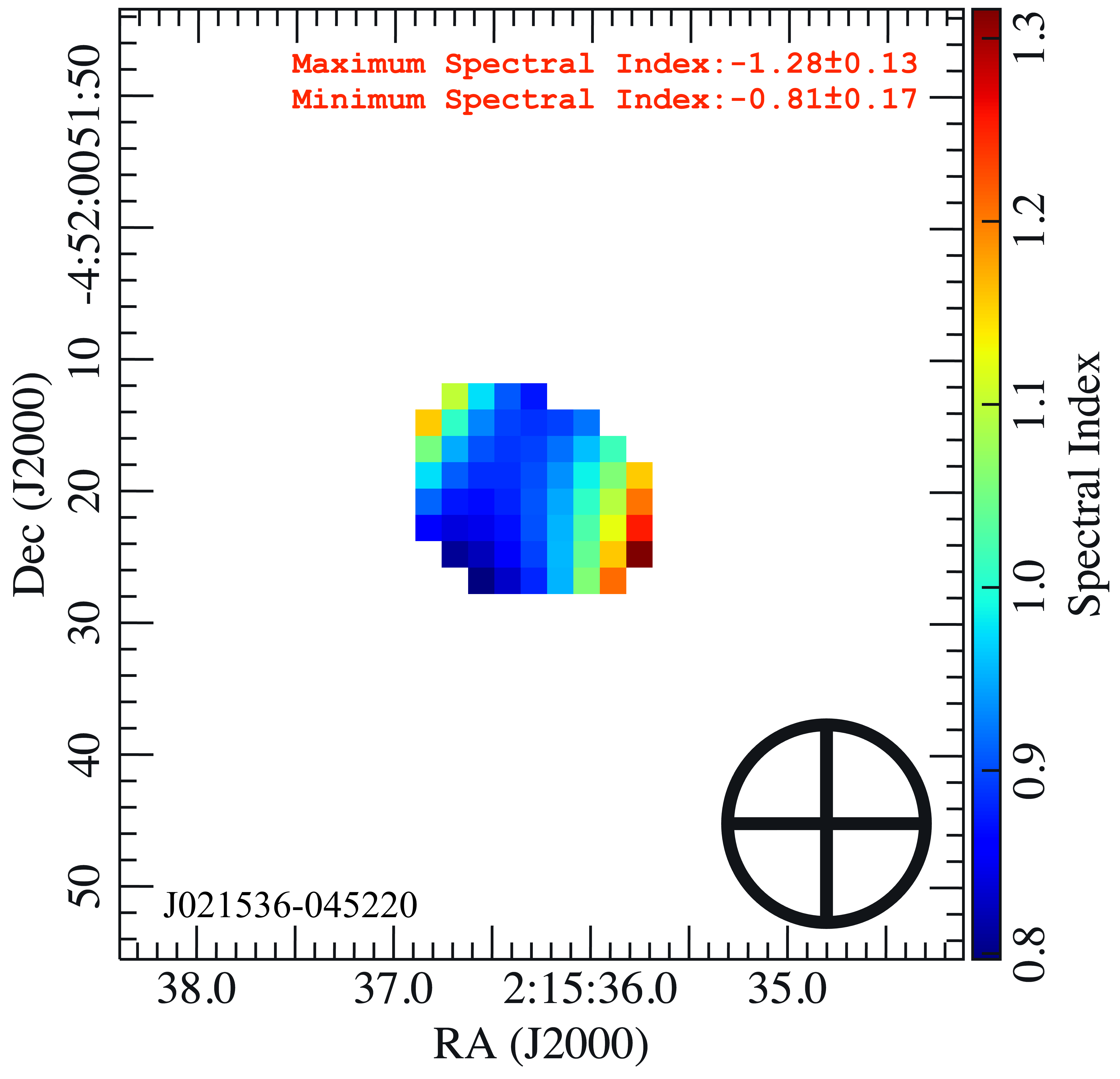}
	\includegraphics[scale=0.035]{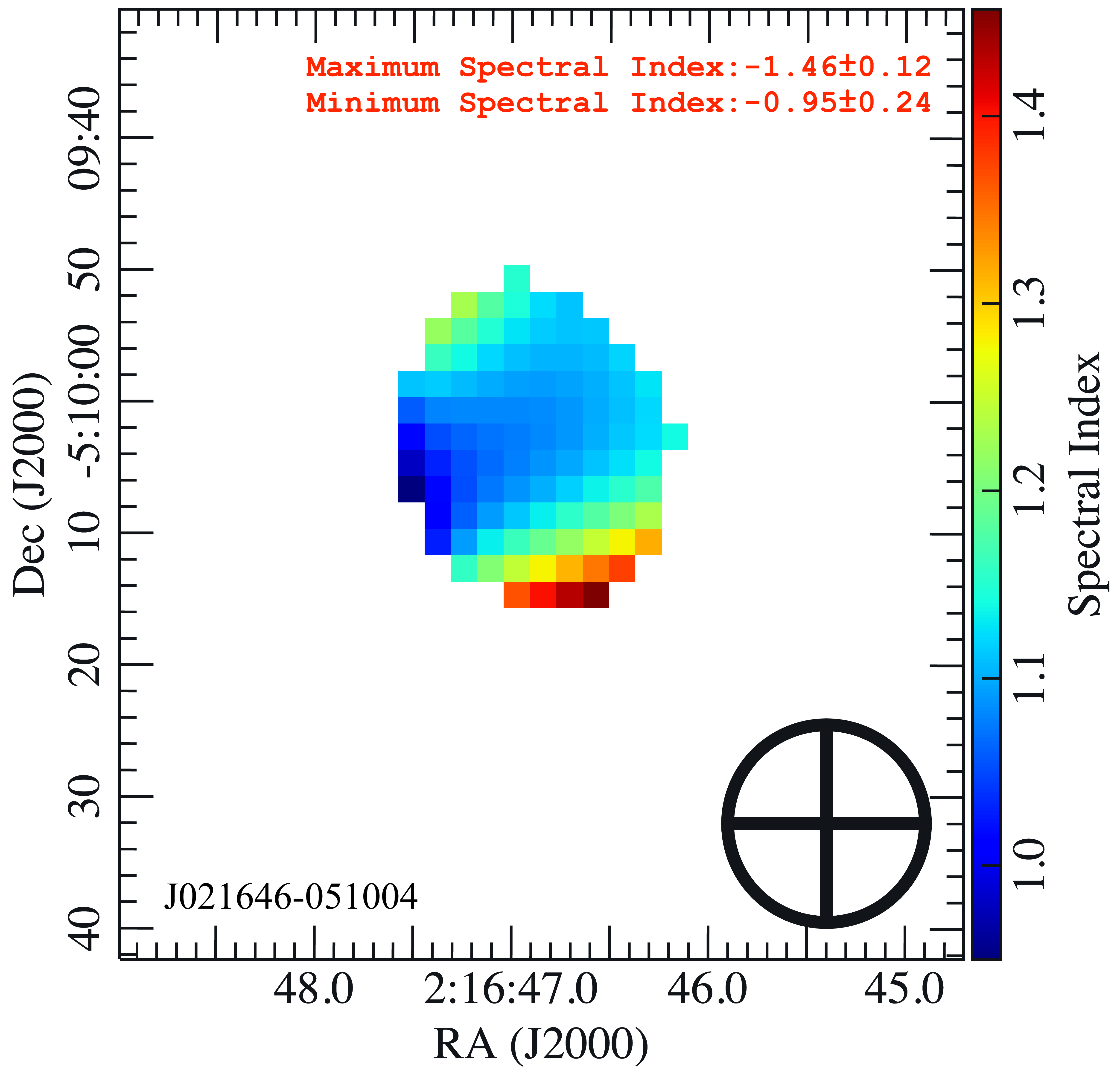}
	\includegraphics[scale=0.068]{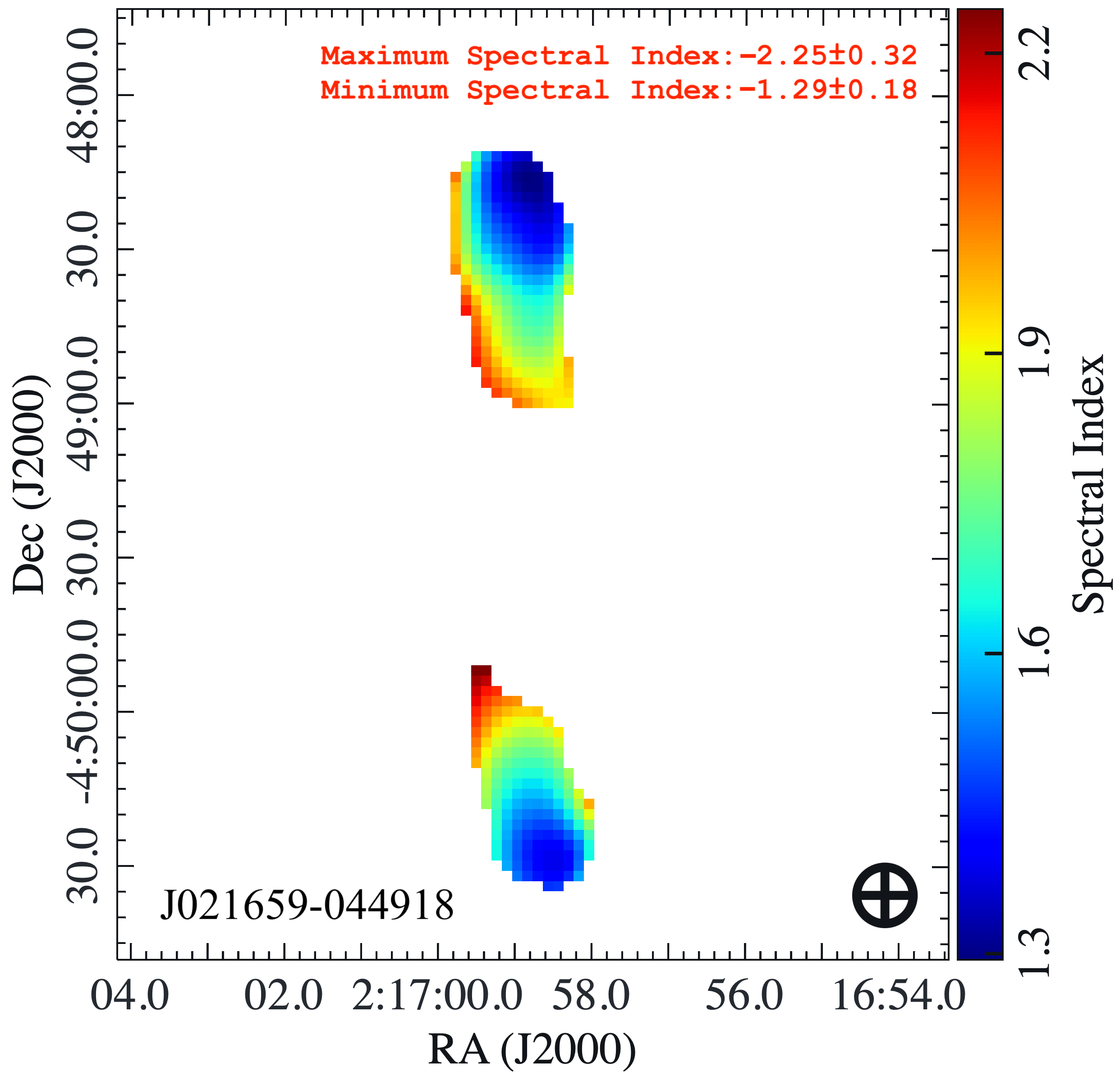}
	\includegraphics[scale=0.070]{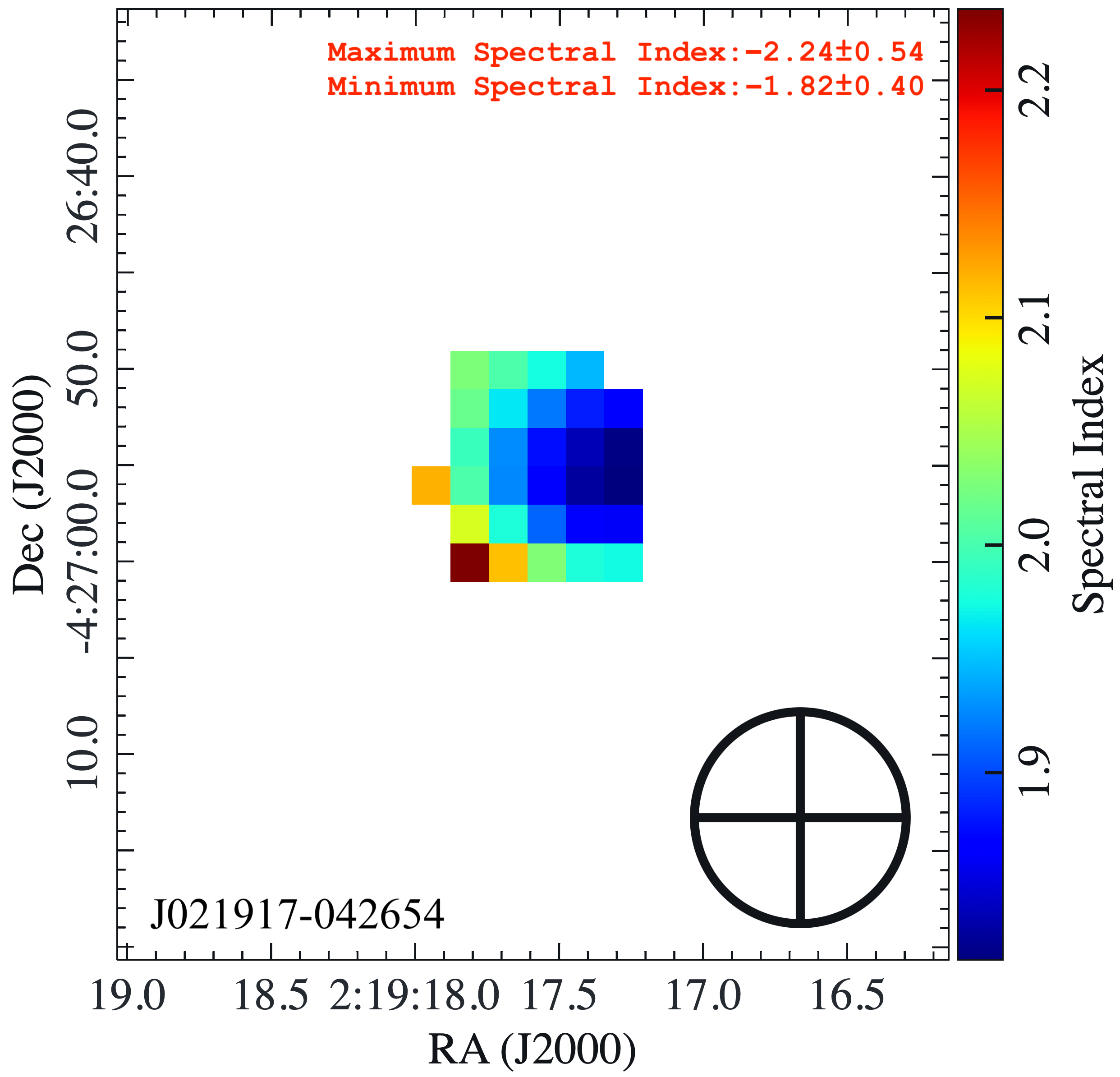}
	\includegraphics[scale=0.068]{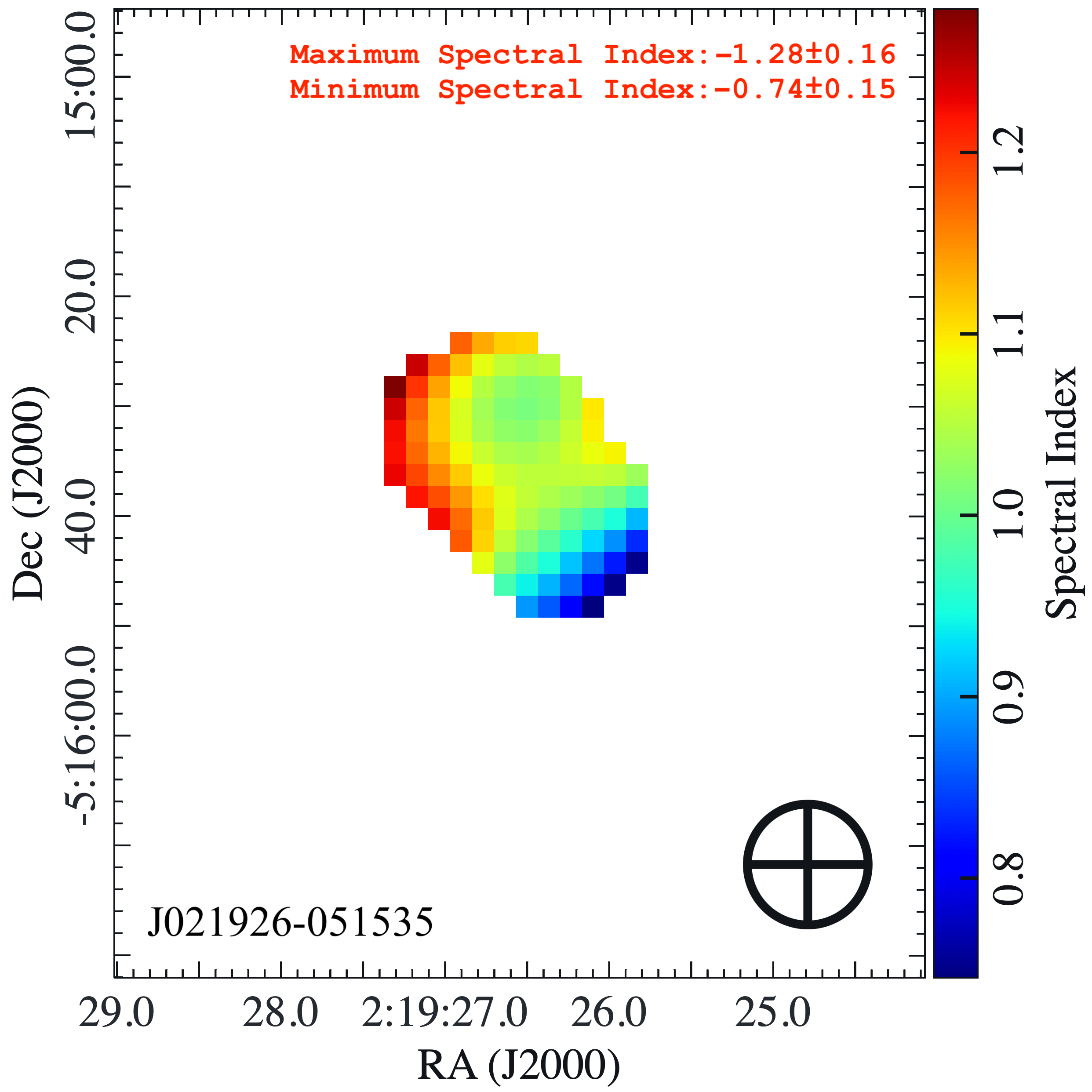}
	\caption{Spectral index maps for our remnant sources. The circle at the right bottom represents a circular PSF beam of 15$^{\prime\prime}$.}
	\label{fig:SpecIndex1}
\end{figure*}

\addtocounter{figure}{-1}

\begin{figure*}
	\centering
	\includegraphics[scale=0.068]{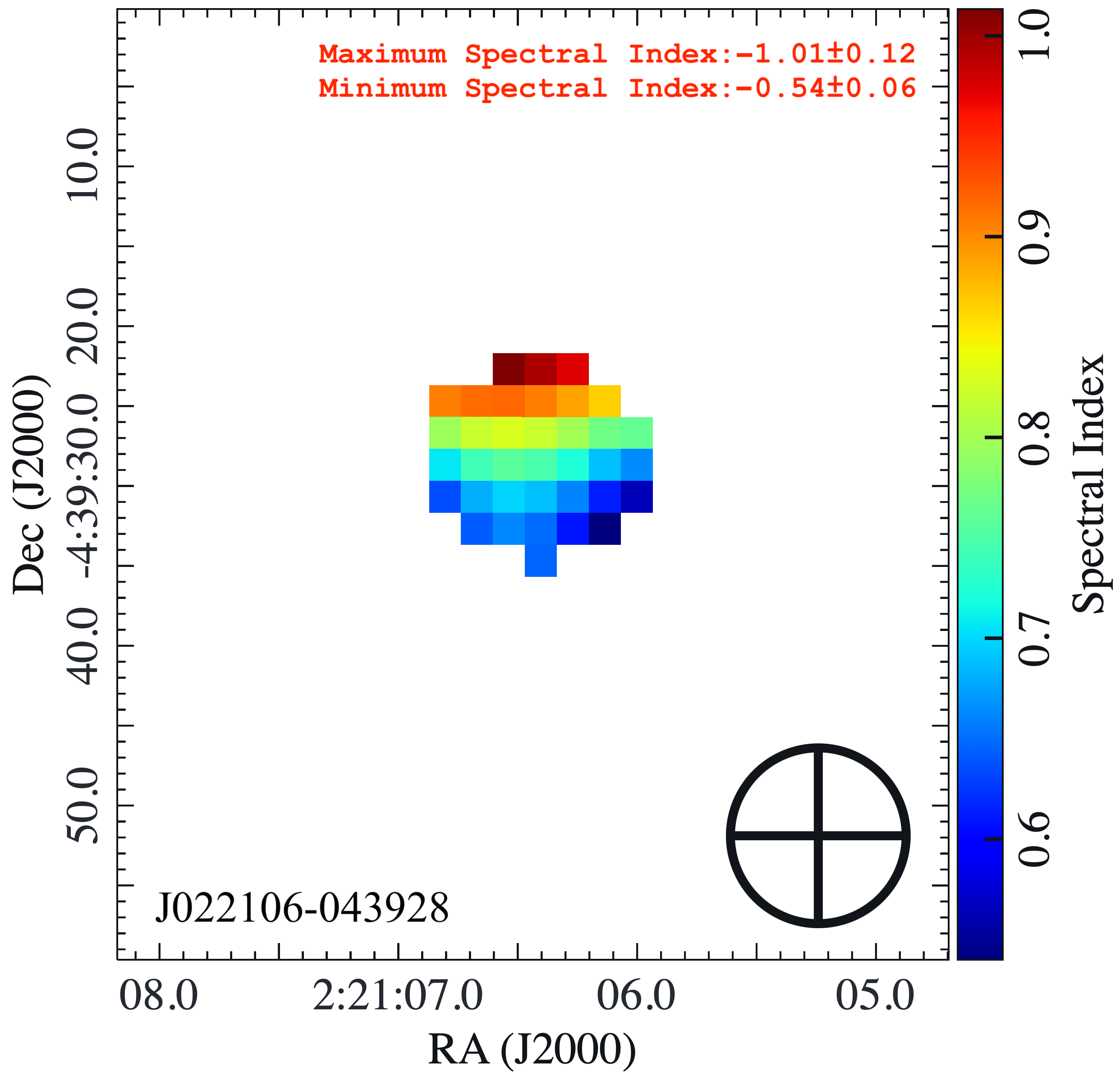}
	\includegraphics[scale=0.068]{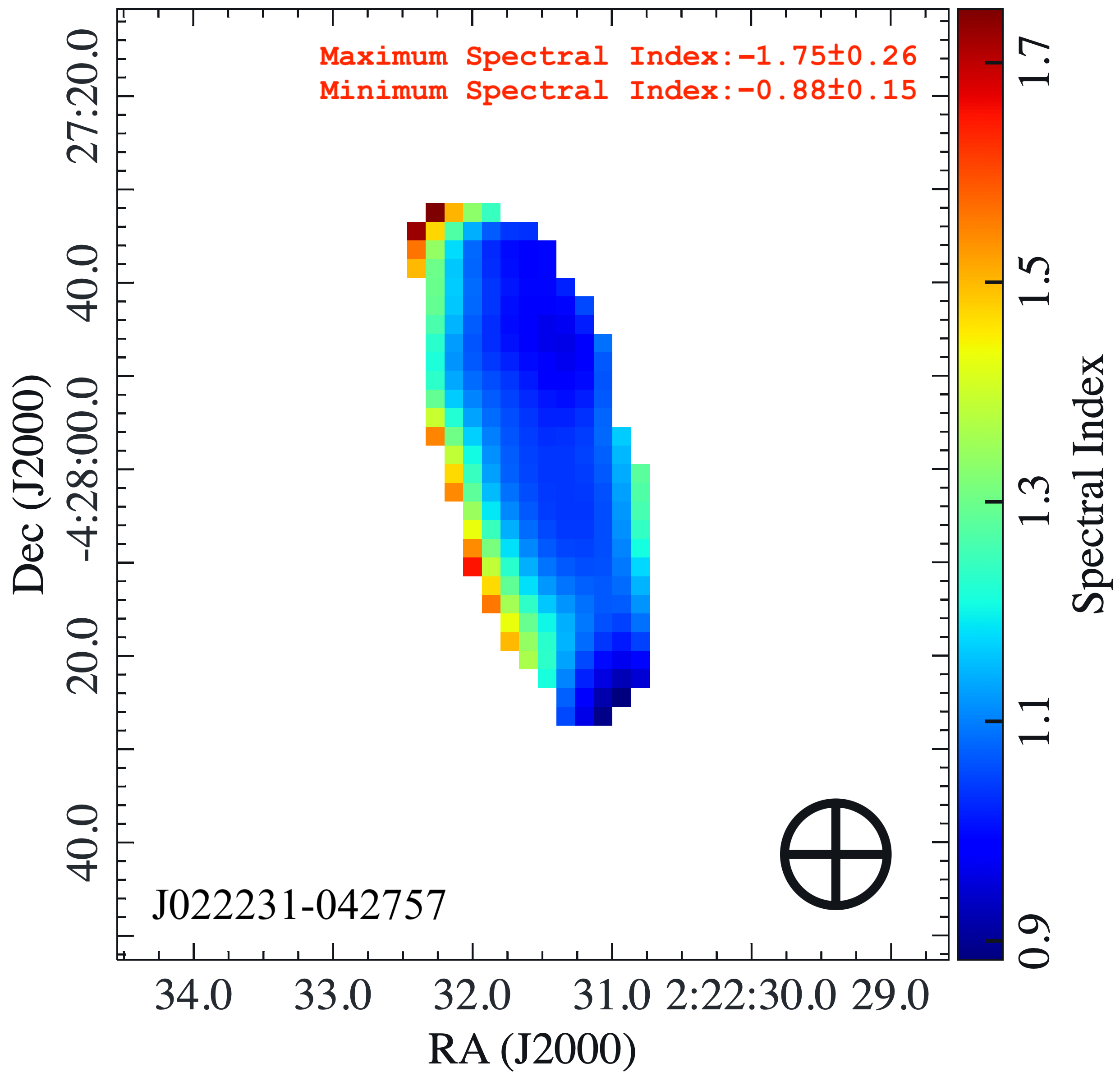}
	\includegraphics[scale=0.068]{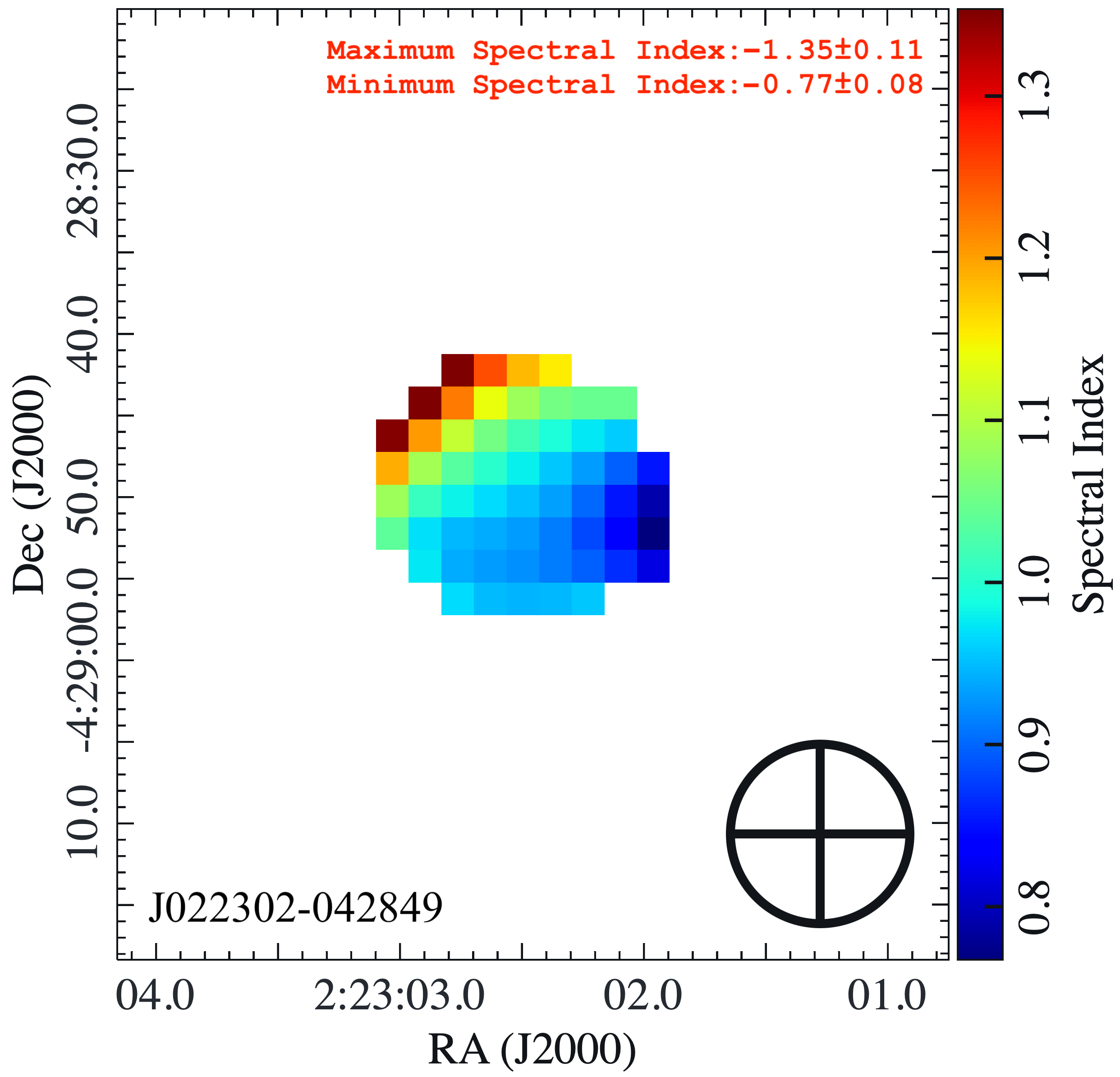}
	\includegraphics[scale=0.068]{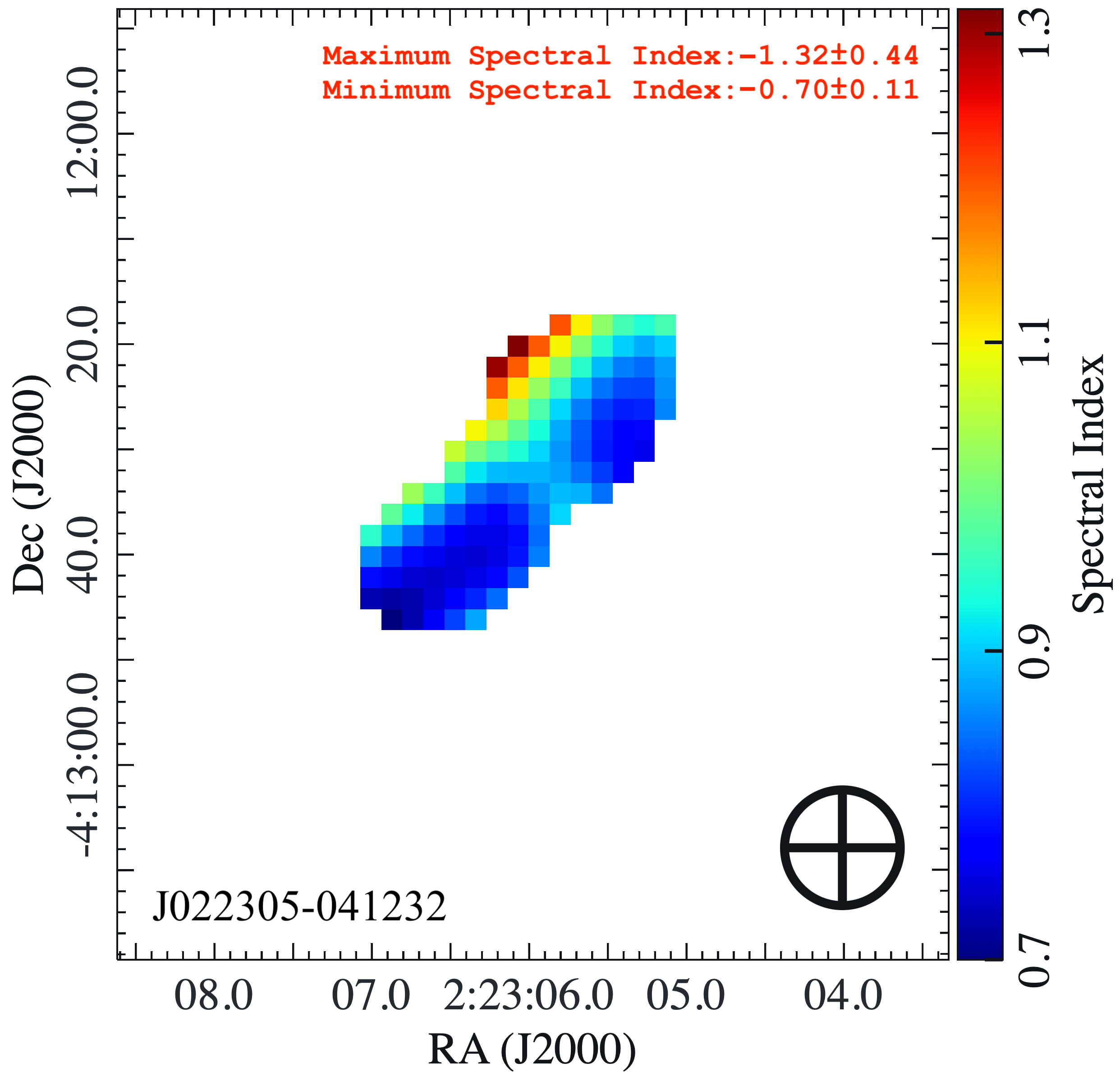}
	\includegraphics[scale=0.068]{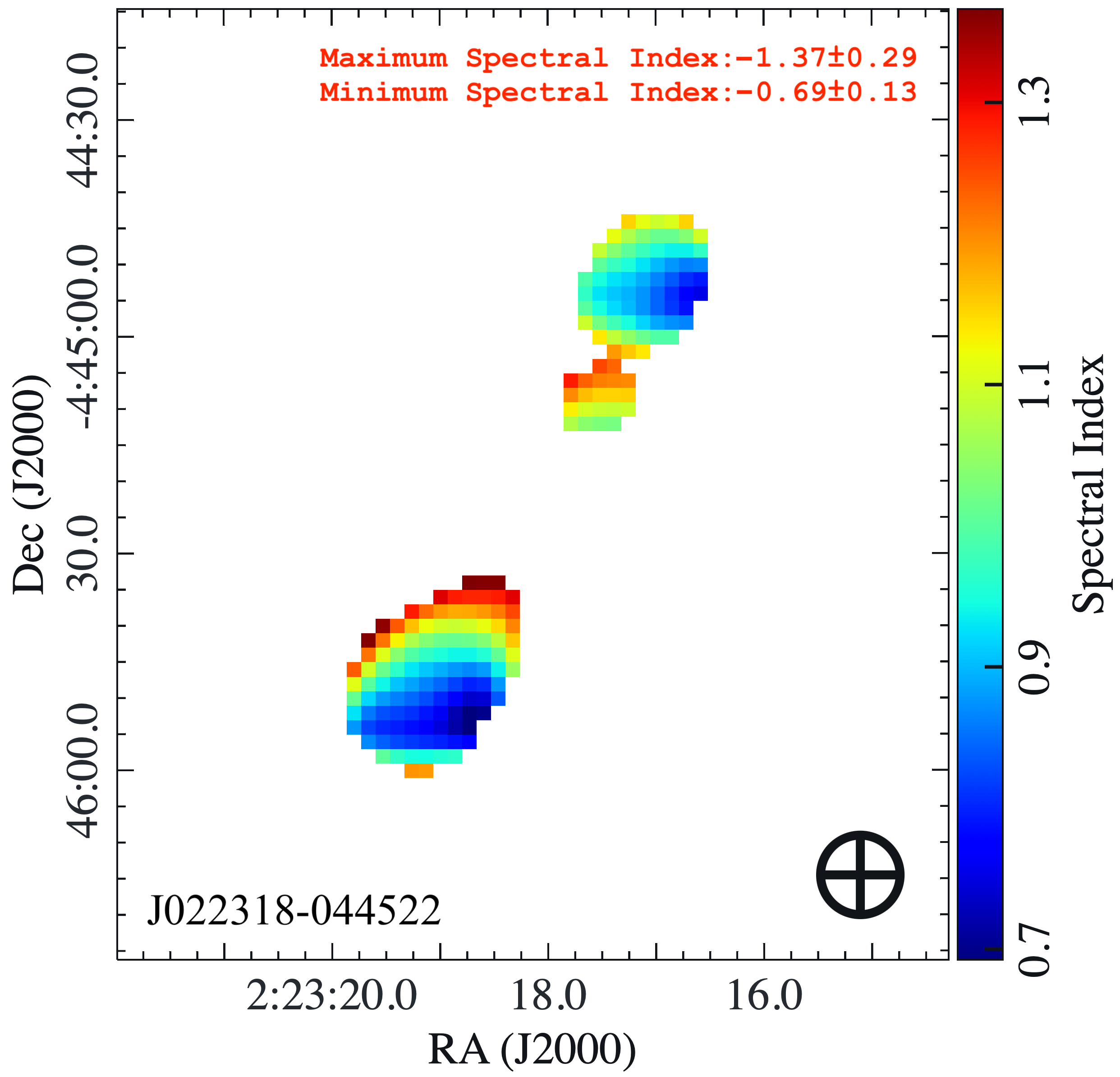}
	\includegraphics[scale=0.068]{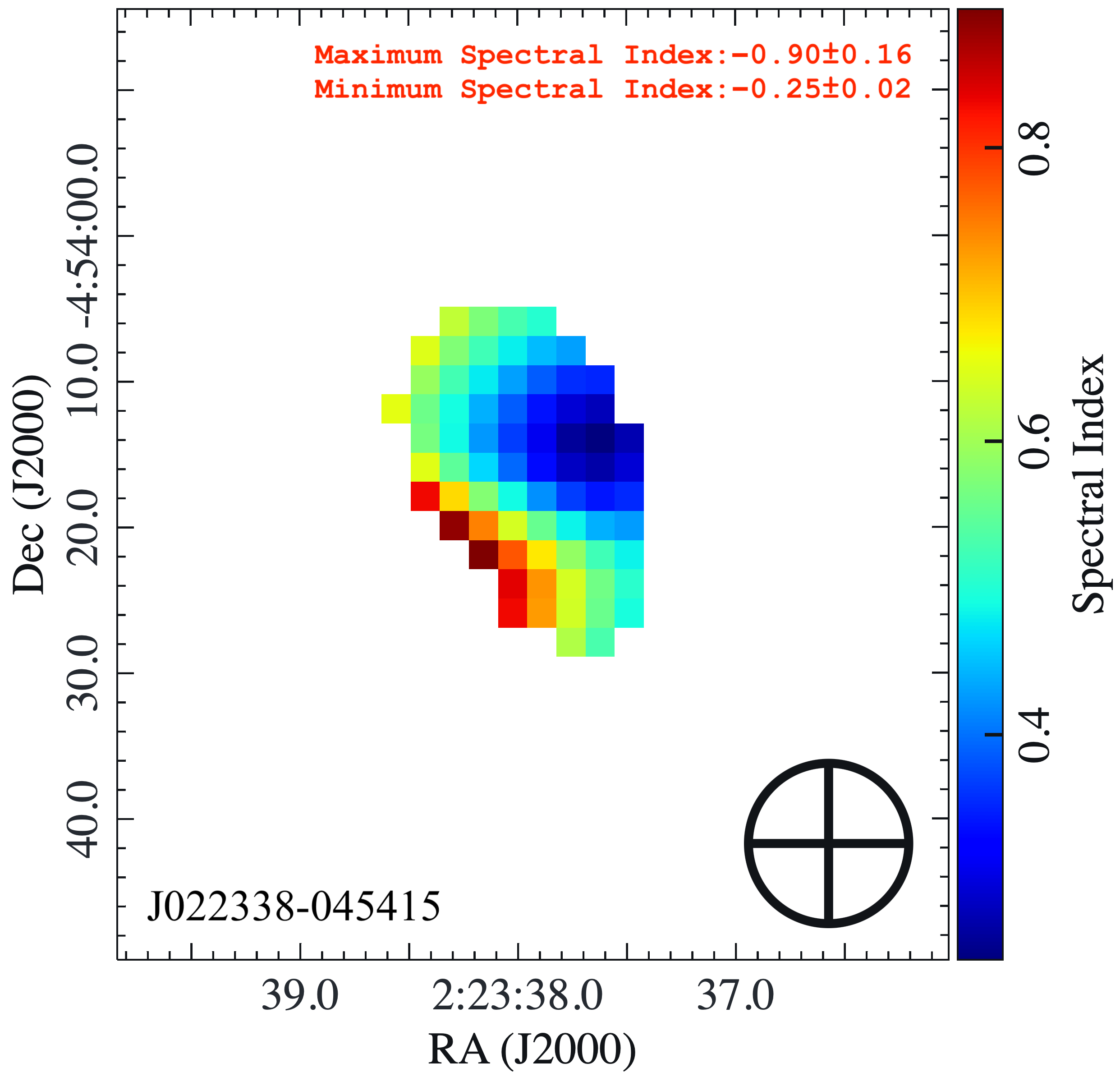}
	\includegraphics[scale=0.068]{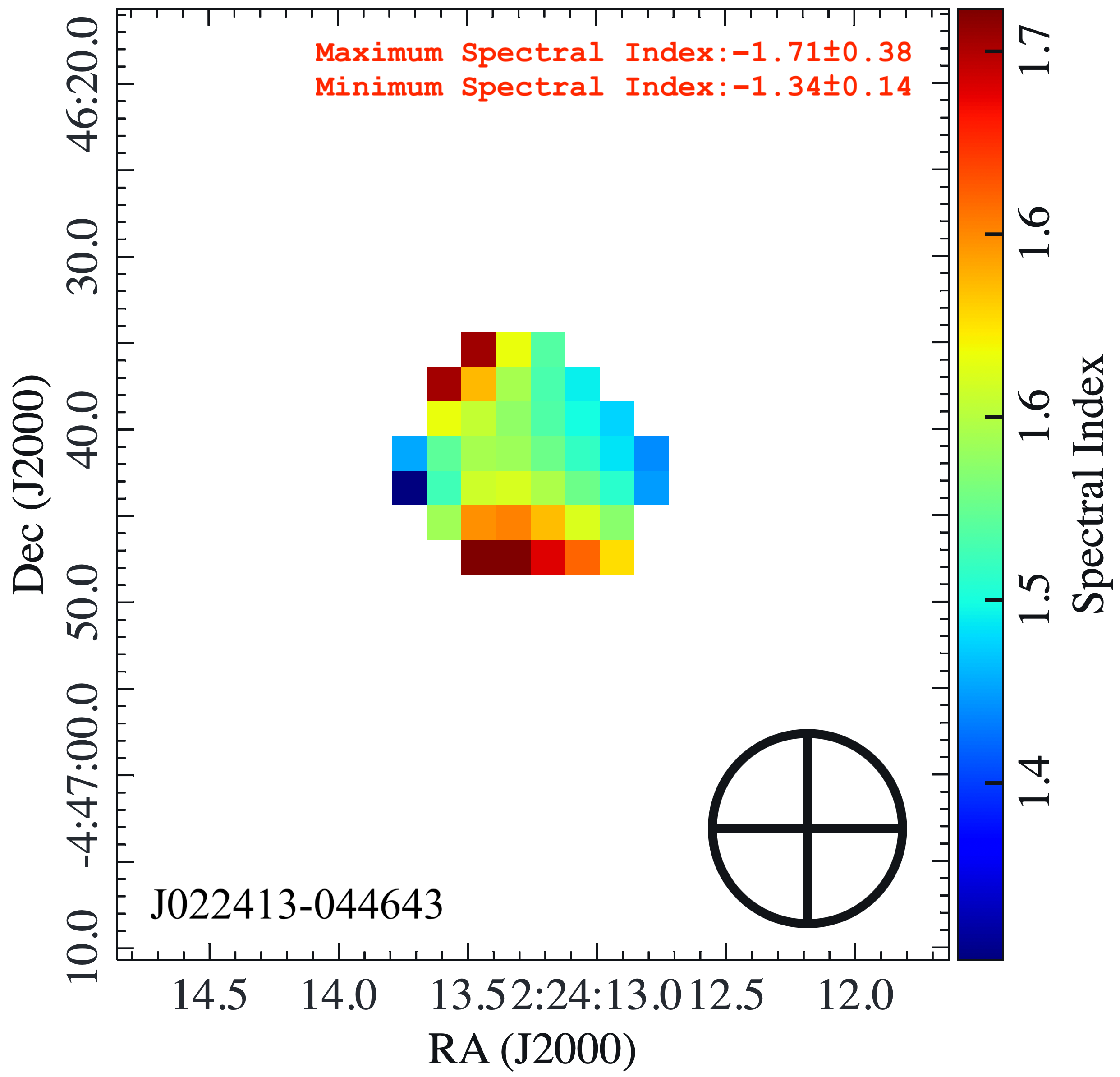}
	\includegraphics[scale=0.035]{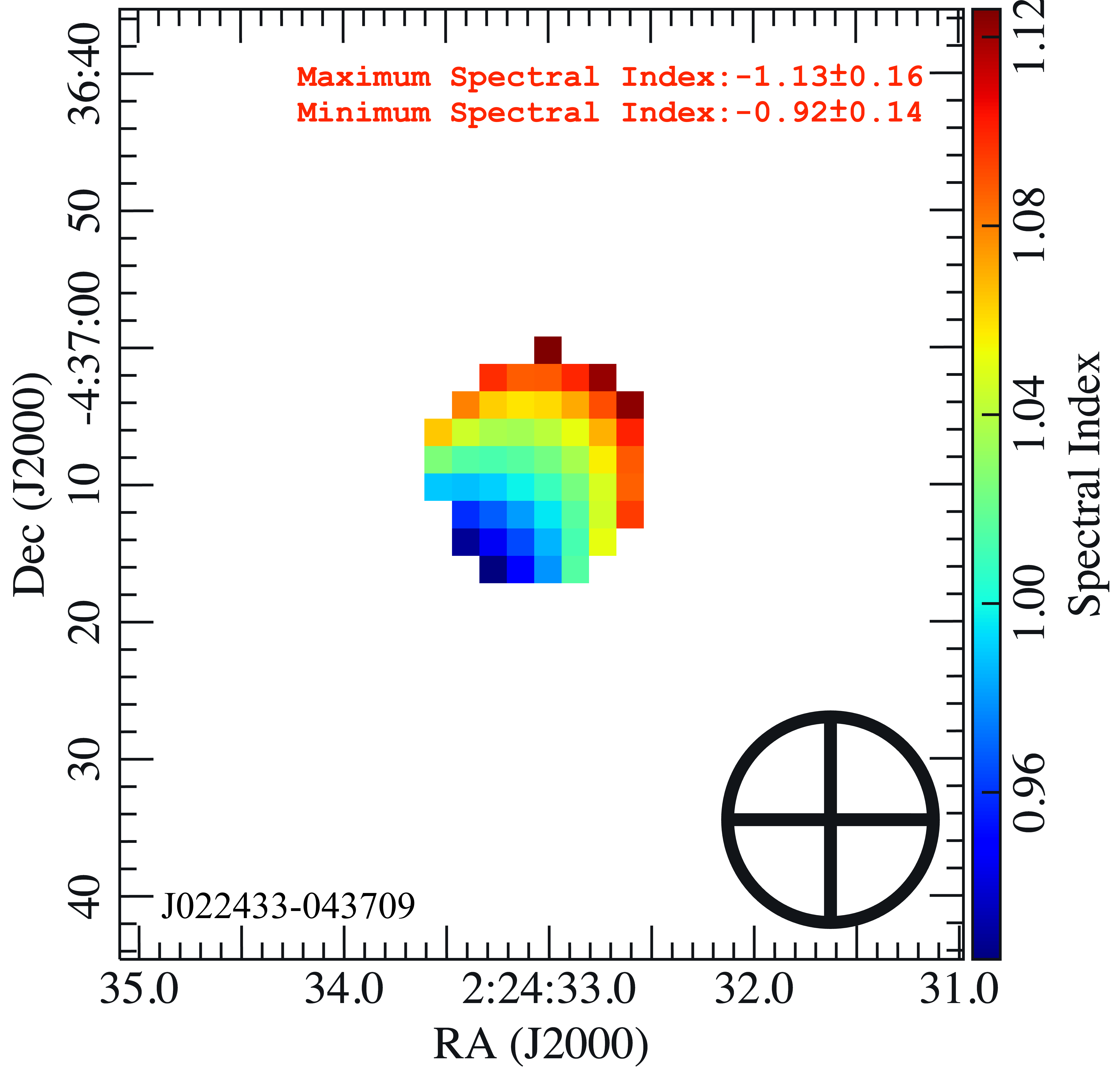}
	\label{fig:SpecIndex2}
\end{figure*} 
\label{lastpage}
\end{document}